\documentclass[a4paper,10pt]{article}

\usepackage[cp1252]{inputenc}
\usepackage[english]{babel}
\usepackage{amsmath,amsfonts,amssymb,amsthm,amstext}
\usepackage{longtable,cite,url,algo,graphicx}
\usepackage{color,relsize}

\usepackage[compact]{titlesec}
\titlespacing{\section}{0pt}{10pt}{0pt}
\titlespacing{\subsection}{0pt}{10pt}{0pt}
\titlespacing{\subsubsection}{0pt}{5pt}{0pt}
\titlespacing{\paragraph}{0pt}{5pt}{0pt}

\newcommand\mysectiona[2]{%
\section
[#1]
{#2}
}

\newcommand\mysectionb[2]{%
\subsection
[#1]
{#2}
}

\newcommand\mysectionc[2]{%
\subsubsection
[#1]
{#2}
}

\newcommand\mysectiond[2]{%
\paragraph
[#1]
{#2}
}

\newcommand{\notion}[1]{{\bf #1}}
\newcommand{\notioninform}[1]{\text{\bf #1}}
\newcommand{\nullmatrix}{\mathsf{O}}
\newcommand{\bigo}{\mathsf{O}}

\newcommand{\overallopera}  {$\bigo\left(2^{C\sigma} t(n)^{68}\right)$}
\newcommand{\overallcompla} {$\bigo\left(2^{C\sigma} t(n)^{136}\right)$}
\newcommand{\overallcomplb} {$\bigo\left(2^{C\sigma} t(n)^{272}\right)$}
\newcommand{\overallcomplc} {$\bigo\left(2^{C\sigma} t(n)^{137}\right)$}

\DeclareMathOperator{\abs}{abs}

\newtheorem{notation}{Notation}[section]
\newtheorem{definition}{Definition}[section]
\newtheorem{proposition}{Proposition}[section]
\newtheorem{lemma}{Lemma}[section]
\newtheorem{theorem}{Theorem}[section]

\newtheorem*{notation*}{Notation}
\newtheorem*{definition*}{Definition}
\newtheorem*{proposition*}{Proposition}
\newtheorem*{lemma*}{Lemma}
\newtheorem*{theorem*}{Theorem}
\newtheorem*{corollary*}{Corollary}

\begin{document}

\def\arxiv{t}


\title{
\notion{P}$\,=\,$\notion{NP}}
\author{
Sergey\,V.\,Yakhontov
\date{}}

{
\def\thefootnote{}
\footnotetext{
\centering
Sergey\,V.\,Yakhontov: Ph.D. in Theoretical Computer Science, Dept. of Computer Science, Faculty of\\
Mathematics and Mechanics, Saint Petersburg State University, Saint Petersburg, Russian Federation;\\
e-mail: SergeyV.Yakhontov@gmail.com, S.Yakhontov@spbu.ru; phone: +7-911-966-84-30;\\
\ifx\journal\undefined
personal Web page: \url{https://sites.google.com/site/sergeyvyakhontov/}; current status of the paper:\\
\url{https://sites.google.com/site/sergeyvyakhontov/home/peqnp-paper-status/}; 19-Mar-2017
\fi
\ifx\arxiv\undefined
\url{https://sites.google.com/site/sergeyvyakhontov/home/peqnp-paper-status/}; 18-Okt-2015
\fi
}
%
\maketitle


\renewcommand{\abstractname}{Abstract}
\abstract{
%
The present paper proves that \notion{P}$\,=\,$\notion{NP}.
The proof, presented in this paper, is a constructive one: The
program of a polynomial time deterministic multi-tape Turing machine
$M\langle \exists AcceptingPath\rangle$, which determines if there exists
an accepting computation path of a polynomial time non-deterministic
single-tape Turing machine $M\langle NP\rangle$, is constructed explicitly
(machine $M\langle \exists AcceptingPath\rangle$ is different for each
machine $M\langle NP\rangle$).

The features of machine $M\langle\exists AcceptingPath\rangle$ are as follows:
\begin{enumerate}
\item[1)]
{the input of machine $M\langle\exists AcceptingPath\rangle$ does not contain
any encoded program of machine $M\langle NP\rangle$, but the program of
machine $M\langle\exists AcceptingPath\rangle$ contains implicitly the program
of machine $M\langle NP\rangle$;}
\item[2)]
{machine $M\langle\exists AcceptingPath\rangle$ is based on reduction
$L\le_{m}^{P}\notioninform{LP}$ (\notion{Linear Programming}) instead of reductions
$L\le_{m}^{P}\notioninform{3-CNF-SAT}\le_{m}^{P}\notioninform{ILP}$
(\notion{Integer Linear Programming}) which are commonly used,
wherein language $L\in\,$\notion{NP}, machine $M\langle NP\rangle$ decides $L$,
and $\le_{m}^{P}$ is polynomial time many-one reduction;
reduction $L\le_{m}^{P}\notioninform{3-CNF-SAT}$ is not used in the present paper;}
\item[3)]
{the reduction to problem \notion{LP} is a set of reductions
$L\le_{m}^{P} \notioninform{TCPE}\le_{m}^{P}\notioninform{LP}$
in fact wherein \notion{TCPE} (Tape-Consistent Path Existence Problem) is
a \notion{NP}-complete problem defined in the present paper;
}
\item[4)]
{problem \notion{TCPE} is reducible to a similar problem \notion{TCPE}$\langle 1\rangle$ that is a
special case of problems \notion{mixed-DHORN-SAT} (dual Horn) and \notion{linear-CNF-SAT};
problem \notion{TCPE}$\langle 1\rangle$ is polynomial time reducible to problem \notion{ILP};
unlike problem \notion{TCPE}$\langle 1\rangle$, a polynomial time algorithm is constructed
for problem \notion{TCPE} in the present paper;
}
\item[5)]
{to determine if there exists an accepting computation path, it is sufficient to find
a fractional solution of the resulting linear program;
}
\item[6)]
{the set of the accepting computation paths of machine $M\langle NP\rangle$ is considered as a
subset of a more general set of all the computation paths in the acyclic control flow graph of
polynomial size of a deterministic computer program that writes values to the tape cells and
reads values from the tape cells;}
\item[7)]
{reduction $L\le_{m}^{P}\notioninform{TCPE}$ is based on the results of reaching
definitions analysis for the deterministic computer program and on the notion of network flow;}
\item[8)]
{the resulting linear program does not express any combinatorial optimization problem
polytope (like \notion{TSP} polytope);}
\item[9)]
{both to accept and reject the input of machine $M\langle NP\rangle$, polynomial $t(n)$,
an upper bound of the time complexity of machine $M\langle NP\rangle$,
is not used in the program of machine $M\langle\exists AcceptingPath\rangle$;
}
\item[10)]
{machine $M\langle\exists AcceptingPath\rangle$ is a pure mathematical construction;
the proof presented in the paper is not based on physics theories (but it seems it should relate
to them in some a way).
}
\end{enumerate}
The time complexity of single-tape Turing machine that corresponds to
multi-tape Turing machine $M\langle\exists AcceptingPath\rangle$ is \overallcomplb;
the time complexity of the pseudocode algorithm of machine $M\langle\exists AcceptingPath\rangle$
on a computer with Von Neumann architecture is \overallopera\ operations ($\sigma$
is a constant depending on transition relation $\Delta$ of machine $M\langle NP\rangle$).

In fact, program analysis (namely, reaching definitions analysis for the special computer program
defined in the present paper) and linear programming are used in the present paper to solve
the \notion{P}~vs.~\notion{NP} Problem.
%
}

\vspace{0.3cm}
{\noindent\bf Keywords:}
%
computational complexity, Turing machine, class \notion{P}, class
\notion{NP}, \notion{P}~vs.~\notion{NP} Problem, class \notion{FP},
accepting computation paths, tape-consistent path existence problem,
program analysis, linear programming.
%
\vspace{0.1cm}

\tableofcontents


\vspace{0.3cm}
%
\mysectiona
{Introduction}
{Introduction}
\thickspace
%
%
This paper
\ifx\arxiv\undefined
\footnote{Preprint of this paper is
available on the Internet at \url{http://arxiv.org/abs/1208.0954v35}}
\fi
concerns the complexity classes of languages over finite
alphabets (wherein the number of symbols is equal to or
more than two) that are decidable by Turing machines.

It follows from the definition of classes \notion{P} and \notion{NP}
\cite{DK00} that \notion{P}$\,\subseteq\,$\notion{NP}
wherein \notion{P} is the shortened indication of \notion{PTIME} and
\notion{NP} is the shortened indication of \notion{NPTIME}.
However, the problem of the strictness of the inclusion, referred to as
{\bf the \notion{P} versus \notion{NP} Problem}, is one of the most
important unsolved problems in the theory of computational
complexity.

The \notion{P}~vs.~\notion{NP} Problem was introduced by
Stephen Cook in 1971 \cite{C71} and independently by Leonid Levin
in 1973 \cite{lL73}. A detailed description of the problem in
\cite{ClayCook} formulates it as follows: Can each language over
a finite alphabet, which is decidable by {\bf a polynomial
time non-deterministic single-tape Turing machine}, also be decided
by {\bf a polynomial time deterministic single-tape Turing
machine}? The shortened formulation of the problem is
\notion{P}$\,=$$?\,$\notion{NP}.

The papers \cite{S92,C03,W07,F09,L10} contain detailed surveys on
the \notion{P}~vs.~\notion{NP} Problem.

The present paper proves that \notion{P}$\,=\,$\notion{NP}.
The proof, suggested in this paper, is a constructive one:
The pseudocode of a polynomial time deterministic multi-tape Turing machine
$M\langle\exists AcceptingPath\rangle$, which determines if there exists
an accepting computation path of a polynomial time non-deterministic
single-tape Turing machine $M\langle NP\rangle$, is constructed explicitly.
More precisely, $M\langle\exists AcceptingPath\rangle$
determines if there exists an accepting computation path of the computation
tree of machine $M\langle NP\rangle$ on the input; at that, machine
$M\langle \exists AcceptingPath\rangle$ is different for each
machine $M\langle NP\rangle$.

It is known that problem \notion{3-CNF-SAT} is \notion{NP}-complete \cite{C71,lL73} (Cook--Levin theorem);
this theorem is usually used as a basis to try to solve the \notion{P}~vs.~\notion{NP} Problem.

Most of the works on the attempts to solve the \notion{P}~vs.~\notion{NP} Problem
can be found on the Internet at \cite{ArxivOrgCC} and \cite{PvsNPPage}.
It seems most of these works use reductions
$$L\le_{m}^{P}\notioninform{3-CNF-SAT}\le_{m}^{P}\ldots\le_{m}^{P}L'$$
wherein language $L\in\,$\notion{NP} and $\le_{m}^{P}$ is polynomial time many-one reduction;
a detailed list of these reductions can be found in \cite{GJ79}.
In particular, reductions to \notion{ILP} (Integer Linear Programming) are often used:
$$L\le_{m}^{P}\notioninform{3-CNF-SAT}\le_{m}^{P}\ldots\le_{m}^{P}\notioninform{ILP};$$
a detailed list of reductions to \notion{ILP} can be found in \cite{S03}.

Regarding the works at \cite{ArxivOrgCC,PvsNPPage}, the author of the present paper
could not find any work that contains a concept similar to the concept suggested
in the present paper. 

The solution suggested in the present paper is completely different from
the well-known approaches to solve the problem; namely, reduction $$L\le_{m}^{P}\notioninform{LP}$$
is used instead of reductions $$L\le_{m}^{P}\notioninform{3-CNF-SAT}\le_{m}^{P}L'$$ in the present
paper. The reason of using of new approach can be partially explained by the fact that there are
a lot of attempts to find a polynomial time algorithm for \notion{NP}-complete problems using reduction
$$L\le_{m}^{P}\notioninform{3-CNF-SAT},$$ and it seems the attempts fail.

The concept of the construction of machine $M\langle\exists AcceptingPath\rangle$ suggested
in the present paper is based on the following general idea:
\begin{enumerate}
\item[1)]
{define the set of the tape-arbitrary paths in the acyclic control flow graph \cite{NNH05}
of polynomial size of a deterministic computer program such that this set is the disjoint
union of the set of the tape-consistent paths and the set of the tape-inconsistent paths;
}
\item[2)]
{using reduction to problem \notion{LP}, determine if there exists a tape-consistent path
in the control flow graph; the reduction is based on the results of reaching definitions analysis
for the deterministic computer program and on the notion of network flow;}
\item[3)]
{there is one-to-one mapping from the set of the tape-consistent accepting paths onto the set of the
accepting computation paths of machine $M\langle NP\rangle$, so one can determine if there exists
an accepting computation path of machine $M\langle NP\rangle$.}
\end{enumerate}
In contrast to problem \notion{ILP}, a fractional solution of problem \notion{LP} can be found
in polynomial time \cite{Kh80,K84}.

The resulting linear program does not express any combinatorial optimization problem
polytope; so, results \cite{Y91,S86,FMPT13} (others papers on this topic could be found
in \cite[references]{FMPT13}), which state that expressing combinatorial optimization
problems requires linear programs of exponential size, are not applicable
to the present paper.

The main feature of the tape-arbitrary paths is that the computations on a path of such
kind starting at a point do not depend on the computations from the start of the path to this point.
This fact is the main reason why exponential time computations are represented by a graph
of polynomial size in the present paper.

To say in more detail, machine $M\langle\exists AcceptingPath\rangle$ works in polynomial
time in $t(n)$ because the space used to compute the computation steps (elements) of a tape-arbitrary
sequence of the computation steps of machine $M\langle NP\rangle$ is logarithmic in $t(n)$ only.

Machine $M\langle\exists AcceptingPath\rangle$ computes also a $t(n)_{\le}$-length accepting
computation path itself of machine $M\langle NP\rangle$ in polynomial time in $t(n)$, wherein $t(n)$
is an upper bound of the time complexity of machine $M\langle NP\rangle$.
%

%
\mysectiona
{Preliminaries}
{Preliminaries}
\thickspace
In the present paper
\begin{enumerate}
\item[1)]
{$t(n)$ is an upper bound of the time complexity of machine $M\langle NP\rangle$,}
\item[2)]
{in the estimations of the time and space complexity of the algorithms, `TM steps' and `TM tape cells'
mean steps and tape cells accordingly of Turing machine,}
\item[3)]
{in the estimations of the time and space complexity of the algorithms, `VN operations' and `VN memory cells'
mean operations and memory cells accordingly of a computer with Von Neumann architecture, and}
\item[4)]
{integer $\mu$ will be used to denote the length of sequence of the computation steps of Turing machine;}
\item[5)]
{direct acyclic graphs are only considered;}
\item[6)]
{all the propositions whose proofs are obvious or follow from the previous text are omitted.}
\end{enumerate}
This section contains general information that is used in all the constructions in the
present paper.
%
%
\mysectionb
{Non-deterministic computations}
{Non-deterministic computations}
\thickspace
Let $$M=\langle Q,\Gamma,b,\Sigma,\Delta,q_{start},F\rangle$$ be a
non-deterministic single-tape Turing machine wherein $Q$ is
the set of states, $\Gamma$ is the set of tape symbols,
$b$ is the blank symbol, $\Sigma$ is the set of input symbols,
$\Delta$ is the transition relation, $q_{start}$ is the initial state,
and $F$ is the set of accepting states. The elements of the set $\{L,R,S\}$
denote, as is usual, the moves of the tape head of machine $M$.

Non-deterministic Turing machines as \notion{decision procedures} (more
precisely, programs for non-deterministic Turing machines as decision
procedures) are usually defined as follows.

\begin{definition}{\normalfont\cite{DK00}}
Non-deterministic Turing machine $M$ accepts input $x$ if there exists
an accepting computation path of machine $M$ on input $x$.
\end{definition}
\begin{definition}{\normalfont\cite{AB09}}
\label{Def:MTRejectsInput}
Non-deterministic Turing machine $M$ rejects input $x$ if all the
computation paths of machine $M$ on input x are finite and these paths
are not accepting computation paths.
\end{definition}
\begin{definition}{\normalfont\cite{DK00}}
Non-deterministic Turing machine $M$ decides a language $L\subseteq\Sigma^*$
if machine $M$ accepts each word $x\in L$ and rejects each word
$x\notin L$.
\end{definition}

The time (space) computational complexity of non-deterministic Turing
machine $M$ is polynomial if there exists a polynomial $t_M(n)$
($s_M(n)$ accordingly) such that for every input $x$
\begin{enumerate}
\item[1)]
{the minimum of the lengths of all the accepting computation paths
of machine $M$ on input $x$ does not exceed $t_M(|x|)$ (accordingly,
the number of the different visited cells on each accepting
computation path does not exceed $s_M(|x|)$) if machine $M$
accepts input $x$, and}
\item[2)]
{the lengths of all the computation paths of machine $M$ on input $x$
do not exceed $t_M(|x|)$ (accordingly, the number of the different
visited cells on each computation path does not exceed $s_M(|x|)$)
if machine $M$ rejects input $x$.}
\end{enumerate}
Here, (as is usual) by means of $|x|$ the length of word $x$ is specified.

Let $\mu$ be an integer.
\begin{definition}
Computation path $p$ of Turing machine $M$ on input $x$ is said to be a
$\mu$-length computation path if the length of $p$ is equal to $\mu$.
Accepting computation path $p$ of machine $M$ on input $x$ is said
to be a $\mu$-length accepting computation path if $p$ is $\mu$-length
computation path.
\end{definition}
\begin{definition}
Computation path $p$ of Turing machine $M$ on input $x$ is said to be a
$\mu_{\le}$-length {\normalfont(}$\mu_{>}$-length{\normalfont)} computation
path if the length of $p$ is less than or equal to $\mu$ {\normalfont(}is
greater than $\mu${\normalfont)}. Accepting computation path $p$ of machine $M$ on
input $x$ is said to be a $\mu_{\le}$-length accepting computation path
if $p$ is a $\mu_{\le}$-length computation path.
\end{definition}

If Turing machine $M$ accepts input $x$ and the time complexity of
machine $M$ is bounded above by polynomial $t_M(n)$, then the computation
tree of machine $M$ on input $x$ has at least one $t_M(|x|)_{\le}$-length
accepting computation path.

If Turing machine $M$ rejects input $x$ and the time complexity
of machine $M$ is bounded above by polynomial $t_M(n)$ then all the computation
paths of machine $M$ on input $x$ are precisely the $t_M(|x|)_{\le}$-length
computation paths, and these paths are not accepting computation paths.

Let's note that there are some differences between the definitions of how
non-deterministic Turing machine rejects the input. Usually, non-deterministic
Turing machines are defined in such a way that it is acceptable that
there are some endless computation paths or $t_M(n)_{>}$-length computation paths
in the case Turing machine rejects the input \cite{DK00,GJ79,Papa94,G08};
sometimes definition \ref{Def:MTRejectsInput}, which is stronger
than the definitions in \cite{DK00,GJ79,Papa94,G08}, is used \cite{AB09}.

Non-deterministic computations are often defined as \notion{guess-and-verify}
computations \cite{K72,DK00} or \notion{search-and-check} computations \cite{lL73,G08}.
In \cite{ClayCook}, the \notion{P}~vs.~\notion{NP} Problem is formulated
precisely in terms of \notion{guess-and-verify} computations, but it is known
\cite{DK00,G08} that these definitions of non-deterministic computations
are equivalent to the definition, which is used in the present paper, of
non-deterministic computations performed by of non-deterministic Turing machines.
%
%
\mysectionb
{Complexity classes \notion{P} and \notion{NP}}
{Complexity classes \notion{P} and \notion{NP}}
\thickspace
Let $t:\mathbb{N}\rightarrow\mathbb{N}$ be a nondecreasing function from integers to
integers, and $C$ be a collection of such functions.
\begin{definition}
{\normalfont \cite{DK00}}
We define $DTIME(t)$ be the class of languages $L$ that are accepted by
deterministic Turing machines $M$ with $t_M(n)\le t(n)$ for almost all $n\ge 0$.
We let
\begin{align*}
DTIME(C)=\bigcup_{t\in C}DTIME(t).
\end{align*}
\end{definition}
\begin{definition}
{\normalfont \cite{DK00}}
We define $NTIME(t)$ be the class of languages $L$ that are accepted by
non-deterministic Turing machines $M$ with $t_M(n)\le t(n)$ for almost all
$n\ge 0$. We let
\begin{align*}
NTIME(C)=\bigcup_{t\in C}NTIME(t).
\end{align*}
\end{definition}
Let $poly$ be the collection of all integer polynomial functions with nonnegative coefficients.
Complexity classes \notion{P} and \notion{NP} in terms of Turing machines are defined as follows.
\begin{definition}
{\normalfont \cite{DK00}}
$\notioninform{P}=DTIME(poly)$.
\end{definition}
\begin{definition}
{\normalfont \cite{DK00}}
$\notioninform{NP}=NTIME(poly)$.
\end{definition}
%
%
\mysectionb
{Notations for graphs}
{Notations for graphs}
\thickspace
Let $G=(V,E)$ be a direct acyclic graph that has one source node $s$ and one
sink node $t$ and (such graphs have no backward edges); let $G$ have no cross edges.
\begin{notation}
Let $Source\langle G\rangle$ be node $s$; let $Sink\langle G\rangle$ be node $t$.
\end{notation}
\begin{notation}
Let $Nodes\langle G\rangle$ be set $V$; let $Edges\langle G\rangle$ be set $E$.
\end{notation}
\begin{notation}
Let $$InnerNodes\langle G\rangle=\{u\in V\ |\ ((u\ne s)\wedge(u\ne t))\}.$$
\end{notation}
\begin{notation}
For each node $u\in V$ let, as is usual,
\begin{align*}
&\delta^{-}(u)=\{(v,u)\ |\ ((v\in V)\wedge ((v,u)\in E))\}\
\text{be the set of all in-edges of node}\ u,\quad\text{and}\\
&\delta^{+}(u)=\{(u,v)\ |\ ((v\in V)\wedge ((u,v)\in E))\}\
\text{be the set of all out-edges of node}\ u.
\end{align*}
\end{notation}
\begin{notation}
Graph $G$ is said to be a $2$-out regular graph if $\delta^{+}(u)\le 2$ for each
node $u\in (V\setminus \{t\})$.
\end{notation}
\begin{notation}
By $$\left(G_1\cap G_2\right)$$ we will denode an ordinary intersection of graphs $G_1$ and $G_2$. Namely,
if $G_1=(V_1,E_1)$ and $G_2=(V_2,E_2)$, then the intersection $$G=(V_1\cap V_2,E_1\cap E_2).$$
\end{notation}
%
%
\mysectionb
{Sets of paths in graphs}
{Sets of paths in graphs}
\thickspace
Let $p$ be a path in graph $G$ (sequence of nodes $(u_1,u_2,\ldots,u_n)$ such that $(u_i,u_{i+1})\in E$
for each $i\in [1..(n-1)]$).
\begin{definition}
Path $p$ in graph $G$ is said to be $s$-$t$ path if $p$ starts with the source node $s$
and ends with the sink node $t$.
\end{definition}
\begin{notation}
Let $AllPaths\langle G\rangle$ be the set of all the $s$-$t$ paths in graph $G$.
\end{notation}
\begin{definition}
Path $p$ in graph $G$ is said to be $u$-$v$ path if $p$ starts with node $u$
and ends with node $v$.
\end{definition}
\begin{definition}
Path $p'$ in graph $G$ is said to be $s$-subpath of $s$-$t$ path $p$ in graph $G$
if $p'$ starts with node $s$ and $p'$ is a subpath of path $p$.
\end{definition}
\begin{notation}
Let $PathSet\langle(u,v)\rangle$, wherein $u$ and $v$ are nodes,
be the set of $u$-$v$ paths in graph $G$.
\end{notation}
\begin{notation}
Let $PathSet\langle(s,u,v,t)\rangle$, wherein $u$ and $v$ are nodes,
be the set of $s$-$t$ paths $p$ in graph $G$ such that $u\in p$ and $v\in p$.
\end{notation}
Let $P$ be a set of $u$-$v$ paths in graph $G$.
\begin{notation}
Let $Subgraph\langle G,P\rangle$ be graph $(V\langle sub\rangle,E\langle sub\rangle)$
wherein
\begin{align*}
&V\langle sub\rangle=\{u\ |\ ((u\in V)\wedge(\exists p\in P :
u\in p))\}\quad \text{and}\\
&E\langle sub\rangle=\{e\ |\ ((e\in E)\wedge(\exists p\in P :
e\in p))\}.
\end{align*}
\end{notation}
\begin{notation}
Let $Subgraph\langle G,(u,v)\rangle$ be graph $Subgraph\langle G,PathSet\langle (u,v)\rangle\rangle$.
\end{notation}
\begin{notation}
Let $p$ is a $s$-$t$ path in graph $V$ and $V'$ is a subgraph of graph $V$. Subpath
$$p'=(u_{1}=s',u_{2},\ldots,u_{n}=t')$$ is denoted by $\left(p\cap V'\right)$ if $$p'\in AllPaths\langle V'\rangle$$
wherein $s'=Source\langle V'\rangle$ and $t'=Sink\langle V'\rangle$.
\end{notation}
\begin{notation}
We write $(u,v)\in p$, wherein $p$ is a path in graph $G$, if $u\in p$ and $v\in p$.
\end{notation}
\begin{notation}
Distance of node $u\in V$ from the source node $s$, denoted by $$Dist(s,u),$$
is defined to be the length of a $s$-$u$ path in graph $G$ if the lengths of such paths
are the same (otherwise, value $Dist(s,u)$ is undefined).
\end{notation}
%
%
\mysectionb
{Network flows}
{Network flows}
\label{SubSec:NetwFlows}
\thickspace
Network flow equations for graph $G$ are defined as follows \cite{CLRS09}:
\begin{enumerate}
\item[1)]
{functions $F$ from $V$ to rationals and $H$ from $E$ to rationals are introduced;}
\item[2)]
{for each node $u\in V$, $u\ne s$,
\begin{align}
\label{Eq:NetwFlowEqs1}
F[u]= \sum_{(v,u)\in \delta^{-}(u)} H[(v,u)];
\end{align}
}
\item[3)]
{for each node $u\in V$, $u\ne t$,
\begin{align}
\label{Eq:NetwFlowEqs2}
F[u]= \sum_{(u,v)\in \delta^{+}(u)} H[(u,v)];
\end{align}
}
\item[4)]
{
$F[u]=0$ if $u\notin V$ and $H[e]=0$ if $e\notin E$ (for the case of subgraph).
}
\end{enumerate}
\begin{notation}
Network flow with equations \eqref{Eq:NetwFlowEqs1} and \eqref{Eq:NetwFlowEqs2} is denoted by $(F,H)$.  
\end{notation}
\begin{definition}
Network flow $(F,H)$ such that $F[u]=0$ for each $u\in V$ is said to be an empty network flow;
otherwise, network flow $(F,H)$ is said to be a non-empty network flow.
\end{definition}
\begin{definition}
Network flow $(F,H)$ such that $F[s]=F[t]=1$ is said to be $1$-$1$ network flow.  
\end{definition}
\begin{definition}
The sum {\normalfont(}subtraction{\normalfont)} of network flows $\mathcal{F}_1=(F_1,H_1)$ and
$\mathcal{F}_2=(F_2,H_2)$, denoted by $\mathcal{F}_1\pm\mathcal{F}_2$, is defined to be the network
flow $\mathcal{F}=(F,H)$ such that $F[u]=F_1[u]\pm F_2[u]$ for each node $u\in V$,
$H[e]=H_1[e]\pm H_2[e]$ for each edge $e\in E$.
\end{definition}
\begin{definition}
We say that $\mathcal{F}_1=\mathcal{F}_2$, wherein $\mathcal{F}_1=(F_1,H_1)$ and
$\mathcal{F}_2=(F_2,H_2)$ are network flows, if $F_1[u]=F_2[u]$ for each node $u\in V$,
$H_1[e]=H_2[e]$ for each edge $e\in E$ (the same for other order relations).
\end{definition}
%
%
\mysectionb
{Path flows in graphs}
{Path flows in graphs}
\thickspace
\begin{definition}
\label{Def:PathFlow}
The flow of a $s$-$t$ path $p$ in graph $G$ is defined to be a network flow $(F,H)$, denoted by
$$\mathcal{PF}\langle p,\theta_p\rangle,$$ such that $H[e]=\theta_p$ for each edge $e\in p$ and $H[e]=0$
otherwise wherein $\theta_p$ is a rational, $0<\theta_p\le 1$.
\end{definition}
Let's note that in that case $F[u]=\theta_p$ for each node $u\in p$.
\begin{definition}
Path flow in graph $G$, corresponding to a path set $P$, $P\ne \emptyset$, is defined to be the
network flow, denoted by $\mathcal{PF}\langle P\rangle$, such that
$$\mathcal{PF}\langle P\rangle=\mathlarger{\sum_{p\in P}\mathcal{PF}\langle p,\theta_p\rangle}.$$
\end{definition}
\begin{definition}
We say that a path set $P$, $P\ne \emptyset$, corresponds to a non-empty path flow $\mathcal{PF}$ if
$\mathcal{PF}=\mathcal{PF}\langle P\rangle$.
\end{definition}
\begin{notation}
Let $$PathSets\langle \mathcal{PF}\rangle$$ be the set of all path sets $P$ such that $P$ corresponds
to path flow $\mathcal{PF}$.
\end{notation}
%
%
\begin{proposition}
\label{Prop:AnyNetwFlowIsPathflow}
For every non-empty network flow $\mathcal{F}$, set $PathSets\langle \mathcal{F}\rangle$
is not empty.
\begin{proof}
Let path set $P:=\emptyset$; let's repeat the following steps until
empty network flow $\mathcal{F}$ is reached:
\begin{enumerate}
\item[1)]
{take a $s$-$t$ path $p$ such that $$\mathcal{PF}\langle p,\theta_p\rangle\le \mathcal{F}$$
for $\theta_p=\mathlarger{\min_{e\in p}H[e]}$ ($\mathcal{PF}\langle p,\theta_p\rangle$
is a non-empty network flow);}
\item[2)]
{$\mathcal{F}:=\mathcal{F}-\mathcal{PF}\langle p,\theta_p\rangle$;}
\item[3)]
{$P:=P\cup \{p\}$.}
\end{enumerate}
The number of such steps is finite because $H[e]=0$ for an edge $e$ after the step is done.
As a result, $P\in PathSets\langle \mathcal{F}\rangle$.
\end{proof}
\end{proposition}
So, every network flow $\mathcal{F}$ is a path flow.
%

%
\mysectiona
{Construction of deterministic multi-tape Turing machine $M\langle\exists AcceptingPath\rangle$}
{Construction of deterministic multi-tape Turing\\
machine $M\langle\exists AcceptingPath\rangle$}
\thickspace
In this section, the components and the program of machine $M\langle\exists AcceptingPath\rangle$
are constructed in detail.
%
%
\mysectionb
{Underlying elements of machine $M\langle\exists AcceptingPath\rangle$}
{Underlying elements of machine $M\langle\exists AcceptingPath\rangle$}
\thickspace
%
%
\mysectionc
{Sequences of computation steps}
{Sequences of computation steps}
\thickspace
The notion of sequences of computation steps is used to define the general set of the tape-arbitrary
paths which includes the set of the tape-consistent paths of machine $M$ (it seems it will be also suitable to say
`tape-less' instead of 'tape-arbitrary').
%
%
\mysectiond
{Computation steps}
{Computation steps.}
\thickspace
\begin{definition}
Computation step $t$ of machine $M$ is defined to be tuple
$$(q,s,q',s',m,\kappa^{(tape)},\kappa^{(step)})$$ such that
$$d=((q,s),(q',s',m))\in\Delta$$ wherein $m\in \{L,R,S\}$,
$\kappa^{(tape)}$ and $\kappa^{(step)}$ are integers.
In that case, we write $d\bigtriangleup t$.
\end{definition}
\begin{notation}
Let computation step $$t=(q,s,q',s',m,\kappa^{(tape)},\kappa^{(step)}).$$
State $q$ in $t$ is denoted by $t.q$ {\normalfont(}the same notation is for
other elements of the tuple{\normalfont)}.
\end{notation}
%

\begin{definition}
\label{Def:CompStepSequentialPair}
Let
\begin{align*}
&t_1=(q_{1},s_{1},q'_{1},s'_{1},m_{1},\kappa^{(tape)}_1,\kappa^{(step)}_1)\quad\text{and}\\
&t_2=(q_{2},s_{2},q'_{2},s'_{2},m_{2},\kappa^{(tape)}_2,\kappa^{(step)}_2)
\end{align*}
be computation steps. Pair $(t_1,t_2)$ is said to be a sequential pair of computation steps
if $q_2=q_1'$, $$\kappa^{(step)}_2=\kappa^{(step)}_1+1,$$ and the following holds:
\begin{enumerate}
\item[1)]
{if $m_{1}=L$ then $\kappa^{(tape)}_2=\kappa^{(tape)}_1-1$;}
\item[2)]
{if $m_{1}=R$ then $\kappa^{(tape)}_2=\kappa^{(tape)}_1+1$;}
\item[3)]
{if $m_{1}=S$ then $\kappa^{(tape)}_2=\kappa^{(tape)}_1$.}
\end{enumerate}
\end{definition}
Only finite sequences of the computation steps, such that each pair
$(t_i,t_{i+1})$ of computation steps is a sequential pair, are considered.
\begin{definition}
Pair of computation steps
\begin{align*}
&t_{i_1}=(q_{i_1},s_{i_1},q'_{i_1},s'_{i_1},m_{i_1},
\kappa^{(tape)}_{i_1},\kappa^{(step)}_{i_1})\quad\text{and}\\
&t_{i_2}=(q_{i_2},s_{i_2},q'_{i_2},s'_{i_2},m_{i_2},
\kappa^{(tape)}_{i_2},\kappa^{(step)}_{i_2})
\end{align*}
is said to be a tape-consistent pair of computation steps
if $$s_{i_2}=s_{i_1}'.$$ Otherwise {\normalfont(}when $s_{i_2}\ne s_{i_1}'$\normalfont{)}
the pair is said to be a tape-inconsistent pair of computation steps.
\end{definition}
Let's note that it is not required in this definition for computation steps
$t_{i_1}$ and $t_{i_2}$ that $t_{i_2}$ follows immediately $t_{i_1}$ in computation
paths of machine $M\langle\exists AcceptingPath\rangle$; there can be a sequence like
$$(\ldots,t_{i_1},\ldots,t_{i_2},\ldots)$$ wherein $i_1<i_2+1$.
%
%
\mysectiond
{Auxiliary notations and definitions}
{Auxiliary notations and definitions.}
\thickspace
Let's place the input $x$ on the tape cells of Turing machine $M$ as follows:
The number of the cell $c_1$, containing the leftmost symbol of input $x$,
is $1$, the number of the cell to the right of $c_1$ is $2$, the number
of the cell to the left of $c_1$ is $0$, and so on.

\begin{notation}
The tape cell with number $\kappa$ is denoted by $c_{\kappa}$.
\end{notation}

\begin{notation}
Let $x$ be an input of machine $M$. The symbol in tape cell $c_{\kappa}$ is
denoted by $Tape\langle x,\kappa\rangle$.
\end{notation}

\begin{notation}
$TapeLBound\langle \mu\rangle=2-\mu$; $TapeRBound\langle\mu\rangle=\mu$.
\end{notation}

\begin{notation}
Integer range $$[TapeLBound\langle \mu\rangle..TapeRBound\langle \mu\rangle]$$
of cell numbers is denoted by $TapeRange\langle \mu\rangle$.
\end{notation}


\begin{definition}
Subsequence $\omega_{sub}=(t_1,\ldots t_{\mu'})$ of sequence $\omega$ of the
computation steps, denoted by $Subseq\langle \omega,\kappa\rangle$, is said to be a
subsequence at cell $c_{\kappa}$ of sequence $\omega$
if $\kappa^{(tape)}=\kappa$ for each computation step
$t=(q,s,q',s',m,\kappa^{(tape)},\kappa^{(step)})$ in $\omega_{sub}$.
\end{definition}

\begin{definition}
We say that sequence $\omega=(t_1,\ldots t_{\mu})$ of the computation steps
starts on input $x$ if $t_1=(q_{start},s,q',s',m,1,1)$
for some $s$, $q'$, $s'$, and $m$.
\end{definition}

\begin{definition}
We say that sequence $\omega=(t_1,\ldots t_{\mu})$ of the computation steps
corresponds to input $x$ at cell $c_{\kappa^{(tape)}}$ if one of the following
holds:
\begin{enumerate}
\item[1)]
{if $$Subseq\langle \omega,\kappa^{(tape)}\rangle=(t_{i_1},\ldots t_{i_k})\
\text{and}\ t_{i_1}=(q,s,q',s',m,\kappa^{(tape)},\kappa^{(step)}),$$
then $s=Tape\langle x,\kappa^{(tape)}\rangle$;} 
\item[2)]
{$Subseq\langle \omega,\kappa^{(tape)}\rangle$ is an empty sequence.}
\end{enumerate}
\end{definition}


\begin{notation}
Let set
$$U=\{j\ |\ (t=(q,s,q',s',m,\kappa^{(tape)},j)\in \omega)\},$$
wherein $\omega$ is a sequence of computation steps, is not empty.
In that case, value $\kappa=\min\{j\ |\ j\in U\}$
is denoted by $$TapeFirst\langle \omega,\kappa^{(tape)}\rangle.$$
\end{notation}

\begin{notation}
Let set
$$U=\{j\ |\ ((j<\kappa^{(step)})
\ \wedge\ (t=(q,s,q',s',m,\kappa^{(tape)},j)\in \omega))\},$$
wherein $\omega$ is a sequence of computation steps, is not empty.
In that case, value $\kappa=\max\{j\ |\ j\in U\}$
is denoted by $$TapePrev\langle \omega,\kappa^{(tape)},\kappa^{(step)}\rangle.$$
\end{notation}


\begin{definition}
Sequence $\omega=(t_1,\ldots t_{\mu})$ of the computation steps
of machine $M$ is said to be $q'$-state sequence of the computation steps if
$$t_{\mu}=(q,s,q',s',m,\kappa^{(tape)},\kappa^{(step)}).$$
\end{definition}

\begin{definition}
Sequence $\omega$ of the computation steps of machine $M$ is said to be an
accepting sequence of the computation steps if $\omega$ is $q$-state sequence
wherein $q\in F$.
\end{definition}

\begin{definition}
Sequence $\omega=(t_1,\ldots t_{\mu})$ of the computation steps
of machine $M$ is said to be a $\mu$-length sequence of the computation steps.
\end{definition}

\begin{definition}
Sequence $\omega=(t_1,\ldots t_{j})$ of the computation steps
of machine $M$ is said to be a $\mu_{\le}$-length sequence of the computation
steps if $j\le\mu$.
\end{definition}

Let's note that $\kappa^{(tape)}\in TapeRange\langle \mu\rangle$ for each
computation step $$t=(q,s,q',s',m,\kappa^{(tape)},\kappa^{(step)})$$
in a $\mu$-length sequence of the computation steps.
%
%
\mysectiond
{Kinds of sequences of computation steps}
{Kinds of sequences of computation steps.}
\thickspace
\begin{definition}
\label{Def:TapeConsistSeq}
Sequence $\omega=(t_1,\ldots t_{\mu})$ of the computation steps
of machine $M$ on input $x$ is said to be a tape-consistent
sequence of the computation steps on input $x$ if the following
holds:
\begin{enumerate}
\item[1)]
{$\omega$ starts on input $x$;}
\item[2)]
{$\omega$ corresponds to input $x$ at each cell $\kappa\in TapeRange\langle \mu\rangle$;}
\item[3)]
{for each $\kappa\in TapeRange\langle \mu\rangle$ the following holds:
  \begin{enumerate}
  \item[3.1)]
  {if subsequence $\omega_{sub}=Subseq\langle \omega,\kappa\rangle$ is not empty, then
  each pair $(t_i,t_{i+1})$ in $\omega_{sub}$ is a tape-consistent
  pair of computation steps.}
  \end{enumerate}
}
\end{enumerate}
\end{definition}

\begin{definition}
\label{Def:TapeInconsistSeq}
Sequence $\omega=(t_1,\ldots t_{\mu})$ of the computation steps
of machine $M$ on input $x$ is said to be a tape-inconsistent at pair
$(\kappa^{(tape)},\kappa^{(step)})$ sequence of the computation steps
on input $x$ if the following holds:
\begin{enumerate}
\item[1)]
{$t=(q,s,q',s',m,\kappa^{(tape)},\kappa^{(step)})\in \omega$;}
\item[2)]
{$\omega$ starts on input $x$;}
\item[3)]
{one of the following holds:
\begin{enumerate}
\item[3.1)]
{if $$\kappa^{(step)}=TapeFirst\langle \omega,\kappa^{(tape)}\rangle,$$ then
$s\ne Tape\langle x,\kappa^{(tape)}\rangle$;}
\item[3.2)]
{if there exists $\kappa$ such that
$$\kappa=TapePrev\langle \omega,\kappa^{(tape)},\kappa^{(step)}\rangle,$$
then pair $(t_{\kappa},t_{\kappa^{(step)}})$ is a
tape-inconsistent pair of the computation steps.}
\end{enumerate}
}
\end{enumerate}
\end{definition}

\begin{definition}
Sequence $\omega=(t_1,\ldots t_{\mu})$ of the computation steps
of machine $M$ on input $x$ is said to be a tape-inconsistent sequence
of the computation steps on input $x$ if $\omega$ is tape-inconsistent
at some pair $(\kappa^{(tape)},\kappa^{(step)})$ sequence on input $x$.
\end{definition}

\begin{definition}
Sequence $\omega=(t_1,\ldots t_{\mu})$ of the computation steps
of machine $M$ on input $x$ is said to be a tape-arbitrary
sequence of the computation steps if $\omega$ just starts on input $x$
{\normalfont(}so $\omega$ is a tape-consistent sequence of the computation steps or
$\omega$ is a tape-inconsistent sequence of the computation steps{\normalfont)}.
\end{definition}


\begin{definition}
Tape-consistent sequence $\omega=(t_1,\ldots t_{\mu})$ of the computation
steps is said to be the sequence corresponding to computation path
$P=\alpha_1\ldots\alpha_{\mu}$ of machine $M$ on input $x$ if
\begin{enumerate}
\item[1)]
{each $d_i$ for $i\in[1..(\mu-1)]$, such that $d_i\bigtriangleup t_i$, is the
transition corresponding to configuration transition $\alpha_i\vdash \alpha_{i+1}$, and}
\item[2)]
{$t_{\mu}=(q,s,q,s,S,\kappa^{(tape)},\kappa^{(\mu)})$; this `extra' computation step $t_{\mu}$ is
added to sequence $\omega$ for simplicity of the definition.
}
\end{enumerate}
\end{definition}
\begin{definition}
Tree $T$ of the computation steps is said to be the $\mu$-length
{\normalfont(}$\mu_{\le}$-length{\normalfont)}
tape-arbitrary tree of the computation steps of machine $M$ on input $x$
if each root-leaves path in $T$ is a tape-arbitrary sequence of the computation
steps of machine $M$ on input $x$, and the tree contains all the
$\mu$-length {\normalfont(}$\mu_{\le}$-length{\normalfont)} tape-arbitrary
sequences of the computation steps.
\end{definition}
%
%
\mysectiond
{A figure to explain the notion}
{A figure to explain the notion.}
\thickspace
The notion of sequences of computation steps is explained in Figure
\ref{Fig:CompStepSeqs}; there
\begin{enumerate}
\item[1)]
{the pair of computation steps
\begin{align*}
s_{2,8}=(q_{i_1},a,q_{i_1}',b,L,2,8)\ \text{and}\ s_{2,14}=(q_{i_2},b,q_{i_2}',c,R,2,14)
\end{align*}
is a tape-consistent pair of the computation steps;}
\item[2)]
{the pair of computation steps
\begin{align*}
s_{4,4}=(q_{i_3},x_{4},q_{i_3}',d,R,4,4)\quad \text{and}\quad s_{4,6}=(q_{i_4},e,q_{i_4}',f,L,4,6),
\end{align*}
wherein $e\ne d$, is a tape-inconsistent pair of the computation steps.}
\end{enumerate}
\begin{figure}
\centering
\includegraphics[height=8cm]{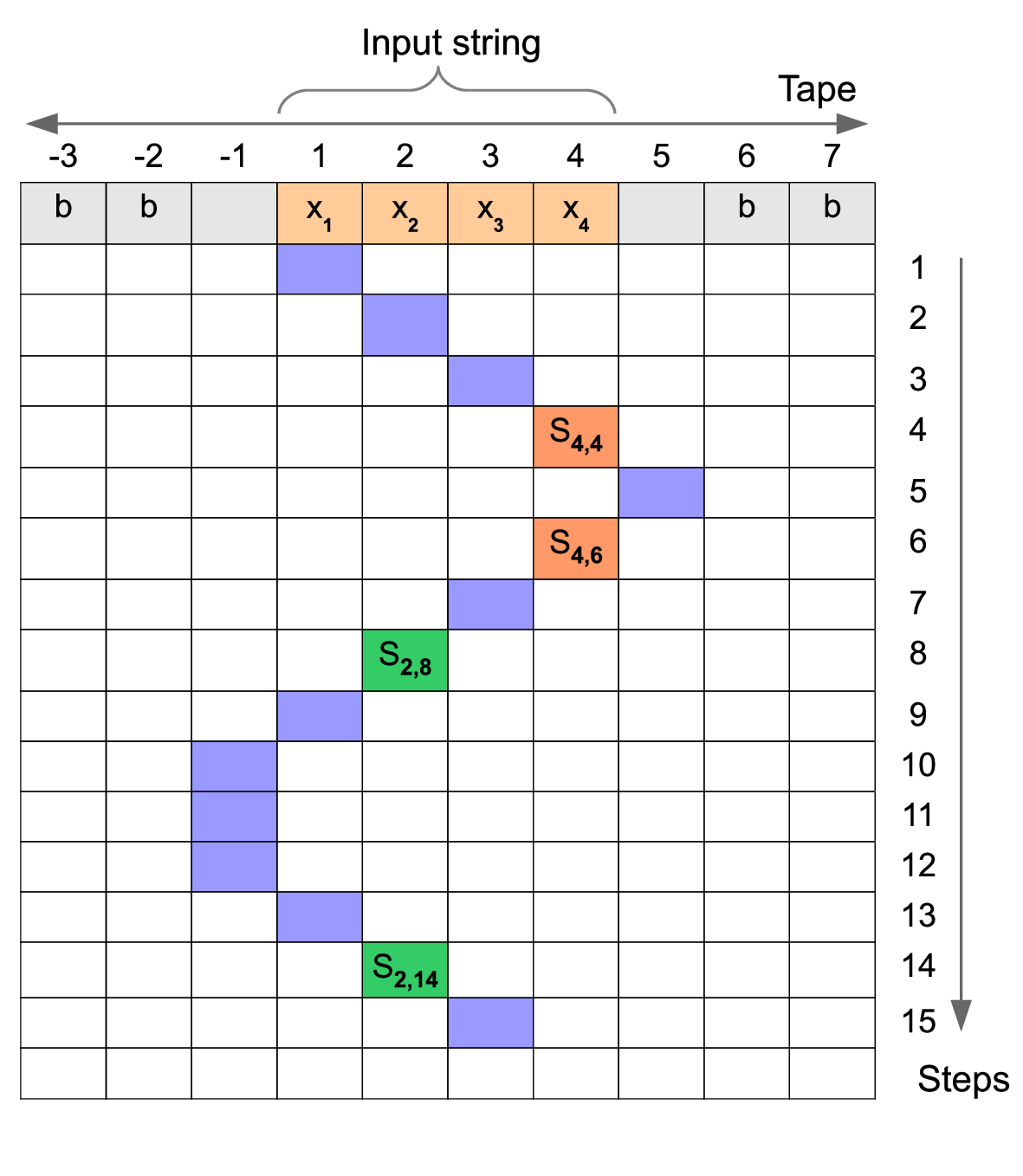}
\caption{A sequence of computation steps.}
\label{Fig:CompStepSeqs}
\end{figure}
%
%
\mysectionc
{Sequences of computation steps in control flow graphs}
{Sequences of computation steps in control flow graphs}
\thickspace
Let $G=(V,E)$ be a control flow graph with one source node $s$ and one
sink node $t$ such that for each node $u\in V$ a computation step of machine
$M\langle NP\rangle$ is associated with node $u$.
\begin{notation}
The computation step of machine $M\langle NP\rangle$ associated with node $u\in V$
is denoted by $u.step$.
\end{notation}
\begin{notation}
Let $p$ be a $s$-$t$ path $(u_1,\ldots,u_m)$ in graph $G$. Sequence of the
computation steps $$(t_2,\ldots,t_{m-1})$$ such that $t_i=u_i.step$ for
$i\in [2..(m-1)]$ is denoted by $\omega\langle \lfloor p\rfloor\rangle$.
\end{notation}
\begin{definition}
$s$-$t$ path $$p=(u_1=s,u_2,\ldots,u_{m-1},u_{m}=t)$$ in graph $G$ is said
to be a tape-consistent {\normalfont (}tape-inconsistent{\normalfont)} path in graph
$G$ if $\omega\langle \lfloor p \rfloor\rangle$ is a tape-consistent
{\normalfont(}tape-inconsistent{\normalfont)} sequence of the computation steps of machine
$M$ on input $x$.
\end{definition}
So, nodes $u_1$ and $u_m$ are artificial constructs from the point of view of accepting
computation path of Turing machine $M$.
%

%
\mysectionb
{Concept of the construction of machine $M\langle\exists AcceptingPath\rangle$}
{Concept of the construction of machine $M\langle\exists AcceptingPath\rangle$}
\thickspace
\label{SubSec:MEAPConcept}

The concept of the construction of machine $M\langle\exists AcceptingPath\rangle$ is
based on the following proposition.
\begin{proposition}
\label{Prop:MapTConsToCompPaths}
There is one-to-one mapping from the set of the $\mu$-length
tape-consistent sequences of the computation steps of machine
$M$ on input $x$ onto the set of the $\mu$-length sequences
of the computation steps of machine $M$ on input $x$ that
correspond to the $\mu$-length computation paths of machine $M$ on
input $x$.
\begin{proof}
The proposition follows directly from the definition of sequences of the computation steps.
\end{proof}
\end{proposition}
%
%
\mysectionc
{Definitions for sets of sequences of computation steps}
{Definitions for sets of sequences of computation steps}
\thickspace
\begin{notation}
Let $TConsistSeqSet\langle x,q,\mu\rangle$ be the set of $\mu$-length
tape-consistent $q$-state sequences of the computation steps of machine
$M$ on input $x$.
\end{notation}
\begin{notation}
Let $TInconsistSeqSet\langle x,q,\mu\rangle$ be the set of the $\mu$-length
tape-inconsistent $q$-state sequences of the computation steps of machine $M$ on input $x$.
\end{notation}
\begin{notation}
Let $TArbitrarySeqSet\langle x,q,\mu\rangle$ be the set of the $\mu$-length
tape-arbitrary $q$-state sequences of the computation steps of machine $M$ on input $x$.
\end{notation}
\begin{notation}
Let $$Q\langle Any\rangle=Q\setminus \{q_{start}\}.$$
\end{notation}
\begin{notation}
Let $$TConsistSeqSet\langle x,S,\mu\rangle=\bigcup_{q\in S}{TConsistSeqSet\langle x,q,\mu\rangle}$$
for some set $S$ of the states of machine $M$.
\end{notation}
%
%
\mysectionc
{Determining if there exists an accepting computation path}
{Determining if there exists an accepting computation path}
\thickspace
\begin{proposition}
\label{Prop:ArbEqConsUnionIncons}
Set $$TArbitrarySeqSet\langle x,q,\mu\rangle$$ is the disjoint union of sets
$$TConsistSeqSet\langle x,q,\mu\rangle$$ and $$TInconsistSeqSet\langle x,q,\mu\rangle.$$
\begin{proof}
The following is to be shown:
\begin{align*}
&(TConsistSeqSet\langle x,q,\mu\rangle\cap
TInconsistSeqSet\langle x,q,\mu\rangle)=\emptyset
\end{align*}
and
\begin{align*}
&TArbitrarySeqSet\langle x,q,\mu\rangle\subseteq
(TConsistSeqSet\langle x,q,\mu\rangle\cup
TInconsistSeqSet\langle x,q,\mu\rangle).
\end{align*}

The first equality follows directly from the definitions of sequences of the computation steps.

Furthermore, inclusions
\begin{align*}
&TConsistSeqSet\langle x,q,\mu\rangle\subseteq
TArbitrarySeqSet\langle x,q,\mu\rangle
\end{align*}
and
\begin{align*}
&TInconsistSeqSet\langle x,q,\mu\rangle\subseteq
TArbitrarySeqSet\langle x,q,\mu\rangle
\end{align*}
also follow directly from the definitions of sequences of the computation steps.

The rest is to show that
$$TArbitrarySeqSet\langle x,q,\mu\rangle\subseteq
(TConsistSeqSet\langle x,q,\mu\rangle\cup
TInconsistSeqSet\langle x,q,\mu\rangle).$$

Let $\omega=(t_1,\ldots t_{\mu})$ be a tape-arbitrary sequence of the computation steps.
Then the following holds:
\begin{enumerate}
\item[1)]
{if one of 3.1) or 3.2) of definition
\ref{Def:TapeInconsistSeq} holds for some $t_i\in \omega$
then $$\omega\in TInconsistSeqSet\langle x,q,\mu\rangle;$$}
\item[2)]
{otherwise, $\omega\in TConsistSeqSet\langle x,q,\mu\rangle$.}
\end{enumerate}
\end{proof}
\end{proposition}
Proposition \ref{Prop:ArbEqConsUnionIncons} is not used directly in the construction of machine
$M\langle\exists AcceptingPath\rangle$; this proposition is used just to show that
the set of the tape-consistent sequences is considered as a subset of the more general set of the
tape-arbitrary sequences.

Let $M\langle NP\rangle$ be a non-deterministic single-tape Turing machine that decides
language $L$ and works in time $t(n)$. To determine if there exists a tape-consistent sequence
of the computation steps of machine $M\langle NP\rangle$ on input $x$, the following is performed:
\begin{enumerate}
\item[1)] 
{construct non-deterministic multi-tape Turing machine $M\langle TArbitrarySeqs\rangle$
such that there is one-to-one mapping from the set of the root-leaves paths
in the computation tree of machine $M\langle TArbitrarySeqs\rangle$,
denoted by $TArbitrarySeqTree$, onto the set of the root-leaves paths in
the $\mu$-length tape-arbitrary tree of the computation steps of machine $M\langle NP\rangle$
on input $x$;}
\item[2)] 
{construct a direct acyclic graph $TArbitrarySeqGraph$
of the nodes of tree $TArbitrarySeqTree$ as a result of deep-first (or breadth-first)
traversal of tree $TArbitrarySeqTree$ such that there is one-to-one mapping from the set
of the root-leaves paths in graph $TArbitrarySeqGraph$ onto the set of
the root-leaves paths in the $\mu$-length tape-arbitrary tree of
the computation steps of machine $M\langle NP\rangle$ on input $x$;
the features of the construction is as follows:
\begin{enumerate}
\item[2.1)]
{Turing machine $M\langle NP\rangle$ does not run explicitly, so the computation tree
of machine $M\langle TArbitrarySeqs\rangle$ is not built explicitly,}
\item[2.2)]
{the size of graph $TArbitrarySeqGraph$ is polynomial in $|x|$ wherein $|x|$ is the length
of the input;}
\end{enumerate}
} 
\item[3)]
{consider graph $TArbitrarySeqGraph$ as a subgraph of the direct acyclic control
flow graph $TArbSeqCFG$ of a deterministic computer program that
writes values to the tape cells and reads values from the tape
cells of machine $M\langle NP\rangle$;}
\item[4)]
{using reaching definitions analysis \cite{NNH05} on graph $TArbSeqCFG$
and on the set of the assignments to the tape cells and the set of the usages
of the tape cells, compute the set of the tape-consistent pairs of the computation steps;}
\item[5)]
{using the results of reaching definitions analysis and the notion of network flow,
reduce the problem of determining if there exists an accepting tape-consistent path
in the control flow graph to problem \notion{LP}; use polynomial time algorithm
to solve problem \notion{LP}
\cite{Kh80,K84};}
\item[6)]
{because proposition \ref{Prop:MapTConsToCompPaths} holds, there is one-to-one mapping from
the set of the tape-consistent accepting paths onto the set of the accepting computation paths of machine
$M\langle NP\rangle$; so one can determine if there exists an accepting computation path
of machine $M\langle NP\rangle$.}
\end{enumerate}
These steps are based on the following key feature of tape-arbitrary sequences of
the computation steps.

To say informally, if a path in computation tree $TArbitrarySeqTree$ starts in some
node then the segment of the path from the node to a leaf node does not depend on the
segment of the path from the source to the node. Therefore, all the subtrees of computation
tree $TArbitrarySeqTree$ that start at the equal nodes are the same, and the set of the paths
in the tree can be represented as the set of the paths in a graph.

So, to construct graph $TArbitrarySeqGraph$, computation tree $TArbitrarySeqTree$ is not built
explicitly; instead, the steps of machine $M\langle TArbitrarySeqs\rangle$ are simulated
to construct the nodes of the tree and to construct the graph at the same time.
If computation tree $TArbitrarySeqTree$ has $r$ subtrees that start with a node $u$
then $(r-1)$ subtrees are cut; it leads to the fact that the paths in the subtrees are not duplicated
in the graph.

The reason why the size of graph $TArbitrarySeqGraph$ is polynomial in $|x|$
is the following: If one computes the elements of a tape-arbitrary sequence of
computation steps of machine $M\langle NP\rangle$, one should know the current
computation step only; therefore, in that case one uses logarithmic space
and polynomial time.

On the contrary, to compute the elements of a tape-consistent sequence of the computation steps of
machine $M\langle NP\rangle$ directly (not using tape-arbitrary sequences), one should keep
all the symbols written on the tape of machine $M\langle NP\rangle$; therefore, in that
case one uses polynomial space and exponential time.

The construction of graph $TArbitrarySeqGraph$ is explained in Figure \ref{Fig:TArbSeqGrConstr};
there
\begin{enumerate}
\item[1)]
{$TAST$ is the shortened indication of $TArbitrarySeqTree$;}
\item[2)]
{$TASG$ is the shortened indication of $TArbitrarySeqGraph$;}
\item[3)]
{a subtree that is cut is orange-colored; nodes $u$ and $u'$ are the same in $TArbitrarySeqGraph$;}
\item[4)]
{new edge in graph $TArbitrarySeqGraph$ is green-colored;}
\item[5)]
{large green `arrow' indicates the transformation from the tree to the graph.}
\end{enumerate}
\begin{figure}
\centering
\includegraphics[height=7cm]{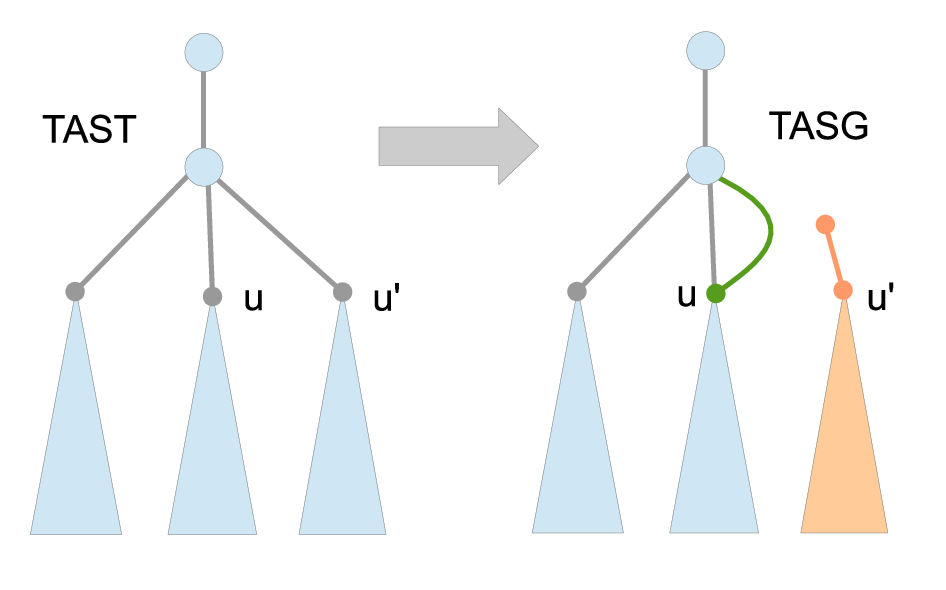}
\caption{Construction of graph $TArbitrarySeqGraph$.}
\label{Fig:TArbSeqGrConstr}
\end{figure}
%
%
\mysectionc
{How machine $M\langle\exists AcceptingPath\rangle$ works}
{How machine $M\langle\exists AcceptingPath\rangle$ works}
\thickspace
Turing machine $M\langle\exists AcceptingPath\rangle$ works as follows.
It performs a loop for $\mu$ from 1 determining at each iteration if
\begin{align*}
\exists\omega\,(\omega\in TConsistSeqSet\langle x,S,\mu\rangle\rangle)
\end{align*}
wherein $S$ is a set of the states of machine $M\langle NP\rangle$.
Since machine $M\langle NP\rangle$ works in time $t(n)$, one of the following happens:
\begin{enumerate}
\item[1)]
{if machine $M\langle NP\rangle$ accepts input $x$, $|x|=n$, then the loop stops
at iteration $\mu\le t(n)$ such that
$$\exists\omega\,(\omega\in TConsistSeqSet\langle x,F,\mu\rangle\rangle);$$
}
\item[2)]
{if machine $M\langle NP\rangle$ rejects input $x$, $|x|=n$, then
the loop stops at iteration $\mu\le (t(n)+1)$ such that
$$|TConsistSeqSet\langle x,Q\langle Any\rangle,\mu\rangle\rangle|=0$$
because there are no $t(n)_{>}$-length computation paths in that case
(here, (as is usual) by means of $|S|$ the cardinality of set $S$ is specified).}
\end{enumerate}
If $t(n)$ is a polynomial, then machine $M\langle\exists AcceptingPath\rangle$
works in polynomial time in $t(n)$ and therefore works in polynomial time in $n$
wherein $n=|x|$.

%
If machine $M\langle NP\rangle$ works according to definition \ref{Def:MTRejectsInput},
both to accept and to reject the input of machine $M\langle NP\rangle$,
polynomial $t(n)$ is not used in the program of machine $M\langle\exists AcceptingPath\rangle$.
Machine $M\langle\exists AcceptingPath\rangle$ should use polynomial $t(n)$ to reject the input
if machine $M\langle NP\rangle$ works according to the weaker definitions
\cite{DK00,GJ79,Papa94,G08}.

In fact, machine $M\langle\exists AcceptingPath\rangle$ is based on a reduction of the initial
string problem to another string problem that is \notion{NP}-complete and decidable in polynomial time
(\notion{TCPE} problem; section \ref{Sec:ProblemTCPE}). 
%
%
\mysectionb
{Differences from reduction $L\le_{m}^{P}\notioninform{3-CNF-SAT}$ in more detail}
{Differences from reduction $L\le_{m}^{P}\notioninform{3-CNF-SAT}$ in more detail}
\thickspace
%
%
Let $L$ be a language from class \notion{NP}; let $L$ be decidable by a non-deterministic
single-tape Turing machine $M$. The features of reduction $L\le_{m}^{P}\notioninform{3-CNF-SAT}$
\cite{C71} in detail compared to the solution suggested in the present paper are the following:
\begin{enumerate}
\item[1)]
{reduction $L\le_{m}^{P}\notioninform{3-CNF-SAT}$ sets in fact one-to-one mapping
from the set of the assignments that satisfy a Boolean formula
onto the set of the tape-consistent sequences of computation steps of machine $M$;}
\item[2)]
{in reduction $L\le_{m}^{P}\notioninform{3-CNF-SAT}$,
the set of the tape-consistent sequences of the computation steps is a subset
of the set of the paths in a graph which is implicitly constructed
($P_{s,t}^{j}$ \cite[page 153]{C71} are some nodes of this graph),
and the set of $s$-$t$ paths in the graph is not the set of tape-arbitrary
paths of machine $M$;}
\item[3)]
{an assignment that does not satisfy a Boolean formula can
correspond to sequences of the computation steps that do not correspond
to computation paths, so there is no one-to-one mapping from the set
of such assignments onto the set of the tape-inconsistent sequences of
the computation steps.}
\end{enumerate}
Thus, the difference between the solution suggested in the present paper and
reduction $L\le_{m}^{P}\notioninform{3-CNF-SAT}$ is as follows.

Reduction $L\le_{m}^{P}\notioninform{3-CNF-SAT}$ is in fact based on the notion of
tape-consistent sequences of the computation steps; in reduction $L\le_{m}^{P}\notioninform{3-CNF-SAT}$,
tape-consistent sequences are not considered as a subset of the more general set
of tape-arbitrary sequences of the computation steps. In contrast, the solution suggested
in the present paper is based on the concept of the set of tape-arbitrary sequences
of the computation steps that consists of the set of tape-consistent
sequences and the set of tape-inconsistent sequences. 

One can construct a graph of the tape-consistent sequences of the computation steps simulating
the moves of Turing machine $M$ (all the $s$-$t$ paths in such graph correspond to the tape-consistent
sequences of the computation steps), but in that case all the visited cells of the tape should be kept;
as a result, exponential time and space is used in that simulation. In contrast, polynomial time
and space is sufficient to construct $TArbitrarySeqGraph$, the graph of the tape-arbitrary sequences
of the computation steps.

Regarding reduction of problem \notion{3-CNF-SAT} to integer linear programming (for example,
problem \notion{0-1 ILP} \cite{K72}, and problem \notion{Simple D2CIF} \cite{EIS76}),
exponential time algorithms for problem \notion{ILP} are only known for now. 


%
\mysectionb
{Program of machine $M\langle\exists AcceptingPath\rangle$}
{Program of machine $M\langle\exists AcceptingPath\rangle$}
\thickspace
%
%
\mysectionc
{Non-deterministic multi-tape Turing machine $M\langle TArbitrarySeqs\rangle$}
{Non-deterministic multi-tape Turing machine $M\langle TArbitrarySeqs\rangle$}
\thickspace
Turing machine $M\langle TArbitrarySeqs\rangle$ is constructed as follows:
\begin{enumerate}
\item[1)]
{the input of the machine is a word $(x,\mu)$, wherein $x$ is a word
in alphabet $\Sigma$ and $\mu$ is a binary positive integer;}
\item[2)]
{the machine has one accepting state $q_{A}$ and state $q_{R}$, $q_{R}\ne q_{A}$,
referred to as rejecting state.}
\end{enumerate}


\noindent\hrulefill
\begin{program}
{Turing machine $M\langle TArbitrarySeqs\rangle$}
{}
\qinput Word $(x,\mu)$\\
\qcom{main loop}\\
\qfor each $\kappa^{(step)}\in [1..\mu]$\\
\qdo\\
  \qif $\kappa^{(step)}=1$\\
  \qthen\\
    {compute non-deterministically computation step
    $$t_1=(q_{start},s,q',s',m,1,1)$$
    of machine $M\langle NP\rangle$ wherein $s=Tape\langle x,1\rangle$}\\
    \qcontinue
  \qfi\\
  \qcom{end of if}\\
  \\
  \qif $\kappa^{(step)}=\mu$\\
  \qthen\\
    \qcom{machine $M\langle NP\rangle$ either stops or does not stop at step $\mu$}\\
    \qstop {\ at accepting state $q_{A}$}
  \qfi\\
  \qcom{end of if}\\
  \\
  {by computation step
  $$t_{\kappa^{(step)}}=(q,s,q',s',m,\kappa^{(tape)},\kappa^{(step)}),$$
  compute non-deterministically computation step
  $$t_{\kappa^{(step)}+1}=(q',s'',q'',s''',m',\kappa^{(tape+1)},\kappa^{(step)}+1)$$
  of machine $M\langle NP\rangle$ such that definition \ref{Def:CompStepSequentialPair} holds}\\
  \\
  \qif {there is no computation step $t_{\kappa^{(step)}+1}$}\\
  \qthen\\
    \qcom{machine $M\langle NP\rangle$ stops at step $\kappa^{(step)}$ such that
    $\kappa^{(step)}<\mu$}\\
    \qstop {\ at rejecting state $q_{R}$}
  \qfi\\
  \qcom{end of if}
\qrof\\ 
\qcom{end of main loop}
\end{program}
\noindent\hrulefill

%
\begin{proposition}
There is one-to-one mapping from the set of the root-leaves paths
in computation tree $TArbitrarySeqTree$, which is the computation tree of machine
$M\langle TArbitrarySeqs\rangle$, onto the set of the root-leaves paths in
the $\mu_{\le}$-length tape-arbitrary tree of the computation steps of machine
$M\langle NP\rangle$ on input $x$.
\end{proposition}
\begin{proposition}
The time complexity of non-deterministic Turing machine
$M\langle TArbitrarySeqs\rangle$ is polynomial in $\mu$, and the
space complexity is logarithmic in $\mu$.
\begin{proof}
Values $\kappa^{(tape)}$ and $\kappa^{(step)}$, contained in the computation steps
of a $\mu$-length sequence of the computation steps, are binary integers
such that $$\abs(\kappa^{(tape)})\le\mu$$ and $$\kappa^{(step)}\le\mu,$$ so the proposition
holds. 
\end{proof}
\end{proposition}
%
%
\mysectionc
{Deterministic algorithm $ConstructTArbitrarySeqGraph$}
{Deterministic algorithm $ConstructTArbitrarySeqGraph$}
\thickspace
To construct graph $TArbitrarySeqGraph$, the algorithm performs deep-first traversal
of computation tree $TArbitrarySeqTree$. The constructed graph is a direct acyclic
graph of polynomial size; it has one source node and a set of bottom node, and the nodes
contain computation steps that are build during the traversal.
As it is explained in subsection \ref{SubSec:MEAPConcept}, computation tree is not
build explicitly in the algorithm of the construction of graph $TArbitrarySeqGraph$.


\noindent\hrulefill
\begin{algorithm}
{$ConstructTArbitrarySeqGraph$}
{}
\qinput  Root node $r$ of tree $TArbitrarySeqTree$
\qoutput Graph $TArbitrarySeqGraph$\\
\qcom{initialization}\\
set $VisitedNodeSet:=\emptyset$\\
graph $G:=(\emptyset,\emptyset)$\\
\\
\qcom{main block}\\
$DFTConstructGraphFromNode(r)$\\
\\
\qreturn (graph $G$)
\end{algorithm}


\noindent\hrulefill
\begin{subalgorithm}
{$DFTConstructGraphFromNode$}
{}
\qinput   Node $u$ of tree $TArbitrarySeqTree$ 
\qupdates Set $VisitedNodeSet$, graph $G$\\
\qcom{check if node $u$ is already visited}\\
\qif $\exists u'\,(u'\in VisitedNodeSet)$ such that $u'.step=u.step$\\
\qthen\\
  \qreturn
\qfi\\
\qcom{end of if}\\
\\
\qcom{update variables}\\
{add $u$ to $VisitedNodeSet$}\\
{add $u$ to $Nodes\langle G\rangle$}\\
\\
\qcom{main loop}\\
\qfor each edge $(u,v)\in \delta^{+}(u)$\\
\qdo\\
  $DFTConstructGraphFromNode(v)$\\
  {add edge $(u,v)$ to $Edges\langle G\rangle$}
\qrof\\
\qcom{end of main loop}
\end{subalgorithm}
\noindent\hrulefill

%
Let's note that deep-first traversal, which is a recursive algorithm,
of tree $TArbitrarySeqTree$ can be simulated on a deterministic
multi-tape Turing machine using a non-recursive algorithm.
Breadth-first traversal of tree $TArbitrarySeqTree$ can be also used to construct the graph.
\begin{proposition}
There is one-to-one mapping from the set of the root-leaves paths in direct acyclic
graph $TArbitrarySeqGraph$ onto the set of the root-leaves paths in the $\mu$-length
tape-arbitrary tree of the computation steps of machine $M\langle NP\rangle$
on input $x$.
\end{proposition}
\begin{proposition}
The count of the nodes in graph $TArbitrarySeqGraph$ is polynomial in $\mu$.
\begin{proof}
Values $\kappa^{(tape)}$ and $\kappa^{(step)}$, contained in the computation steps
of a $\mu_{\le}$-length sequence of the computation steps, are binary integers
such that $\abs(\kappa^{(tape)})\le\mu$ and $\kappa^{(step)}\le\mu$.
Therefore, the count of the nodes in graph $TArbitrarySeqGraph$ is
$$\bigo\left(2^{C\cdot\log_2(\mu)}\right)$$ (total count of different computation steps
of $\mu_{\le}$-length sequences) which is $\bigo\left(\mu^{C}\right)$. So the proposition holds. 
\end{proof}
\end{proposition}
So, the count of the nodes in computation $TArbitrarySeqTree$ can be exponential in $\mu$,
but the count of the nodes in graph  $TArbitrarySeqGraph$ is polynomial in $\mu$.
\begin{proposition}
The time complexity of deterministic algorithm $$ConstructTArbitrarySeqGraph$$
is polynomial in $\mu$.
\end{proposition}
%
%
\mysectionc
{Control flow graph $TArbSeqCFG$}
{Control flow graph $TArbSeqCFG$}
\thickspace
Graph $TArbitrarySeqGraph$ is considered as a subgraph of the acyclic control flow graph
$$TArbSeqCFG$$ of a deterministic computer program that writes values to the tape cells and
reads values from the tape cells of machine $M\langle NP\rangle$.
Namely, each computation step $$t=(q,s,q',s',m,\kappa^{(tape)},\kappa^{(step)})$$
in nodes $$Nodes\langle TArbitrarySeqGraph\rangle,$$ is treated as the usage of symbol
$s$ in the tape cell with number $\kappa^{(tape)}$ and the assignment of symbol $s'$
to this tape cell. 

Let $S$ be a set of the states of machine $M\langle NP\rangle$.
Graph $TArbSeqCFG$ is constructed as follows:
\begin{enumerate}
\item[1)]
{let $$Nodes\langle TArbSeqCFG\rangle=Node\langle TArbitrarySeqGraph\rangle$$ and
$$Edges\langle TArbSeqCFG\rangle=Edges\langle TArbitrarySeqGraph\rangle;$$}
\item[2)]
{create in graph $TArbSeqCFG$ source node $s$, and add edge $(s,r)$
wherein $r$ is the root node of graph $TArbSeqCFG$; add to node $s$ a special assignment
which is treated as the assignment of the following symbols to each cell of the tape
of machine $M\langle NP\rangle$ when the machine starts:
\begin{enumerate}
\item[2.1)]
{input symbols for cells $c_{\kappa}$ if $\kappa\in[1..|x|]$, and}
\item[2.2)]
{blank symbol for cells $c_{\kappa}$ if $\kappa\notin[1..|x|]$;}
\end{enumerate}
}
\item[3)]
{
create in graph $TArbSeqCFG$ sink node $t$; connect $t$ with the bottom nodes $u$
such that $u.step$ contains a state $q\in S$; it is used so that the
computation paths of machine $M\langle NP\rangle$ ending with the states from $S$
are only considered;
}
\item[4)]
{add to node $t$ a special `extra' usage in such a way that if an assignment in node $u$ reaches
node $t$ then there is a tape consistent pair $$(u.step,t.step);$$ this usage is just a technical
solution and used for simplicity of linear program \notion{TCPEPLP} defined below; 
}
\item[5)]
{remove all the simple chains in graph $TArbSeqCFG$ (they do not end with the sink node $t$)
as it is shown in Figure \ref{Fig:TASGCutChains}; there $TASCFG$ is the shortened indication of
$TArbSeqCFG$, $q_{A}$ is an accepting state, $q_{R}$ is a rejecting state, and the elements
of the graph that are removed are red-colored.}
\end{enumerate}
\begin{figure}
\centering
\includegraphics[height=7cm]{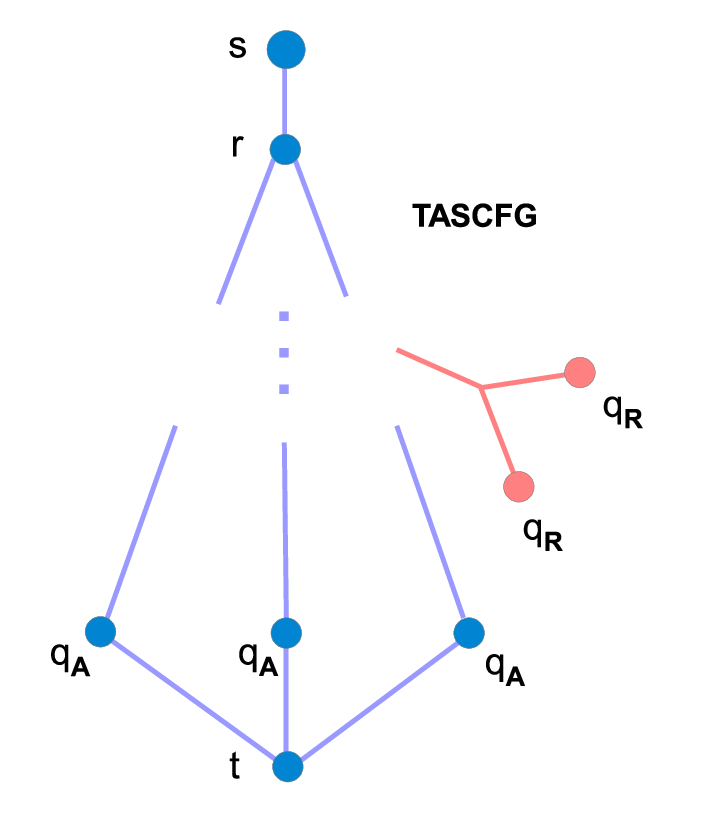}
\caption{Removing the simple chains in graph $TArbSeqCFG$.}
\label{Fig:TASGCutChains}
\end{figure}
%
%
\mysectionc
{Deterministic algorithm $ComputeTConsistPairSet$}
{Deterministic algorithm $ComputeTConsistPairSet$}
\thickspace
\begin{notation}
The set of pairs $(u,v)$ of the nodes of graph $TArbSeqCFG$, such
that pair $$(u.step,v.step)$$ is a tape-consistent pair of computation steps,
is denoted by $TConsistPairSet$.
\end{notation}
Algorithm $ComputeTConsistPairSet$ computes set $TConsistPairSet$.


\noindent\hrulefill
\begin{algorithm}
{$ComputeTConsistPairSet$}
{}
\qinput  Graph $TArbSeqCFG$
\qoutput Set $TConsistPairSet$\\
\qcom{initialization}\\
set $TConsistPairSet:=\emptyset$\\
\\
\qcom{main block}\\
{enumerate all the assignments to the tape cells in nodes
$Nodes\langle TArbSeqCFG\rangle$}\\
{enumerate all the usages of the tape cells in nodes
$Nodes\langle TArbSeqCFG\rangle$}\\
\\
{using the reaching definitions analysis on control flow graph
$TArbSeqCFG$ and on the sets of assignments and usages,
compute set $DefUsePairSet$ of the def-use pairs}\\
{call $ProcessDefUsePairSet$}\\
\\
\qreturn $TConsistPairSet$
\end{algorithm}


\noindent\hrulefill
\begin{subalgorithm}
{$ProcessDefUsePairSet$}
{}
\quses    Graph $TArbSeqCFG$, set $DefUsePairSet$
\qupdates Set $TConsistPairSet$\\
\qfor each pair $(def,use)\in DefUsePairSet$\\
\qdo\\
  \qcom{let $node_{def}$ and $node_{use}$ be nodes in
  $Nodes\langle TArbSeqCFG\rangle$ containing assignment
  and usage accordingly}\\
  \\
  \qif $node_{def}=s$\\
  \qthen\\
    \qcom{let $\kappa=node_{use}.step.\kappa^{(tape)}$}\\
    \qif $node_{use}.step.s=Tape\langle x,\kappa\rangle$\\
    \qthen\\
      {add pair $(node_{def},node_{use})$ to $TConsistPairSet$}
    \qfi\\
    \qcom{end of if}\\
    \qcontinue
  \qfi\\
  \qcom{end of if}\\
  \\
  \qif $node_{use}=t$\\
  \qthen\\
    {add pair $(node_{def},node_{use})$ to $TConsistPairSet$}\\
    \qcontinue
  \qfi\\
  \qcom{end of if}\\
  \\
  \qif pair $(node_{def}.step,node_{use}.step)$ is a pair such that
       definition \ref{Def:TapeConsistSeq} holds\\
  \qthen\\
    {add pair $(node_{def},node_{use})$ to $TConsistPairSet$}
  \qfi\\
  \qcom{end of if}
\qrof\\
\qcom{end of for loop}
\end{subalgorithm}
\noindent\hrulefill


\begin{proposition}
The time complexity of deterministic algorithm $$ComputeTConsistPairSet$$
is polynomial in $\mu$.
\begin{proof}
The time complexity of the reaching definition analysis is polynomial
in the count of the nodes and the count of the edges in the control
flow graph, so the proposition holds.
\end{proof}
\end{proposition}
%
%
\mysectionc
{Pseudocode of machine $M\langle\exists AcceptingPath\rangle$}
{Pseudocode of machine $M\langle\exists AcceptingPath\rangle$}
\thickspace
Deterministic multi-tape Turing machine $M\langle\exists AcceptingPath\rangle$
is constructed using deterministic algorithms
\begin{align*}
&ConstructTArbitrarySeqGraph,\ ComputeTConsistPairSet,\ and\\ 
&DetermineIfExistsTConsistPath;
\end{align*}
algorithm $$DetermineIfExistsTConsistPath,$$ which is defined in section \ref{Sec:ProblemTCPE},
determines if there exists a tape-consistent path in control flow graph $TArbSeqCFG$.
%

\noindent\hrulefill
\begin{program}
{The pseudocode of Turing machine $M\langle\exists AcceptingPath\rangle$}
{}
\qinput  Word $x$
\qoutput If there exists an accepting computation path of machine
         $M\langle NP\rangle$ on input $x$\\
\qcom{initialization}\\
integer $\mu:=1$\\
\\
\qcom{main loop}\\
\qwhile \qtrue\\
\qdo\\
  {graph $TArbitrarySeqGraph:=ConstructTArbitrarySeqGraph(x,\mu)$}\\
  \\
  {construct control flow graph $TArbSeqCFG$ wherein $S=F$}\\
  {set $TConsistPairSet:=ComputeTConsistPairSet(TArbSeqCFG)$}\\
  \\
  \qcom{$\mathcal{E}_F=\exists p\left(\omega\langle \lfloor p\rfloor\rangle\in
    TConsistSeqSet\langle x,F,\mu\rangle\right)$}\\
  {$\mathcal{E}_F:=DetermineIfExistsTConsistPath(TArbSeqCFG,TConsistPairSet)$}\\
  \\
  \qif $\mathcal{E}_F$\\
  \qthen\\
    {write $\mathbf{True}$ to the output}\\
    \qstop
  \qfi\\
  \qcom{end of if}\\
  \\
  {construct control flow graph $TArbSeqCFG$ wherein $S=Q\langle Any\rangle$}\\
  {set $TConsistPairSet:=ComputeTConsistPairSet(TArbSeqCFG)$}\\
  \\
  \qcom{$\mathcal{E}_{Any}=\exists p\left(\omega\langle \lfloor p\rfloor\rangle\in
    TConsistSeqSet\langle x,Q\langle Any\rangle,\mu\rangle\right)$}\\
  {$\mathcal{E}_{Any}:=DetermineIfExistsTConsistPath(TArbSeqCFG,TConsistPairSet)$}\\
  \\
  \qif $\neg (\mathcal{E}_{Any})$\\
  \qthen\\
    {write $\mathbf{False}$ to the output}\\
    \qstop
  \qfi\\
  \qcom{end of if}\\
  \\
  $\mu+:=1$
\qelihw\\
\qcom{end of main loop}\\
\\
\qcom{at this point, there is no $t(n)_{\le}$ computation paths wherein $n=|x|$}\\
{write $\mathbf{False}$ to the output}
\end{program}
\noindent\hrulefill

%
\begin{proposition}
If $M\langle NP\rangle$ is a non-deterministic single-tape
Turing machine that decides a language $L$ then deterministic
multi-tape Turing machine $M\langle\exists AcceptingPath\rangle$ determines
if there exists an accepting computation path of machine $M\langle NP\rangle$
on input $x$.
\begin{proof}
Machine $M\langle\exists AcceptingPath\rangle$ works as explained in subsection
\ref{SubSec:MEAPConcept}, so the machine determines if there exists an accepting computation path of
machine $M\langle NP\rangle$.
\end{proof}
\end{proposition}
\begin{proposition}
The time complexity of machine $M\langle\exists AcceptingPath\rangle$ is
polynomial in $t(n)$.
\begin{proof}
The time complexity of the algorithms, used in the program of machine
$M\langle\exists AcceptingPath\rangle$, is polynomial in $\mu$, and $\mu\in [1..(t(n)+1)]$,
wherein $n=|x|$; therefore, the time complexity of the machine is polynomial in $t(n)$.
\end{proof}
\end{proposition}
%
%
\mysectionb
{Time and space complexity of machine $M\langle\exists AcceptingPath\rangle$}
{Time and space complexity of machine $M\langle\exists AcceptingPath\rangle$}
\thickspace
Let $$n\_count\ \text{be}\ |Nodes\langle TArbitrarySeqGraph\rangle|$$ and
$$e\_count\ \text{be}\ |Edges\langle TArbitrarySeqGraph\rangle|.$$
\begin{notation}
Let constant
\begin{align}
\label{Eq:SigmaDef}
\sigma=\max_{q\in Q}|U_{(q)}|
\end{align}
wherein sets
$$U_{(q)}=\{s\ |\ ((q,s),(q',s',m))\in\Delta\}$$ and
$\Delta$ is the transition relation of machine $M\langle NP\rangle$.
\end{notation}
%
%
The estimations of the time and space complexities of the constructed Turing machines are shown
in Tables \ref{Tab:MEAPTimeCompl} and \ref{Tab:MEAPSpaceCompl}.
\begin{table}
\centering
\begin{tabular}{|l|l|c|}
\hline
\parbox[t][][c]{5.4cm}{
{\bf Algorithm/machine}} &
\parbox[t][][c]{4cm}{
{\bf Used algorithm}} &
\parbox[t][][c]{3.8cm}{
{\bf Overall time\\
complexity}}
\\ \hline
\parbox[t][][c]{5.4cm}{machine\\
$M\langle TArbitrarySeqs\rangle$\\
(not run explicitly)} &  &
\parbox[t][][c]{3.8cm}{
$\bigo(t(n)\log(t(n)))$\\
TM steps} \\ \hline
\parbox[t][][c]{5.4cm}{algorithm\\
$ConstructTArbitrarySeqGraph$} &  &
\parbox[t][][c]{3.8cm}{
$\bigo(t(n)^2\log(t(n)))$\\
VN operations} \\ \hline
\parbox[t][][c]{5.4cm}{algorithm\\
$ComputeTConsistPairSet$} &
\parbox[t][][c]{4cm}{
reaching definitions\\
analysis with time\\
complexity\\
$\bigo((n\_count_1)^2)$} &
\parbox[t][][c]{3.8cm}{
$\bigo(t(n)^6)$\\
VN operations} \\ \hline
\parbox[t][][c]{5.4cm}{algorithm\\
$DetermineIfExistsTConsistPath$} &
\parbox[t][][c]{4cm}{
Karmarkar's algorithm\\
with time complexity\\
$\bigo\left((v\_count)^{3.5}D\cdot M(D)\right)$} &
\parbox[t][][c]{3.8cm}{
$\bigo(2^{C\sigma} t(n)^{67})$\\
VN operations} \\ \hline
\parbox[t][][c]{5.4cm}{machine\\
$M\langle \exists AcceptingPath\rangle$} & &
\parbox[t][][c]{3.8cm}
{\overallcompla\\
TM steps} \\ \hline
\parbox[t][][c]{5.4cm}{
pseudocode algorithm of\\
machine $M\langle \exists AcceptingPath\rangle$} & &
\parbox[t][][c]{3.8cm}
{\overallopera\\
VN operations} \\ \hline
\end{tabular}
\caption{The time complexity of machine $M\langle \exists AcceptingPath\rangle$.}
\label{Tab:MEAPTimeCompl}
\end{table}
%
%
\begin{table}
\centering
\begin{tabular}{|l|l|c|}
\hline
\parbox[t][][c]{5.4cm}{
{\bf Algorithm/machine}} &
\parbox[t][][c]{4cm}{
{\bf Used algorithm}} &
\parbox[t][][c]{4cm}{
{\bf Overall space\\
complexity}} \\ \hline
\parbox[t][][c]{5.4cm}{machine\\
$M\langle TArbitrarySeqs\rangle$\\
(not run explicitly)} &  &
\parbox[t][][c]{3.8cm}{
$\bigo(\log(t(n)))$\\
TM tape cells} \\ \hline
\parbox[t][][c]{5.4cm}{algorithm\\
$ConstructTArbitrarySeqGraph$} &  &
\parbox[t][][c]{3.8cm}{
$\bigo(t(n)^2\log(t(n))^2)$\\
VN memory cells} \\ \hline
\parbox[t][][c]{5.4cm}{algorithm\\
$ComputeTConsistPairSet$} &
\parbox[t][][c]{4cm}{
reaching definitions\\
analysis with time\\
complexity\\
$\bigo((n\_count_1)^2)$} &
\parbox[t][][c]{3.8cm}{
$\bigo(t(n)^6)$\\
VN memory cells} \\ \hline
\parbox[t][][c]{5.4cm}{algorithm\\
$DetermineIfExistsTConsistPath$} &
\parbox[t][][c]{4cm}{
Karmarkar's algorithm\\
with time complexity\\
$\bigo\left((v\_count)^{3.5}D\cdot M(D)\right)$} &
\parbox[t][][c]{3.8cm}{
$\bigo(2^{C\sigma} t(n)^{39})$\\
VN memory cells} \\ \hline
\parbox[t][][c]{5.4cm}{machine\\
$M\langle \exists AcceptingPath\rangle$} & &
\parbox[t][][c]{3.8cm}
{$\bigo\left(2^{C\sigma} t(n)^{39}\right)$\\
TM tape cells} \\ \hline
\parbox[t][][c]{5.4cm}{
pseudocode algorithm of\\
machine $M\langle \exists AcceptingPath\rangle$} & &
\parbox[t][][c]{3.8cm}
{$\bigo\left(2^{C\sigma} t(n)^{39}\right)$\\
VN memory cells} \\ \hline
\end{tabular}
\caption{The space complexity of machine $M\langle \exists AcceptingPath\rangle$.}
\label{Tab:MEAPSpaceCompl}
\end{table}
%

The following is taken into account in the estimations of the time and space complexities
of the machine:
\begin{enumerate}
\item[1)]
{the count $n\_count$ of the nodes in graph $TArbitrarySeqGraph$ is
$\bigo(2^{\sigma}t(n)^2\log(t(n)))$ because the number of steps
that are needed to compute the next computation step is
$\bigo(\log(t(n)))$;}
\item[2)]
{the count $n\_count_1$ of the nodes, that correspond  to the computation
steps of Turing machine $M\langle NP\rangle$, in graph $TArbitrarySeqGraph$ is
$\bigo(2^{\sigma}t(n)^2)$;}
\item[3)]
{the count $e\_count$ of the edges in graph $TArbitrarySeqGraph$ is
$\bigo(2^{\sigma}\cdot n\_count)$;}
\item[4)]
{the length of the record of graph $TArbSeqCFG$ is $\bigo(n\_count\cdot \log_2(n\_count))$;
}
\item[5)]
{value $\eta=|TConsistPairSet|$ is $\bigo((n\_count_1)^2)$;}
\item[6)]
{the count $r$ of the paths in graph $TArbitrarySeqGraph$ is
$2^{\bigo\left(2^{\sigma}\cdot n\_count\right)}$;}
\item[7)]
{the count of nodes in each graph $ConnG\langle h',h''\rangle$ is $\bigo\left(\eta^2\right)$;}
\item[8)]
{integer $m$, which is declared in subsection \ref{SubSec:R2LP}, is $\bigo\left(\log(n\_count_1)\right)$;
}
\item[9)]
{the matrix of the equations of linear program \notion{TCPEPLP} is
$$\bigo\left(n\_count\cdot t(n)\cdot \eta)\times\bigo(e\_count\cdot t(n)\cdot \eta\right)$$ matrix;} 
\item[10)]
{the count of the equations in linear program \notion{TCPEPLP}
is $\bigo\left(n\_count\cdot t(n)\cdot \eta\right)$;}
\item[11)]
{the count $v\_count$ of the variables in linear program \notion{TCPEPLP}
is $\bigo\left(e\_count\cdot t(n)\cdot \eta\right)$;}
\item[12)]
{the length $D$ of the input of linear program \notion{TCPEPLP} is
$$\bigo\left(n\_count\cdot e\_count\cdot t(n)^2\cdot \eta^2\right);$$}
\item[13)]
{Karmarkas's algorithm performs $\bigo(v\_count^{3.5}D)$ operations with complexity $M(D)$
on $\bigo(D)$ digits numbers wherein $v\_count$ is the number of the variables and $D$ is
the length of the input of linear program \notion{TCPEPLP} ($M(D)$ denotes the time complexity
of multiplication of $\bigo(D)$ digits numbers);}
\item[14)]
{the number of the iterations in the loops of algorithm $DetermineIfExistsTConsistPath$
is $\bigo\left(t(n)^2\right)$;}
\item[15)]
{the number of the iterations in the main loop of machine $M\langle\exists AcceptingPath\rangle$
is $\bigo\left(t(n)\right)$;}
\item[16)]
{$\bigo\left(p^2 q\right)$ steps of multi-tape Turing machine are needed to get randomly
the elements of an array with $p$ elements;}
\item[17)]
{deep-first traversal, which is a recursive algorithm, of tree $TArbitrarySeqTree$
can be simulated on a deterministic multi-tape Turing machine using a stack of depth $\bigo(t(n))$.}
\end{enumerate}
%

%
\mysectiona
{\notion{NP}-complete problem \notion{TCPE}}
{\notion{NP}-complete problem \notion{TCPE}}
\thickspace
\label{Sec:ProblemTCPE}
Let $S$ be a set of the states of machine $M\langle NP\rangle$; let $\mu$ denote the length
of a sequence of the computation steps.
%
%
\mysectionb
{Setting up the problem}
{Setting up the problem}
\thickspace
\begin{notation}
The set of the tape-consistent paths of machine $M\langle NP\rangle$ on input $x$ in
control flow graph $TArbSeqCFG$ {\normalfont(}$s$-$t$ paths $p$ in graph $TArbSeqCFG$ such that
$\omega\langle \lfloor p\rfloor\rangle\in TConsistSeqSet\langle x,S,\mu\rangle${\normalfont)}
is denoted by $P\langle tcon\rangle$.
\end{notation}
\begin{definition}
The problem of determining if set $P\langle tcon\rangle$ is not empty is denoted by
{\normalfont \notion{TCPE} (Tape-Consistent Path Existence problem)}.
\end{definition}
Reduction $L\le_{m}^{P} \notioninform{TCPE}$, wherein machine $M\langle NP\rangle$ decides language $L$,
is provided (according to the definition of such reduction in \cite{DK00}) as follows:
\begin{enumerate}
\item[1)]
{Computable function $f:\Sigma^*\to\Gamma_1^*$ transfers each input string $x\in \Sigma^*$ of machine $M\langle NP\rangle$
to a string representation of graph $TArbSeqCFG$ and set $TConsistPairSet$ as descibed in subsection
\ref{SubSec:MEAPConcept}. Here an appopriate set $\Gamma_1$ and a reasonable encoding of the resulting objects are used;
for example, integers are recorded in binary notation.}
\item[2)]
{It is a polynomial time reduction because the construction of graph $TArbSeqCFG$ and set $TConsistPairSet$
works in polynomial time in $|x|$.}
\end{enumerate}
Because we have a computable function $f:\Sigma^*\to\Gamma_1^*$ such that for each $x\in \Sigma^*$,
$x\in L$ if and only if $f(x)\in \notioninform{TCPE}$, the following theorem holds.

\begin{theorem}
Problem \notion{TCPE} is \notion{NP}-complete.
\end{theorem}

Algorithm $DetermineIfExistsTConsistPath$, described in this section, solves problem \notion{TCPE}
finding a solution of linear program \notion{TCPEPLP} (see subsection \ref{SubSec:TCPEPLP})
for a network flow $\mathcal{PF}\langle G\rangle$ in graph $TArbSeqCFG$ (the network flow
is similar to multi-commodities network flow \cite{EIS76}, but not the same); there
exists a fractional solution of the linear program iff there exists a tape-consistent path
in graph $TArbSeqCFG$.

\begin{notation}
Let $V$ be $Nodes\langle TArbSeqCFG\rangle$; let $E$ be $Edges\langle TArbSeqCFG\rangle$.
\end{notation}
%
%
\mysectionb
{Making $2$-out-regular graph $TArbSeqCFG$}
{Making $2$-out-regular graph $TArbSeqCFG$}
\label{SubSec:Making2OutG}
\thickspace
Using $2$-out-regular graph is a key for the proof of proposition \ref{Prop:TCPEPLPProof},
so graph $TArbSeqCFG$ is preliminary transformed to a $2$-out-regular graph as follows.
%

\noindent\hrulefill

\begin{algorithm}
{$Make2OutRegularGraph$}
{}
\qinput   Graph $TArbSeqCFG$
\qoutput  $2$-out-regular graph $TArbSeqCFG$\\
\qfor each node $u\in V$, $u\ne t$, such that $\sigma^{+}(u)>2$\\
\qdo\\
  {let $\sigma^{+}(u)=\{(u,v_i)\ |_{i\in [1..m]}\}$}\\
  {remove all edges $e\in \sigma^{+}(u)$}\\
  \\
  {add `fake' nodes $f_{u,i,j}$ for each $i\in [2..(m-1)]$ and $j\in [1..i]$}\\
  \\
  \qfor each `fake' node $u$\\
  \qdo\\
    {$f_{u,i,j}.step.\kappa^{(tape)} := u.step.\kappa^{(tape)}$}
  \qrof\\
  \\
  {array $L:=[v_1,v_2,\ldots,v_m]$}\\
  \qfor each integer $i\in [(m-1)..2]$\\
  \qdo\\
    {add `fake' edge $(f_{u,i,i},L[i+1])$}\\
    \qfor each integer $j\in [i..1]$\\
    \qdo\\
      {add `fake' edge $(f_{u,i,j},L[j])$}
    \qrof\\
    \qcom{end of for loop}\\
    \\
    {array $L:=[f_{u,i,1},f_{u,i,2},\ldots,f_{u,i,i}]$}
  \qrof\\
  \qcom{end of for loop}
\qend\\
\qcom{end of for loop}\\
\\
\qreturn the transformed graph
\end{algorithm}

\noindent\hrulefill

Algorithm $Make2OutRegularGraph$ is explained in Figure \ref{Fig:Make2OutRG}; there
\begin{enumerate}
\item[1)]
{the removed edges are red-colored, 
}
\item[2)]
{the added `fake' nodes are blue-colored,
}
\item[3)]
{the added `fake' edges are green-colored, and
}
\item[4)]
{large green `arrow' indicates the transformation from the one graph to another graph.}
\end{enumerate}
\begin{figure}
\centering
\includegraphics[height=6.5cm]{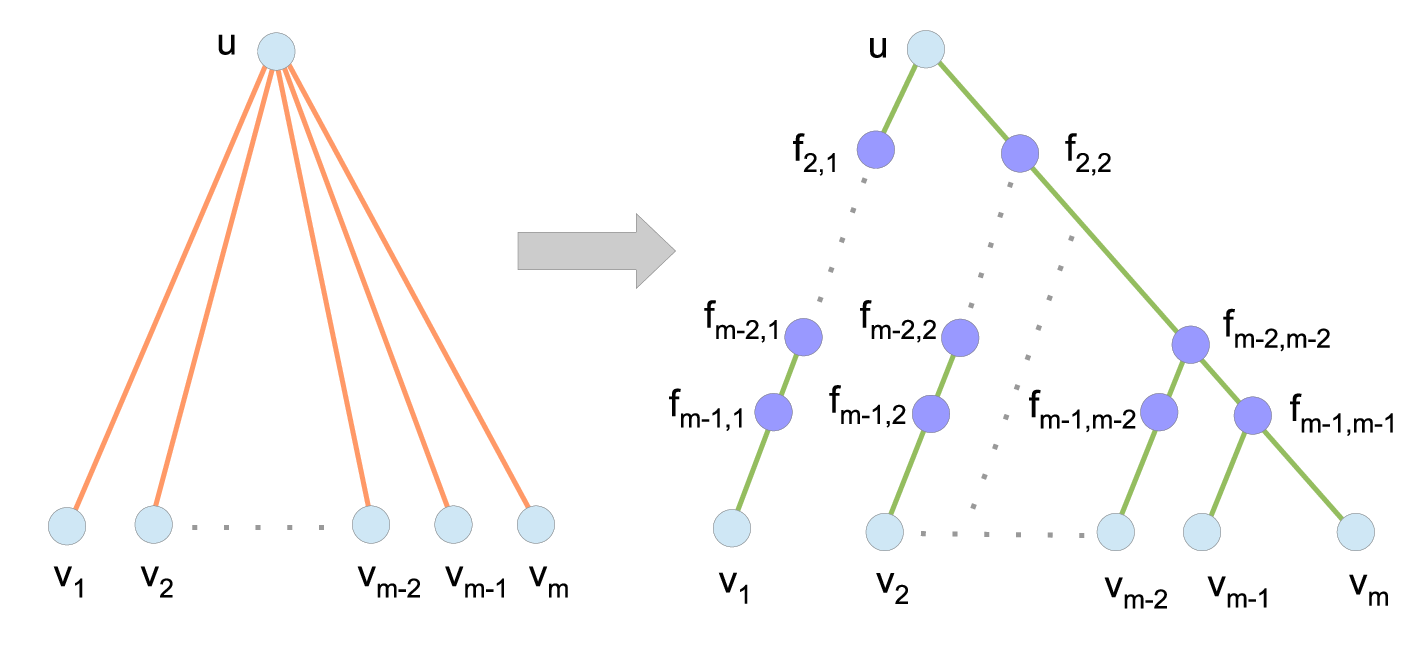}
\caption{Making $2$-out-regular graph $TArbSeqCFG$.}
\label{Fig:Make2OutRG}
\end{figure}
Let's note that $2$-out-regular graph $TArbSeqCFG$ has $\bigo(d^2\cdot |V|)$ nodes
and $\bigo(d^2\cdot |E|)$ edges wherein $$d=\max_{u\in V}{\sigma^{+}(u)}.$$
\begin{notation}
Let $FakeNodes$ be the set of `fake' nodes $f_{u,i,j}$ for each $i\in [2..(m-1)]$
and $j\in [1..i]$ wherein $m=\sigma^{+}(u)$.
\end{notation}
Let $p$ be a $s$-$t$ path in $2$-out-regular graph $TArbSeqCFG$.
\begin{notation}
Let's path $p$ without `fake' nodes $f_{u,i,j}$ denote by $$RemFakeNodes\langle p\rangle.$$
\end{notation} 
%
%
\mysectionb
{Commodities for tape-consistent pairs}
{Commodities for tape-consistent pairs}
\thickspace
%
%
\mysectionc
{Definition for the commodities}
{Definition for the commodities}
\thickspace
Let
\begin{enumerate}
\item[1)]
{integer $\eta=|TConsistPairSet|$;}
\item[2)]
{integer segment $CommSeg$ be $[1..\eta]$ (elements of $CommSeg$ are referred to as `commodity indeces');}
\item[3)]
{subgraphs $$G_i=Subgraph\langle TArbSeqCFG, (u_i,v_i)\rangle$$ for each $i\in CommSeg$ wherein
$(u_i,v_i)\in TConsistPairSet$;}
\item[4)]
{$V_i$ be $Nodes\langle G_i\rangle$ and $E_i$ be $Edges\langle G_i\rangle$.}
\end{enumerate}
Commodities $$K_i(s_i,t_i),\quad i\in CommSeg,$$ in graph $TArbSeqCFG$ are defined as follows:
\begin{enumerate}
\item[1)]
{the set of the nodes and the set of the edges of commodity $K_i$ are
$V_i$ and $E_i$ accordingly (excluding some nodes and edges as explained in paragraph
\ref{Par:RemoveHDefs});}
\item[2)]
{$s_i=u_i$ and $t_i=v_i$ (so $s_i=Source\langle V_i\rangle$ and $t_i=Sink\langle V_i\rangle$).}
\end{enumerate}
\begin{notation}
Let $i\in CommSeg$ and $p$ be a $s$-$t$ path in graph $TArbSeqCFG$.
If there exists path $\left(p\cap V_i\right)$ then we write $$\left(p\between K_i(s_i,t_i)\right).$$
\end{notation}
\begin{notation}
Let $p$ be a $s$-$t$ path in graph $TArbSeqCFG$; let's suppose that for each node $u\in p$, $u\ne t$,
there exist commodity $K_i(s_i,t_i)$ such that $u=s_i$ and $$\left(p\between K_i(s_i,t_i)\right)$$ holds. In that case,
the set of such integers $i$ is denoted by $K\langle p\rangle$.
\end{notation}
In other words, $K\langle p\rangle$ denotes the set of commodity indeces $i$ such that path $p$ intersects with
commodity $K_i(s_i,t_i)$.
%
%
\mysectionc
{Removing `hiding' definitions}
{Removing `hiding' definitions}
\thickspace
\label{Par:RemoveHDefs}
Let's consider node $d$ in $V_i$, $d\ne u_i$ and $d\ne v_i$, such that definition in $d.step$ `hides'
the definition in $u_i$ as follows:
\begin{enumerate}
\item[1)]
{such node is contained in a path $$p\in PathSet\langle(u_i,v_i)\rangle;$$}
\item[2)]
{$$d.step.\kappa^{(tape)}=u_i.step.\kappa^{(tape)}.$$}
\end{enumerate}
Excluding nodes from graph $G_i$ is shown in Figure \ref{Fig:CommExcludeDefs}; there the
excluded elements of graph $G_i$ are red-colored.
\begin{figure}
\centering
\includegraphics[height=7cm]{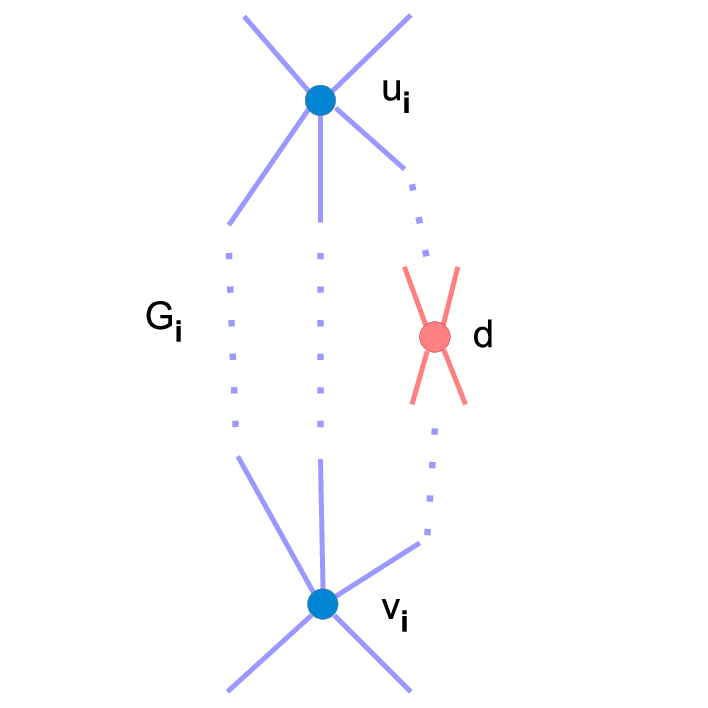}
\caption{Excluding the definitions that hide $u_{i}$.}
\label{Fig:CommExcludeDefs}
\end{figure}
Graphs $G_i$ without the excluded nodes can be easily computed using label propagate algorithm.
It is sufficient to do the following:
\begin{enumerate}
\item[1)]
{propagate a label $A$ from node $u_i$ to node $v_i$, excluding nodes
$d$ that hides the definition in $u_i$;}
\item[2)]
{propagate a label $B$ from node $v_i$ to node $u_i$;}
\item[3)]
{get the intersection of the subgraphs such that their nodes are marked by both
labels $A$ and $B$.}
\end{enumerate}
Removing `hiding' definitions in the commodities does not reduce the number of the
tape-consistent paths in graph $TArbSeqCFG$ (see proposition \ref{Prop:TapeConsToKSet}).
%
%
\mysectionc
{Tape segment for commodities}
{Tape segment for commodities}
\thickspace
Let integer segment $$TapeSeg$$ be $[L..R]$ wherein
$$TapeLBound\langle \mu\rangle\le L\le R\le TapeRBound\langle \mu\rangle;$$
let integer $\zeta \in TapeSeg$.
\begin{notation}
Let set
\begin{align*}
TConsistPairSet_{\zeta}=&\{(u,v)\ |\
(((u,v)\in TConsistPairSet) \wedge\\
&\wedge (u.step.\kappa^{(tape)}=\zeta))\}.
\end{align*}
\end{notation}
\begin{notation}
Let set
\begin{align*}
KSet_{\zeta}=\{K_i(s_i,t_i)\ | \ (s_i,t_i)\in TConsistPairSet_{\zeta}\}.
\end{align*}
\end{notation}
\begin{notation}
Let set
\begin{align*}
KSet\langle TapeSeg\rangle=\bigcup_{\zeta\in TapeSeg}KSet_{\zeta}.
\end{align*}
\end{notation}
\begin{notation}
Let set
$$CommSeg_{\zeta}=\{i\ |\ K_i(s_i,t_i)\in KSet_{\zeta}\}.$$
\end{notation}
\begin{notation}
Let set
\begin{align*}
CommSeg\langle TapeSeg\rangle=\bigcup_{\zeta\in TapeSeg}CommSeg_{\zeta}.
\end{align*}
\end{notation}
\begin{notation}
Let set $$CommSegPairs=CommSeg\langle TapeSeg\rangle\times CommSeg\langle TapeSeg\rangle.$$
\end{notation}
%
%
\mysectionb
{Linear equations $(X\sqcup X)^{(m)}$}
{Linear equations $(X\sqcup X)^{(m)}$}
\label{SubSec:R2LP}
\thickspace
Let
\begin{enumerate}
\item[1)]
{$m$ be an integer, $m > 5$;}
\item[2)]
{$\lambda$ be a rational;}
\item[3)]
{$\nu$ be an integer, $\nu\ge 2^{4}$;}
\item[4)]
{rational
\begin{align*}
\alpha=\sum_{i=1}^{\nu}\frac{1}{2^{i}};
\end{align*}
}
\item[5)]
{rational
\begin{align*}
\beta=\sum_{i=2}^{\nu}\frac{1}{2^{i}};
\end{align*}
}
\item[6)]
{integer pair set
$$R=\{(i,j)\ |\ ((1\le i\le \nu)\wedge(2\le j\le \nu+1))\};$$
}
\item[7)]
{integer pair set $$R'=R\setminus \{(1,2)\};$$
}
\item[8)]
{rational
\begin{align*}
\gamma=\sum_{(i,j)\in R'}\left(\frac{1}{2^i}\cdot \frac{1}{2^{j-1}}\right).
\end{align*}
}
\end{enumerate}
%
%
\mysectionc
{Set of linear equations $R2LPEqSet\langle X\times X\rangle$}
{Set of linear equations $R2LPEqSet\langle X\times X\rangle$}
\thickspace
Let $X$ be a rational, $0\le X$; let's represent $X$ as follows:
\begin{align}
\label{Eq:XDef1}
X = 2^{-1}X_1+2^{-1}\beta^{-1}\left(\sum_{i=2}^{\nu}{\frac{1}{2^{i}}X_i}\right)
\end{align}
wherein $X_i$, $i\in [1..\nu]$, are rationals, $0\le X_i$. Let's introduce set of linear equations,
denoted by $$R2LPEqSet\langle X\times X\rangle,$$ as follows:
\begin{enumerate}
\item[1)]
{rational variables $X_i$ for $i\in [1..\nu]$;}
\item[2)]
{linear equation \eqref{Eq:XDef1};
}
\item[3)]
{rational variable $B$;}
\item[4)]
{linear equations $X_{i}=B$ for $i\in [2..\nu]$;
}
\item[5)]
{rational variables $B_{i,j}$ for $(i,j)\in R$;
}
\item[6)]
{linear equations $B_{i,j}= B$ for $(i,j)\in R'$.
}
\end{enumerate}
The variables of this set of linear equations can be represented by the following matrix:
\begin{align*}
\begin{bmatrix}
1 & 2 & \hdotsfor{1} & \nu & \nu+1 & \hdotsfor{1} & 2\nu-1&\\
X_{1} & X_{2} & \hdotsfor{1} & X_{\nu} & & & &\\
B_{1,2} & B_{1,3} & \hdotsfor{1} & B_{1,\nu+1} & & & &\\ 
        & B_{2,2} & B_{2,3} & \hdotsfor{1} & B_{2,\nu+1} & & &\\ 
        &         & B_{3,2} & B_{3,3} & \hdotsfor{1} & B_{3,\nu+1} & &\\ 
& & \ddots & \hdotsfor{3} & &\\
& & & \hdotsfor{4} &\\
        &         &         &
B_{\nu,2} & B_{\nu,3} & \hdotsfor{1} & B_{\nu,\nu+1} &\\ 
\end{bmatrix}
\end{align*}
%
%
\mysectionc
{Set of linear equations $R2LPEqSet\langle X\sqcup X\rangle$}
{Set of linear equations $R2LPEqSet\langle X\sqcup X\rangle$}
\thickspace
%
%
The set of linear equations, denoted by $$R2LPEqSet\langle X\sqcup X\rangle,$$
are introduced as follows:
\begin{enumerate}
\item[1)]
{set of linear equations $$R2LPEqSet\langle X\times X\rangle;$$
}
\item[2)]
{
linear equations $$X_1 = B_{1,2} = 1 + 2(X-1);$$
}
\item[3)]
{rational variables $W_i$ for $i\in [1..\nu]$;
}
\item[4)]
{linear equations
\begin{align*}
&W_1=2^{-2}B_{1,2}+\frac{3}{4}\gamma^{-1}\left(\sum_{j=2}^{\nu+1}{\frac{1}{2^{1+j-1}}B_{1,j}}\right)
\end{align*}
and for $i\in[2..\nu]$ linear equations
\begin{align*}
&W_i=\frac{3}{4}\gamma^{-1}\left(\sum_{j=2}^{\nu+1}{\frac{1}{2^{i+j-1}}B_{i,j}}\right);
\end{align*}
}
\item[5)]
{rational variable $W\langle X\sqcup X\rangle$;}
\item[6)]
{linear equation 
\begin{align*}
W\langle X\sqcup X\rangle = \left(\sum_{i=1}^{\nu}{W_i}\right) + \lambda.
\end{align*}
}
\end{enumerate}
%
%
\mysectionc
{Set of linear equations $R2LPEqSet\langle (X\sqcup X)^{(m)}\rangle$}
{Set of linear equations $R2LPEqSet\langle (X\sqcup X)^{(m)}\rangle$}
\thickspace
%
%
\begin{definition}
Let $$R2LPEqSet\langle (X\sqcup X)\rhd Y\rangle$$ be set of linear equations
$$R2LPEqSet\langle Y\sqcup Y\rangle$$ wherein $Y=W\langle X\sqcup X\rangle$.
\end{definition}
\begin{definition}
Let $$R2LPEqSet\langle (X\sqcup X)^{(m)}\rangle,$$ wherein integer $m\ge 2$, be set of linear equations
$$R2LPEqSet\langle Z_{m}\sqcup Z_{m}\rangle$$ such that there are sets of linear equations
$$R2LPEqSet\langle (Z_{i-1}\sqcup Z_{i-1})\rhd Z_i \rangle$$ for $i\in [2..m]$ wherein $Z_1=X$.
In case $m=1$, $$R2LPEqSet\langle (X\sqcup X)^{(m)}\rangle$$ is defined to be
$R2LPEqSet\langle X\sqcup X\rangle$.
\end{definition}
\begin{notation}
Let rational variable $$W\langle X_{m}\sqcup X_{m}\rangle$$ of set of linear equations
$$R2LPEqSet\langle (X\sqcup X)^{(m)}\rangle$$ be denoted by $$W\langle (X\sqcup X)^{(m)}\rangle.$$
\end{notation}
%
%
\mysectionc
{Proposition on $X=1+\delta$}
{Proposition on $X=1+\delta$}
\thickspace
%
%
\begin{proposition}
\label{Prop:R2SqcupDeltaEQ}
For every solution of set of linear equations $$R2LPEqSet\langle (X\sqcup X)^{(m)}\rangle:$$
If $$X=1+\delta,$$ $\delta$ is a rational, $\delta\ge 0$, and $m\ge 1$,
then
\begin{align*}
W\langle (X\sqcup X)^{(m)}\rangle =
1+2^{-m}\delta + \lambda\left(\sum^{0}_{i=m-1}{2^{-i}}\right).
\end{align*}
\begin{proof}
By mathematical induction.

Base case. Let $m=1$; in that case,
\begin{enumerate}
\item[1)]
{
\begin{align*}
X_1 = B_{1,2} 
& = 1+2(X-1) = \\
& = 1+2\delta;
\end{align*}
}
\item[2)]
{
$X_i = 1$ for each $i\in [2..\nu]$;
}
\item[3)]
{
\begin{align*}
X
& = 2^{-1}X_1+2^{-1}\beta^{-1}\left(\sum_{i=2}^{\nu}{\frac{1}{2^{i}}X_i}\right) = \\
& = 2^{-1}(1+2\delta)+2^{-1}(1) = \\
& = 1+\delta;
\end{align*}
}
\item[4)]
{
\begin{align*}
W\langle (X\sqcup X)^{(m)}\rangle 
& = 2^{-2}B_{1,2}+\frac{3}{4}\gamma^{-1}\left(\sum_{(i,j)\in R'}{\frac{1}{2^{i+j-1}}B_{i,j}}\right) + \lambda = \\
& = 2^{-2}(1+2\delta)+\frac{3}{4}(1) + \lambda = \\
& = 1+2^{-1}\delta + \lambda = \\
& = 1+2^{-1}\delta + \lambda\left(\sum^{0}_{i=0}{2^{-i}}\right).
\end{align*}
}
\end{enumerate}

Inductive step. Let $m \ge 1$; let
\begin{align*}
X = 1+2^{-m}\delta + \lambda\left(\sum^{0}_{i=m-1}{2^{-i}}\right).
\end{align*}
In that case,
\begin{enumerate}
\item[1)]
{
\begin{align*}
X_1 = B_{1,2}
& = 1+2(X-1) = \\
& = 1+2\left(1 + 2^{-m}\delta + \lambda\sum^{0}_{i=m-1}{2^{-i}} - 1\right) = \\
& = 1+2^{-m+1}\delta + \lambda\left(\sum^{0}_{i=m-1}{2^{-i+1}}\right);
\end{align*}
}
\item[2)]
{
$X_i = 1$ for each $i\in [2..\nu]$;
}
\item[3)]
{
\begin{align*}
X
& = 2^{-1}X_1+2^{-1}\beta^{-1}\left(\sum_{i=2}^{\nu}{\frac{1}{2^{i}}X_i}\right) = \\
& = 2^{-1}\left(1+2^{-m+1}\delta + \lambda\sum^{0}_{i=m-1}{2^{-i+1}}\right) + 2^{-1}(1) = \\
& = 1+2^{-m}\delta + \lambda\left(\sum^{0}_{i=m-1}{2^{-i}}\right);
\end{align*}
}
\item[4)]
{
\begin{align*}
W\langle (X\sqcup X)^{(m+1)}\rangle
& = 2^{-2}B_{1,2}+\frac{3}{4}\gamma^{-1}\left(\sum_{(i,j)\in R'}{\frac{1}{2^{i+j-1}}B_{i,j}}\right) + \lambda = \\
& = 2^{-2}\left(1+2^{-m+1}\delta + \lambda\sum^{0}_{i=m-1}{2^{-i+1}}\right)+\frac{3}{4}(1) + \lambda = \\
& = 1+2^{-m-1}\delta + \left(\lambda\left(\sum^{0}_{i=m-1}{2^{-i-1}}\right) + \lambda\right) = \\
& = 1+2^{-(m+1)}\delta + \lambda\left(\sum^{0}_{i=m}{2^{-i}}\right).
\end{align*}
}
\end{enumerate}
So, one gets that point 1 holds for $m+1$.
\end{proof}
\end{proposition}
%

%
\mysectionb
{Connectors for commodities}
{Connectors for commodities}
\label{SubSec:ConnForComms}
\thickspace
%
%
\mysectionc
{Commodity layers}
{Commodity layers}
\label{Par:CommLayers}
\thickspace
%
%
Let's use the following auxiliary notations to define commodity layers:
\begin{enumerate}
\item[1)]
{let integer $H=\mu+2$; this value is the length of each $s$-$t$ path in graph $TArbSeqCFG$
(all the lengths of $s$-$t$ paths in graph $TArbSeqCFG$ are the same because graph $TArbSeqCFG$
has no backward and cross edges);
}
\item[2)]
{let integer segment $HSeg$ be $[1..(H-1)]$;
}
\item[3)]
{let set $VLevel_h$ be the set of the nodes $u\in V$ such that $Dist(s,u)=h$ in graph $TArbSeqCFG$.
}
\end{enumerate}
Also, let's add `fake' commodities in graph $TArbSeqCFG$ which correspond to `fake' nodes as follows:
\begin{enumerate}
\item[1)]
{for each `fake' node $f_{u,i,j}\in FakeNodes$, add `fake' pair $(u,f_{u,i,j})$ to set
$TConsistPairSet$;
}
\item[2)]
{for each `fake' node $f_{u,i,j}\in FakeNodes$, add commodity $K_i(s_i,t_i)$ to the set of
the commodities wherein $s_i=u$ and $t_i=f_{u,i,j}$;
}
\item[3)]
{set integer $$\eta=|TConsistPairSet|+|FakeNodes|;$$ let $CommSeg$ be $[1..\eta]$ for new value $\eta$.
}
\end{enumerate}
\begin{notation}
Let
$$\overset{\longleftrightarrow}{G_{i}}=Subgraph\langle TArbSeqCFG, PathSet\langle (s,s_i,t_i,t)\rangle\rangle.$$
\end{notation}  
\begin{notation}
Let $\overset{\longleftrightarrow}{V_{i}}$ be $Nodes\langle \overset{\longleftrightarrow}{G_{i}}\rangle$;
let $\overset{\longleftrightarrow}{E_{i}}$ be $Edges\langle \overset{\longleftrightarrow}{G_{i}}\rangle$.
\end{notation}
\begin{definition}
Commodity layer for $h\in HSeg$, denoted by $CommLayer\langle h\rangle$, is defined to be the set
of graphs $\overset{\longleftrightarrow}{G_{i}}$ sush that $s_i\in VLevel_h$.
\end{definition}
Commodity layer can include commodity indeces for `fake' commodities.
%
%
Commodity layers are explained in Figure \ref{Fig:CommLayers}.
\begin{figure}
\centering
\includegraphics[height=8cm]{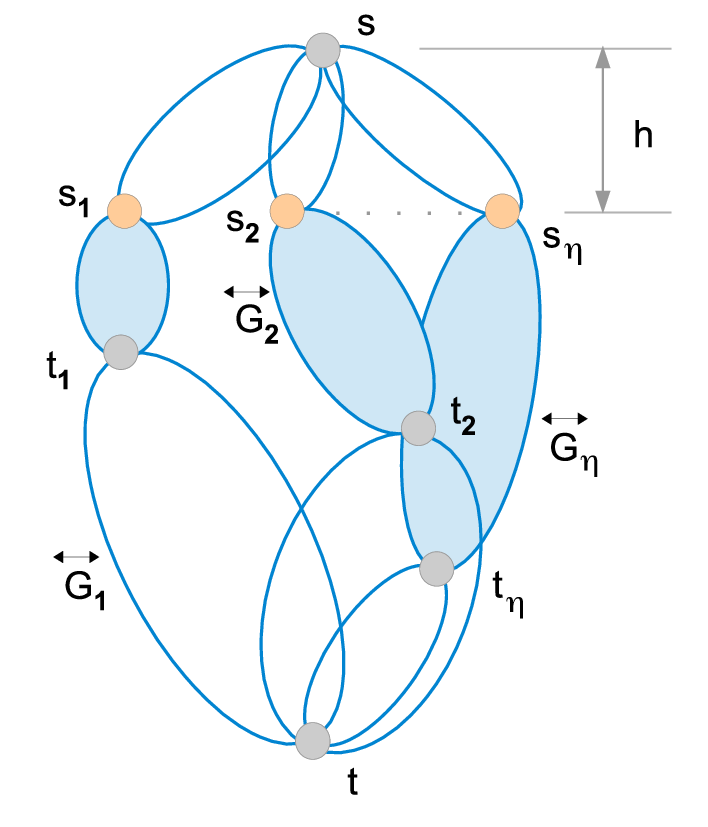}
\caption{Commodity layer $CommLayer\langle h\rangle$.}
\label{Fig:CommLayers}
\end{figure}
%
%
\mysectionc
{Connector graphs $ConnG\langle h',h''\rangle$}
{Connector graphs $ConnG\langle h',h''\rangle$}
\thickspace
%
%
Let integers $h'\in HSeg$ and $h''\in HSeg$. Graph $ConnG\langle h',h''\rangle$ is used to
define linear equations for connector graphs $ConnG\langle h',h''\rangle$ in paragraph \ref{Par:ConnGEqs}.
\begin{definition}
Pair $(i,j)\in CommSegPairs$, such that $i,j\in CommSeg\langle TapeSeg\rangle$ and $i\ne j$, is said
to be a pair of common path commodities if there exists $s$-$t$ path $p$ in graph $TArbSeqCFG$ such that
there exists paths $\left(p\cap \overset{\longleftrightarrow}{G_i}\right)$ and
$\left(p\cap \overset{\longleftrightarrow}{G_j}\right)$.
\end{definition}
\begin{definition}
Set $CPCPairSet$ is defined to be the set of the pairs of common path commodities.
\end{definition}
Pairs of common path commodities are explained in Figure \ref{Fig:CPCPairs}; there path $p$
is gray-colored.
\begin{figure}
\centering
\begin{tabular}{cc}
\includegraphics[height=7cm]{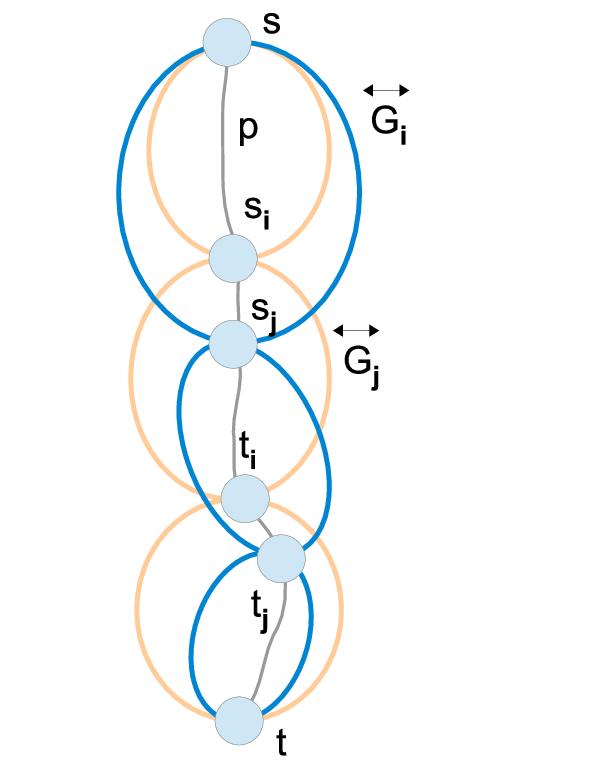}
\includegraphics[height=7cm]{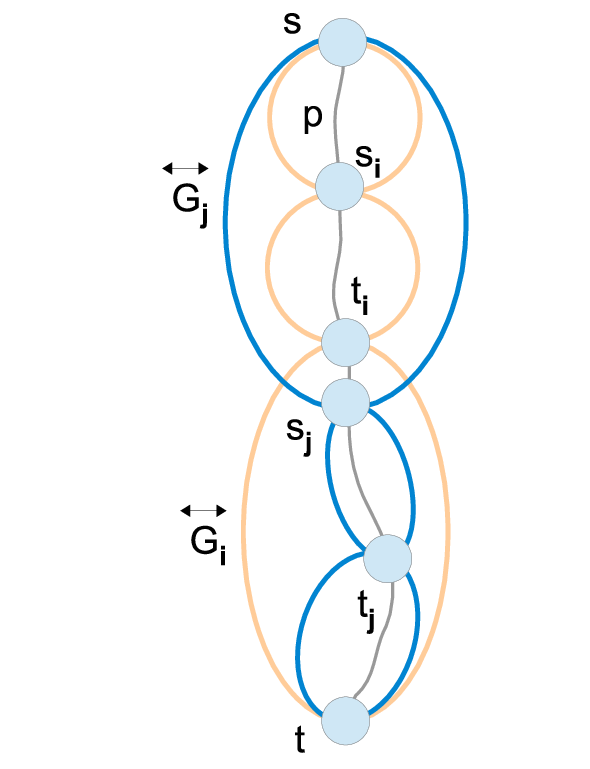}
\end{tabular}
\caption{Pairs of common path commodities.}
\label{Fig:CPCPairs}
\end{figure}
%
%
\begin{definition}
Let set
\begin{align*}
ConnElemSet\langle h\rangle=
\{i\ |\ (&(i\in CommSeg\langle TapeSeg\rangle)\wedge\\
&\wedge(\overset{\longleftrightarrow}{G_i}\in CommLayer\langle h\rangle))\}.
\end{align*}
\end{definition}
\begin{definition}
Let set
\begin{align*}
ConnPairSet\langle h',h''\rangle=
\{(i,j)\ |\ &((i,j)\in CommSegPairs)\wedge\\
&(\overset{\longleftrightarrow}{G_i}\in CommLayer\langle h'\rangle)\wedge\\
&(\overset{\longleftrightarrow}{G_j}\in CommLayer\langle h''\rangle))\}.
\end{align*}
\end{definition}
\begin{definition}
Set $ConnPairSet$ is defined to be set
\begin{align*}
\bigcup_{
\begin{array}{l}
h'\in HSeg,\\
h''\in HSeg,\\
h'< h''
\end{array}
}
ConnPairSet\langle h',h''\rangle.
\end{align*}
\end{definition}
\begin{notation}
Let set $$ConnPairElems\langle h',h''\rangle_1=\{i\ |\ (i,j)\in ConnPairSet\langle h',h''\rangle\}.$$
\end{notation}
\begin{notation}
Let set $$ConnPairElems\langle h',h''\rangle_2=\{j\ |\ (i,j)\in ConnPairSet\langle h',h''\rangle\}.$$
\end{notation}
\begin{notation}
Let set $$ConnPairElems=\{i\ |\ (i,j)\in ConnPairSet\}\cup \{j\ |\ (i,j)\in ConnPairSet\}.$$
\end{notation}
\begin{notation}
Let set $$hhPairs=\{(h',h'')\ |\ ((h'\in HSeg)\wedge (h''\in HSeg)\wedge (h'<h''))\}.$$
\end{notation}
Using algorithms $MakeConnG$, let's construct a special connector graph $ConnG\langle h',h''\rangle$.
%

\noindent\hrulefill

\begin{algorithm}
{$MakeConnG$}
{}
\qoutput  Connector graph $ConnG\langle h',h''\rangle$\\
{add nodes $r_s$ and $r_t$}\\
{set $Source\langle ConnG\langle h',h''\rangle\rangle=r_s$ and
$Sink\langle ConnG\langle h',h''\rangle\rangle=r_t$}\\
\\
\qfor each $i\in ConnPairElems\langle h',h''\rangle_1$\\
\qdo\\
  {add node $S_{i}$}
\qrof\\
\qcom{end of for loop}\\
\\
\qfor each $j\in ConnPairElems\langle h',h''\rangle_2$\\
\qdo\\
  {add node $T_{j}$}
\qrof\\
\qcom{end of for loop}\\
\\
\qfor each pair $(i,j)\in ConnPairSet\langle h',h''\rangle$\\
\qdo\\
  {add edge $(S_{i},T_{j})$}\\
  {add edge $(r_s,S_{i})$}\\
  {add edge $(T_{j},r_t)$}
\qrof\\
\qcom{end of for loop}\\
\\
\qreturn the constructed graph
\end{algorithm}

\noindent\hrulefill

Connector graph $ConnG\langle h',h''\rangle$ is shown in Figure \ref{Fig:ConnG};
there $k=|ConnPairSet\langle h',h''\rangle|$.
\begin{figure}
\centering
\includegraphics[height=8cm]{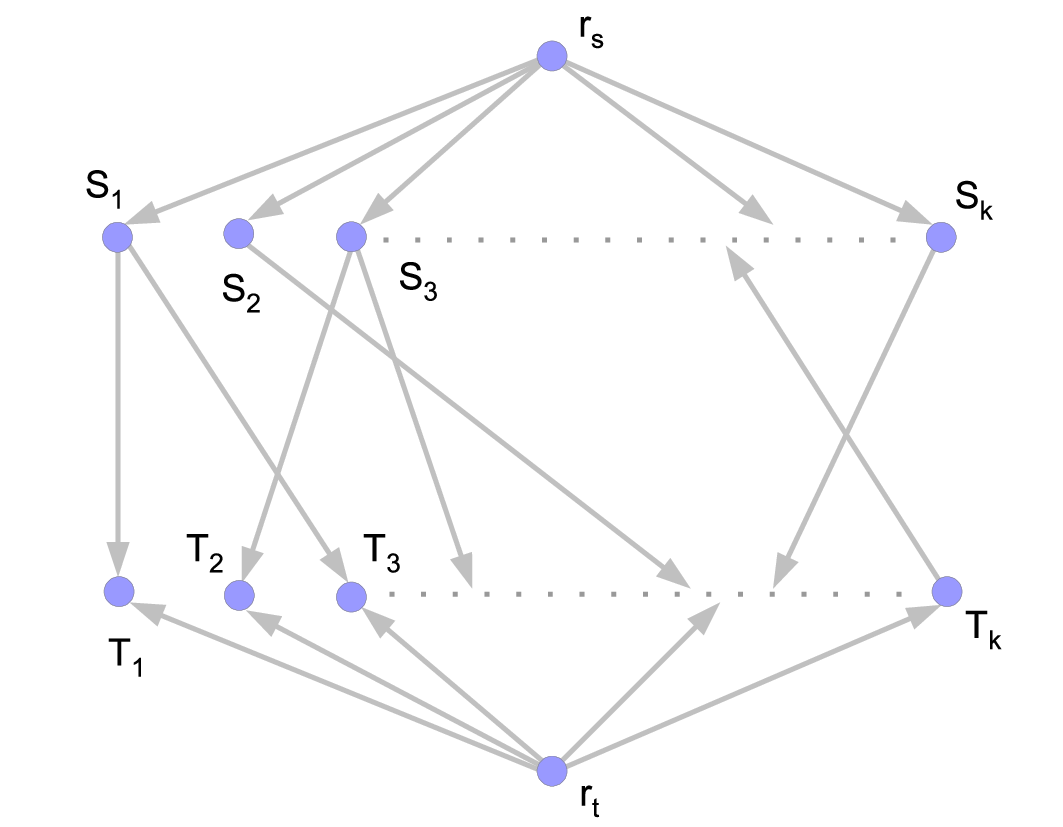}
\caption{Connector graph $ConnG\langle h',h''\rangle$.}
\label{Fig:ConnG}
\end{figure}
%
%
\mysectionc
{Linear equations for connector graphs}
{Linear equations for connector graphs}
\label{Par:ConnGEqs}
\thickspace
Let integer pair $(h',h'')\in hhPairs$; let
\begin{enumerate}
\item[1)]
{integers
\begin{align*}
&k_{h'} = |ConnPairElems\langle h',h''\rangle_1|\quad\text{and}\\
&k_{h''} = |ConnPairElems\langle h',h''\rangle_2|;
\end{align*}
}
\item[2)]
{rationals
\begin{align*}
&\xi_{h'}=(k_{h'})^{-1}\quad \text{and}\quad\\
&\xi_{h''}=(k_{h''})^{-1};
\end{align*}
}
\item[3)]
{rational
\begin{align*}
\xi=\min\left(\xi_{h'},\xi_{h''}\right);
\end{align*}
}
\item[4)]
{rational $$\overline{\delta}=\xi^{-1}$$ (let's note that $\overline{\delta}\ge 1$).
}
\end{enumerate}
%
%
Let's introduce set of linear equations, denoted by $$ConnLPEqSet\langle h',h''\rangle_0,$$ as follows:
\begin{enumerate}
\item[1)]
{for each $$i\in ConnPairElems\langle h',h''\rangle_1,$$ rational variable $A_{i}$;
for each $$j\in ConnPairElems\langle h',h''\rangle_2,$$ rational variable $B_{j}$;
for each pair $$(i,j)\in ConnPairSet\langle h',h''\rangle,$$ rational variable
$C_{(i,j)}$;
}
\item[2)]
{for each $$i\in ConnPairElems\langle h',h''\rangle_1,$$ rational variable $\delta_{a,i}$;
for each $$j\in ConnPairElems\langle h',h''\rangle_2,$$ rational variable $\delta_{b,j}$;
}
\item[3)]
{for each $i\in ConnPairElems\langle h',h''\rangle_1$, linear equations
\begin{align*}
\delta_{a,i}=\xi_{h'}^{-1} A_{i};
\end{align*}
for each $j\in ConnPairElems\langle h',h''\rangle_2$, linear equations
\begin{align*}
\delta_{b,j}=\xi_{h''}^{-1} B_{j};
\end{align*}\
let's note that $\delta_{a,i}\ge 1$ for $A_{i}\ge \xi_{h'}$ and
$\delta_{b,j}\ge 1$ for $B_{j}\ge \xi_{h''}$;
}
\item[4)]
{for each $$i\in ConnPairElems\langle h',h''\rangle_1,$$ rational variable $\overset{\triangleleft}{A}_{i}$;
for each $$j\in ConnPairElems\langle h',h''\rangle_2,$$ rational variable $\overset{\triangleleft}{B}_{j}$;
}
\item[5)]
{for each $i\in ConnPairElems\langle h',h''\rangle_1$, linear equations
\begin{align*}
\overset{\triangleleft}{A}_{i} = \delta_{a,i} + 1;
\end{align*}
for each $j\in ConnPairElems\langle h',h''\rangle_2$, linear equations
\begin{align*}
\overset{\triangleleft}{B}_{j} = \delta_{b,j} + 1;
\end{align*}      
}
%
%
\item[6)]
{for each $$i\in ConnPairElems\langle h',h''\rangle_1,$$ rational variable $(\delta\uparrow)_{a,i}$;
for each $$j\in ConnPairElems\langle h',h''\rangle_2,$$ rational variable $(\delta\uparrow)_{b,j}$;
}
\item[7)]
{
for each $i\in ConnPairElems\langle h',h''\rangle_1$, linear equations
$$(\delta\uparrow)_{a,i}=\delta_{a,i} + 2^{-8};$$
for each $i\in ConnPairElems\langle h',h''\rangle_1$, linear equations
and $$(\delta\uparrow)_{b,j}=\delta_{b,j} + 2^{-8};$$
}
\item[8)]
{for each $$i\in ConnPairElems\langle h',h''\rangle_1,$$ rational variable $\lambda_{a,i}$;
for each $$j\in ConnPairElems\langle h',h''\rangle_2,$$ rational variable $\lambda_{b,j}$;
}
\item[9)]
{
for each $i\in ConnPairElems\langle h',h''\rangle_1$, linear equations
\begin{align*}
\lambda_{a,i} = \frac{\left((\delta\uparrow)_{a,i}-\frac{7}{8}\delta_{a,i}\right)}{(\delta\uparrow)_{a,i}};
\end{align*}
for each $j\in ConnPairElems\langle h',h''\rangle_2$, linear equations
\begin{align*}
\lambda_{b,j} = \frac{\left((\delta\uparrow)_{b,j}-\frac{7}{8}\delta_{b,j}\right)}{(\delta\uparrow)_{b,j}};
\end{align*}
}
%
%
%
\item[10)]
{for each $i\in ConnPairElems\langle h',h''\rangle_1$, linear equations
\begin{align*}
R2LPEqSet\langle (\overset{\triangleleft}{A}_{i}\sqcup \overset{\triangleleft}{A}_{i})^{(m)}\rangle
\end{align*}
wherein $\delta=\delta_{a,i}$ and $\lambda=\lambda_{a,i}$;
for each $j\in ConnPairElems\langle h',h''\rangle_2$, linear equations
\begin{align*}
R2LPEqSet\langle (\overset{\triangleleft}{B}_{j}\sqcup \overset{\triangleleft}{B}_{j})^{(m)}\rangle
\end{align*}
wherein $\delta=\delta_{b,j}$ and $\lambda=\lambda_{b,j}$;
}
\item[11)]
{for each $$i\in ConnPairElems\langle h',h''\rangle_1,$$ rational variable $\overset{\triangleleft}{A}_{i}^{(m)}$;
for each $$j\in ConnPairElems\langle h',h''\rangle_2,$$ rational variable $\overset{\triangleleft}{B}_{j}^{(m)}$;
}
\item[12)]
{for each $i\in ConnPairElems\langle h',h''\rangle_1$, linear equations
\begin{align*}
\overset{\triangleleft}{A}_{i}^{(m)} = W\langle (\overset{\triangleleft}{A}_{i}\sqcup \overset{\triangleleft}{A}_{i})^{(m)}\rangle;
\end{align*}
for each $j\in ConnPairElems\langle h',h''\rangle_2$, linear equations
\begin{align*}
\overset{\triangleleft}{B}_{j}^{(m)} = W\langle (\overset{\triangleleft}{B}_{j}\sqcup \overset{\triangleleft}{B}_{j})^{(m)}\rangle;
\end{align*}
}
%
%
\item[13)]
{let's take integer $m$ (introduced above) in such a way that
\begin{align*}
2^{-m}\overline{\delta} < 2^{-4};
\end{align*}
in that case,
\begin{align*}
&2^{-m}\delta_{a,i} < 2^{-4}\quad\text{and}\quad\\
&2^{-m}\delta_{b,j} < 2^{-4}.
\end{align*}
}
\end{enumerate}
%
%
Let's consider non-empty network flow $$\mathcal{F}\langle ConnG\langle h',h''\rangle\rangle=
(F\langle ConnG\langle h',h''\rangle\rangle, H\langle ConnG\langle h',h''\rangle\rangle)$$
in connector graph $ConnG\langle h',h''\rangle$; let's define set of linear equation, denoted by
$$ConnLPEqSet\langle h',h''\rangle,$$ as follows:
\begin{enumerate}
\item[1)]
{set of linear equations $ConnLPEqSet\langle h',h''\rangle_0$;}
\item[2)]
{linear equations \eqref{Eq:NetwFlowEqs1} and \eqref{Eq:NetwFlowEqs2} for network flow
$\mathcal{F}\langle ConnG\langle h',h''\rangle\rangle$;
}
\item[3)]
{for each pair $(i,j)\in ConnPairSet\langle h',h''\rangle$, linear equations
\begin{align*}
C_{(i,j)}
&\ge \left(2^{-1} - \left(\overset{\triangleleft}{A}_{i}^{(m)} - 1\right)\right) +
\left(2^{-1} - \left(\overset{\triangleleft}{B}_{j}^{(m)} - 1\right)\right);
\end{align*}
}
\item[4)]
{
for each $i\in ConnPairElems\langle h',h''\rangle_1$, linear equations
\begin{align*}
F\langle ConnG\langle h',h''\rangle\rangle[S_i] = A_{i};
\end{align*}
for each $j\in ConnPairElems\langle h',h''\rangle_2$, linear equations
\begin{align*}
F\langle ConnG\langle h',h''\rangle\rangle[T_j] = B_{j};
\end{align*}
for each pair $(i,j)\in ConnPairSet\langle h',h''\rangle$, linear equations
\begin{align*}
H\langle ConnG\langle h',h''\rangle\rangle[(S_i,T_j)] \ge C_{(i,j)}.
\end{align*}
}
\end{enumerate}
%
%
\mysectionc
{Some lemmas for $W\langle (X\sqcup X)^{(m)}\rangle$}
{Some lemmas for $W\langle (X\sqcup X)^{(m)}\rangle$}
\label{Par:WLemmas}
\thickspace
%
%
Below are some lemmas that follow from proposition \ref{Prop:R2SqcupDeltaEQ}
and that are needed for the proofs in paragraph \ref{Par:ConnGEqsProp}.
At that, let's take into account that
\begin{align*}
1 < \left(\sum^{0}_{i=m-1}{2^{-i}}\right) < 2
\end{align*}
(due to $m>4$).

%
\begin{lemma}
For every solution of set of linear equations $R2LPEqSet\langle (X\sqcup X)^{(m)}\rangle$:
If $$\delta_{a,i} \ge 1$$ then
\begin{align*}
W\langle (X\sqcup X)^{(m)}\rangle =
1+2^{-m}\delta_{a,i} + \lambda_{a,i}\left(\sum^{0}_{i=m-1}{2^{-i}}\right);
\end{align*}
the same for $\delta_{b,j}$.
\end{lemma}
%
%
\begin{lemma}
For every solution of set of linear equations $R2LPEqSet\langle (X\sqcup X)^{(m)}\rangle$:
If $$\delta_{a,i}=0$$ then
\begin{align*}
W\langle (X\sqcup X)^{(m)}\rangle = 1+\left(\sum^{0}_{i=m-1}{2^{-i}}\right);
\end{align*}
the same for $\delta_{b,j}$.
\begin{proof}
If $\delta_{a,i}=0$ then
\begin{align*}
\lambda_{a,i} = \frac{\left((\delta\uparrow)_{a,i}-\frac{7}{8}\delta_{a,i}\right)}{(\delta\uparrow)_{a,i}} = 1,
\end{align*}
and
\begin{align*}
W\langle (X\sqcup X)^{(m)}\rangle = 1+\left(\sum^{0}_{i=m-1}{2^{-i}}\right).
\end{align*}
\end{proof}
\end{lemma}
%
%
\begin{lemma}
For every solution of set of linear equations $R2LPEqSet\langle (X\sqcup X)^{(m)}\rangle$:
If $$\delta_{a,i} \ge 1$$ then
\begin{align*}
1 + 2^{-2} < W\langle (X\sqcup X)^{(m)}\rangle < 1 + 2^{-1};
\end{align*}
the same for $\delta_{b,j}$.
\begin{proof}
We have:
\begin{align*}
W\langle (X\sqcup X)^{(m)}\rangle =
1+2^{-m}\delta_{a,i} + \lambda_{a,i}\left(\sum^{0}_{i=m-1}{2^{-i}}\right);
\end{align*}
\begin{align*}
\lambda_{a,i}
& = \frac{\left((\delta\uparrow)_{a,i}-\frac{7}{8}\delta_{a,i}\right)}{(\delta\uparrow)_{a,i}} = \\
& = \frac{\left(\delta_{a,i} + 2^{-8} -\frac{7}{8}\delta_{a,i}\right)}{\delta_{a,i} + 2^{-8}};
\end{align*}
\begin{align*}
\lambda_{a,i}
& = 1 - \frac{\left(\frac{7}{8}\delta_{a,i} + \frac{7}{8}2^{-8} - \frac{7}{8}2^{-8}\right)}{\delta_{a,i} + 2^{-8}} = \\
& = 1 - \frac{7}{8} + \frac{7}{8}2^{-8}\left(\delta_{a,i} + 2^{-8}\right)^{-1} < \\
& < 1 - \frac{7}{8} + 2^{-8} = \\
& < 2^{-3} + 2^{-8}.
\end{align*}
here $\delta_{a,i}\ge 1$ is taken into account. At that, $\lambda_{a,i}>2^{-3}$. Therefore,
\begin{align*}
W\langle (X\sqcup X)^{(m)}\rangle
& = 1+2^{-m}\delta_{a,i} + \lambda_{a,i}\left(\sum^{0}_{i=m-1}{2^{-i}}\right) < \\
& < 1 + 2^{-4} + 2\cdot(2^{-3} + 2^{-8}) < \\
& = 1 + 2^{-1};
\end{align*}
and
\begin{align*}
W\langle (X\sqcup X)^{(m)}\rangle &> 1+\lambda_{a,i}\left(\sum^{0}_{i=m-1}{2^{-i}}\right) > \\
& > 1 + 2^{-3}\cdot 2 = 1 + 2^{-2}.
\end{align*}
\end{proof}
\end{lemma}
%
%
\begin{lemma}
For every solution of set of linear equations $R2LPEqSet\langle (X\sqcup X)^{(m)}\rangle$:
If $$\delta_{a,i} \ge 1$$ then
\begin{align*}
0 < 2^{-1} - \left(W\langle (X\sqcup X)^{(m)}\rangle - 1\right) < 2^{-2};
\end{align*}
the same for $\delta_{b,j}$.
\begin{proof}
We have:
\begin{align*}
2^{-2} < W\langle (X\sqcup X)^{(m)}\rangle - 1 < 2^{-1},
\end{align*}
and one gets the result.
\end{proof}
\end{lemma}
%
%
\begin{lemma}
For every solution of set of linear equations $R2LPEqSet\langle (X\sqcup X)^{(m)}\rangle$:
If $$\delta_{a,i}=0$$ then
\begin{align*}
2^{-1} - \left(W\langle (X\sqcup X)^{(m)}\rangle - 1\right) < -2^{-1};
\end{align*}
the same for $\delta_{b,j}$.
\begin{proof}
We have:
\begin{align*}
2^{-1} - \left(\sum^{0}_{i=m-1}{2^{-i}}\right) < 2^{-1} - 1 = -2^{-1}.
\end{align*}
\end{proof}
\end{lemma}
%
%
\mysectionc
{Main property of linear equations for connector graphs}
{Main property of linear equations for connector graphs}
\label{Par:ConnGEqsProp}
\thickspace

\begin{proposition}
\label{Prop:ConnEqsProof1}
For every solution of set of linear equations $$ConnLPEqSet\langle h',h''\rangle,$$ the following holds
for each pair $(i,j)\in ConnPairSet$:
\begin{align}
\label{Eq:AxiBxiC}
(((A_{i}\ge \xi_{h'})\wedge (B_{j}\ge \xi_{h''}))\Rightarrow (C_{(i,j)}> 0).
\end{align}
\begin{proof}
In that case,
\begin{enumerate}
\item[1)]
{rationals $\delta_{a,i} \ge 1$ and $\delta_{b,j} \ge 1$;
}
\item[2)]
{
\begin{align*}
&2^{-1} - \left(\overset{\triangleleft}{A}_{i}^{(m)} - 1\right) > 0\quad\text{and}\\
&2^{-1} - \left(\overset{\triangleleft}{B}_{j}^{(m)} - 1\right) > 0
\end{align*}
(lemmas from paragraph \ref{Par:WLemmas});
}
\item[3)]
{                          
\begin{align*}
C_{(i,j)} > 0.
\end{align*}
}
\end{enumerate}
\end{proof}
\end{proposition}
%
%
\begin{proposition}
\label{Prop:ConnEqsProof2}
There exists solution of set of linear equations $ConnLPEqSet\langle h',h''\rangle$ such that
there exists a pair $(i,j)\in ConnPairSet\langle h',h''\rangle$ such that
\begin{enumerate}
\item[1)]
{
\begin{align*}
((A_{i}=1)\wedge (B_{j}=1)\wedge(C_{(i,j)}>0);
\end{align*}
}
\item[2)]
{
\begin{align*}
((A_{i'}=0)\wedge (B_{j'}=0)\wedge(C_{(i',j')}=0)
\end{align*}
for each pair $(i',j')\in ConnPairSet\langle h',h''\rangle$, $i'\ne i$ and $j'\ne j$;
}
\item[3)]
{
\begin{align*}
((A_{i}=1)\wedge (B_{j'}=0)\wedge(C_{(i,j')}=0)
\end{align*}
for each pair $(i,j')\in ConnPairSet\langle h',h''\rangle$, $j'\ne j$;
}
\item[4)]
{
\begin{align*}
((A_{i'}=0)\wedge (B_{j}=1)\wedge(C_{(i',j)}=0)
\end{align*}
for each pair $(i',j)\in ConnPairSet\langle h',h''\rangle$, $i'\ne i$.
}
\end{enumerate}
\begin{proof}
Let $A_{i}=1$ and $B_{j}=1$; let $A_{i'}=0$ for $i'\in ConnPairElems\langle h',h''\rangle_1$, $i'\ne i$
and $B_{j'}=0$ for $j'\in ConnPairElems\langle h',h''\rangle_2$, $j'\ne j$.

Point 1. In case of pair $(i,j)$, the following hold:
\begin{enumerate}
\item[1)]
{rationals $\delta_{a,i} \ge 1$ and $\delta_{b,j} \ge 1$;
}
\item[2)]
{
\begin{align*}
&2^{-1} - \left(\overset{\triangleleft}{A}_{i}^{(m)} - 1\right) > 0\quad\text{and}\\
&2^{-1} - \left(\overset{\triangleleft}{B}_{j}^{(m)} - 1\right) > 0
\end{align*}
(lemmas from paragraph \ref{Par:WLemmas});
}
\item[3)]
{                          
\begin{align*}
C_{(i,j)} > 0.
\end{align*}
}
\end{enumerate}
Point 2. In case of pair $(i',j')$, $i'\ne i$ and $j'\ne j$, the following hold:
\begin{enumerate}
\item[1)]
{rationals $\delta_{a,i} = 0$ and $\delta_{b,j} = 0$;
}
\item[2)]
{
\begin{align*}
&2^{-1} - \left(\overset{\triangleleft}{A}_{i}^{(m)} - 1\right) < (-2^{-1})\quad\text{and}\\
&2^{-1} - \left(\overset{\triangleleft}{B}_{j}^{(m)} - 1\right) < (-2^{-1})
\end{align*}
(lemmas from paragraph \ref{Par:WLemmas});
}
\item[3)]
{
\begin{align*}
C_{(i,j)} \ge 2\cdot(-2^{-1}) < 0;
\end{align*}
so, one can take $C_{(i',j')}=0$.
}
\end{enumerate}
Point 3. In case of pair $(i,j')$, $j'\ne j$, the following hold:
\begin{enumerate}
\item[1)]
{rationals $\delta_{a,i} \ge 1$ and $\delta_{b,j} = 0$;
}
\item[2)]
{
\begin{align*}
&2^{-1} - \left(\overset{\triangleleft}{A}_{i}^{(m)} - 1\right) < 2^{-2}\quad\text{and}\\
&2^{-1} - \left(\overset{\triangleleft}{B}_{j}^{(m)} - 1\right) < (-2^{-1})
\end{align*}
(lemmas from paragraph \ref{Par:WLemmas});
}
\item[3)]
{
\begin{align*}
C_{(i,j)} \ge 2^{-2} - 2^{-1} < 0;
\end{align*}
so, one can take $C_{(i,j')}=0$.
}
\end{enumerate}
Point 4. The same for pairs $(i',j)$, $i'\ne i$.
\end{proof}
\end{proposition}
Graphically proposition \ref{Prop:ConnEqsProof2} mean that there exists an edge $(S_i,T_j)$
in connector graph $ConnG\langle h',h''\rangle$ such that
\begin{align*}
(((F\langle ConnG&\langle h',h''\rangle\rangle[S_i]\ge \xi_{h'})\wedge
(F\langle ConnG\langle h',h''\rangle\rangle[T_{j}]\ge \xi_{h''}))\Rightarrow\\
&(H\langle ConnG\langle h',h''\rangle\rangle[(S_i,T_j)] > 0)).
\end{align*}
Let's note that $m$ is $\bigo\left(|V|\right)$ (here $|V|$ is the count of the nodes in graph
$TArbSeqCFG$).
%

%
\mysectionb
{Linear program \notion{TCPEPLP}}
{Linear program \notion{TCPEPLP}}
\thickspace
\label{SubSec:TCPEPLP}
%
%
\mysectionc
{Graphs $G\langle tcon\rangle_{\zeta}$}
{Graphs $G\langle tcon\rangle_{\zeta}$}
\thickspace
After `hiding' definitions are excluded, graphs of commodities are constructed as follows.
Let integer $\zeta \in TapeSeg$.
\begin{notation}
Let sets
\begin{align*}
&V\langle tcon,s\rangle_{\zeta}=\{s_i\ |\ K_i(s_i,t_i)\in KSet_{\zeta}\};\\
&V\langle tcon,t\rangle_{\zeta}=\{t_i\ |\ K_i(s_i,t_i)\in KSet_{\zeta}\}\}.
\end{align*}
\end{notation}
\begin{notation}
Let set
$V\langle tcon\rangle_{\zeta}=V\langle tcon,s\rangle_{\zeta}\cup V\langle tcon,t\rangle_{\zeta}$.
\end{notation}
\begin{notation}
Let set
$E\langle tcon\rangle_{\zeta}=TConsistPairSet_{\zeta}$.
\end{notation}
\begin{notation}
Let graph $$G\langle tcon\rangle_{\zeta}=(V\langle tcon\rangle_{\zeta},E\langle tcon\rangle_{\zeta}).$$
\end{notation}
Each graph $G\langle tcon\rangle_{\zeta}$, $\zeta\in TapeSeg$, is acyclic as a subgraph of
direct acyclic graph $TArbSeqCFG$.
\begin{definition}
\label{Def:KSetZetaPath}
$s$-$t$ path $$p=(s_1=s,..,t_1=s_2,..,t_2=s_3,..,t_{m}=t)$$ in graph $TArbSeqCFG$
is said to be $KSet_{\zeta}$-path if subpath $$p'=(s_1=s,t_1=s_2,t_2=s_3,..,t_m=t)$$ is
a path in graph $G\langle tcon\rangle_{\zeta}$ and there exists path
$$\left(p'\between K_{i_{\xi}}(s_{i_{\xi}},t_{i_{\xi}})\right)$$ for each subpath
$(s_{i_{\xi}},..,t_{i_{\xi}})$, $\xi\in [1..m]$, of path $p$ (at that $i_{\xi}\in CommSeg_{\zeta}$).
\end{definition}
\begin{definition}
$s$-$t$ path $p$ in graph $TArbSeqCFG$ is said to be $KSet$-path if $p$ is a
$KSet_{\zeta}$-path for each $\zeta\in CommSeg\langle TapeSeg\rangle$.
\end{definition}
Graph $G\langle tcon\rangle_{\zeta}$ and a $KSet_{\zeta}$-path $p$ are shown in Figure
\ref{Fig:GTConsZetaGraph}; there commodities $K_{i_{\xi}}(s_{i_{\xi}},t_{i_{\xi}})$ are blue-colored, and
path $p$ is green-colored.
\begin{figure}
\centering
\includegraphics[height=7cm]{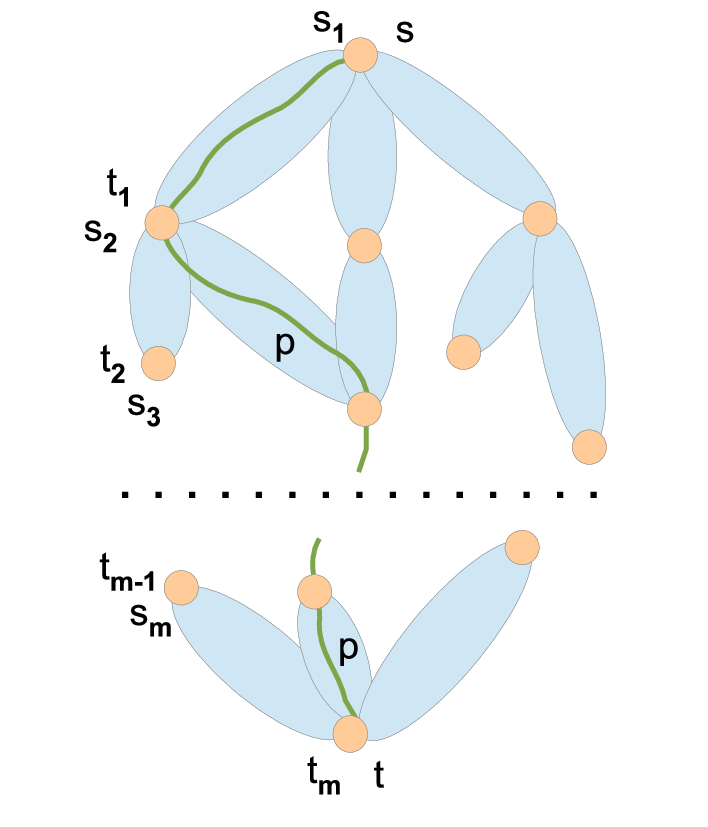}
\caption{Graph $G\langle tcon\rangle_{\zeta}$ and $KSet_{\zeta}$-path $p$.}
\label{Fig:GTConsZetaGraph}
\end{figure}

Algorithm $DetermineIfExistsTConsistPath$ is based on the following propositions.
\begin{proposition}
\label{Prop:TapeConsToKSet}
There is one-to-one mapping from the set of the tape-consistent paths in graph $TArbSeqCFG$
onto the set of $KSet$-paths in graph $TArbSeqCFG$.
\begin{proof}
It follows from the fact that subpaths $p'$ in definition \ref{Def:KSetZetaPath} of
$KSet_{\zeta}$-path correspond to subsequences $$\omega_{sub}=Subseq\langle \omega,\kappa\rangle$$
in point 3.1 of definition \ref{Def:TapeConsistSeq} if $\zeta$ is set to be equal to $\kappa$.
\end{proof}
\end{proposition}
\begin{proposition}
There is one-to-one mapping from the set of $KSet$-paths in the initial graph $TArbSeqCFG$
{\normalfont(}when the initial set of commodities is considered {\normalfont)} onto the set of
$KSet$-paths in the $2$-out-regular graph $TArbSeqCFG$ {\normalfont(}when the set of commodities
with `fake' commodities added is considered {\normalfont)}.
\begin{proof}
It follows from the definition of `fake' commodities in paragraph \ref{Par:CommLayers}.
\end{proof}
\end{proposition}
\begin{definition}
Commodity $$K_i(s_i,t_i)\in KSet_{\zeta}$$ is said to be `orphan' commodity if
$$\delta^{-}(s_i)=0\quad (s_i\ne s)$$ or $$\delta^{+}(t_i)=0\quad (t_i\ne t)$$ in
graph $G\langle tcon\rangle_{\zeta}$.
\end{definition}
Let's note that there may exist `orphan' commodities in graphs $G\langle tcon\rangle_{\zeta}$.
%
%
\mysectionc
{Definition for the linear program}
{Definition for the linear program}
\thickspace
\begin{notation}
Let set
$$ZLS(u)=\{\zeta\ |\ ((\zeta\in TapeSeg)\wedge
(u\in V\langle tcon\rangle_{\zeta}))\}.$$
\end{notation}
Let's introduce set of linear equations, denoted by $$TCPEPLPEqSet,$$ as follows:
\begin{enumerate}
\item[1)]
{for each graph $G_i$ of commodities $K_i(s_i,t_i)$, $i\in CommSeg\langle TapeSeg\rangle$,
network flow (linear equations \eqref{Eq:NetwFlowEqs1} and \eqref{Eq:NetwFlowEqs2})
$$\mathcal{PF}\langle K\rangle_i=(F\langle K\rangle_i,H\langle K\rangle_i)$$
in the graph;
}
\item[2)]
{
for each graph $G\langle tcon\rangle_{\zeta}$, $\zeta\in TapeSeg$, network flow
(linear equations \eqref{Eq:NetwFlowEqs1} and \eqref{Eq:NetwFlowEqs2})
$$\mathcal{F}\langle tcon\rangle_{\zeta}=(F\langle tcon\rangle_{\zeta},H\langle tcon\rangle_{\zeta})$$
in the graph;
}
\item[3)]
{for each $\zeta\in TapeSeg$, functions $F\langle tcon,sum\rangle_{\zeta}$
from nodes to rationals and linear equations
\begin{align*}
F\langle tcon,sum\rangle_{\zeta}[u]=
\sum_{j\in CommSeg_{\zeta}}F\langle K\rangle_j[u]
\end{align*}
for each node $u\in V$, $u\notin V\langle tcon\rangle_{\zeta}$;
\begin{align*}
F\langle tcon,sum\rangle_{\zeta}[u]=
F\langle tcon\rangle_{\zeta}[u]
\end{align*}
for each node $u\in V\langle tcon\rangle_{\zeta}$ (according to subsection \ref{SubSec:NetwFlows},
$F\langle K\rangle_j[u]=0$ if $u\notin V_j$);
}
\item[4)]
{
for the whole graph $TArbSeqCFG$, $1$-$1$ network flow
(linear equations \eqref{Eq:NetwFlowEqs1} and \eqref{Eq:NetwFlowEqs2})
$$\mathcal{PF}\langle G\rangle=(F\langle G\rangle,H\langle G\rangle)$$
in the graph;
}
\item[5)]
{for each $\zeta\in TapeSeg$, linear equations
\begin{align*}
F\langle tcon,sum\rangle_{\zeta}[u]=F\langle G\rangle[u]
\end{align*}
for each node $u\in V$;
}
\item[6)]
{in case $$|ZLS(u)|<|TapeSeg|,$$ $u\ne s$ and $u\ne t$, linear equations
$$F\langle tcon,sum\rangle_{\zeta}[u]=0;$$
}
\item[7)]
{in case $\delta^{+}(u)=\emptyset$ or $\delta^{-}(u)=\emptyset$ in graph $G\langle tcon\rangle_{\zeta}$,
$u\ne s$ and $u\ne t$, linear equations $$F\langle tcon,sum\rangle_{\zeta}[u]=0$$
(the case of `orphan' commodities); 
}
\item[8)]
{for each graph $\overset{\longleftrightarrow}{G_i}$, $i\in CommSeg\langle TapeSeg\rangle$,
network flow (linear equations \eqref{Eq:NetwFlowEqs1} and \eqref{Eq:NetwFlowEqs2})
$$\mathcal{PF}\langle \overset{\longleftrightarrow}{K}\rangle_i=
(F\langle \overset{\longleftrightarrow}{K}\rangle_i,H\langle \overset{\longleftrightarrow}{K}\rangle_i);$$
in the graph;
}
\item[9)]
{for each $i\in CommSeg\langle TapeSeg\rangle$, linear equations
$$F\langle K\rangle_i[u]=F\langle \overset{\longleftrightarrow}{K}\rangle_i[u]$$
for each node $u\in V_i$;
}
\item[10)]
{intersection network flows (linear equations \eqref{Eq:NetwFlowEqs1} and \eqref{Eq:NetwFlowEqs2})
$$\mathcal{IGF}\langle K\rangle_{i,j}=(F\langle IGF\rangle_{i,j},H\langle IGF\rangle_{i,j})$$
for intersection graphs $IG_{i,j}$ (defined below in paragraph \ref{Par:IntersectGraphs});}
\item[11)]
{for every pair $(i,j)\in ConnPairSet$, linear equations
\begin{align*}
&F\langle \overset{\longleftrightarrow}{K}\rangle_i[u]\ge F\langle IGF\rangle_{i,j}[u],\\
&F\langle \overset{\longleftrightarrow}{K}\rangle_j[u]\ge F\langle IGF\rangle_{i,j}[u],
\end{align*}
for each node $u\in Nodes\langle IG_{i,j}\rangle$;
}
\item[12)]
{for every pair $(i,j)\in ConnPairSet$, linear equations
\begin{align*}
&H\langle \overset{\longleftrightarrow}{K}\rangle_i[e]\ge H\langle IGF\rangle_{i,j}[e],\\
&H\langle \overset{\longleftrightarrow}{K}\rangle_j[e]\ge H\langle IGF\rangle_{i,j}[e],
\end{align*}
for each node $e\in Edges\langle IG_{i,j}\rangle$;
}
\item[13)]
{for each pair $(h',h'')\in hhPairs$, set of linear equations $ConnLPEqSet\langle h',h''\rangle$;
}
\item[14)]
{for each $i\in ConnPairElems\langle h',h''\rangle_1$, linear equations
\begin{align*}
F\langle \overset{\longleftrightarrow}{K}\rangle_i[s]=A_{i};
\end{align*}
for each $j\in ConnPairElems\langle h',h''\rangle_2$, linear equations
\begin{align*}
F\langle \overset{\longleftrightarrow}{K}\rangle_j[s]=B_{j};
\end{align*}
}
\item[15)]
{for each pair $(i,j)\in CPCPairSet$, linear equations
\begin{align*}
&F\langle IGF\rangle_{i,j}[s] \ge C_{(i,j)},\\
&F\langle IGF\rangle_{i,j}[s] \le 1;
\end{align*}
otherwise, $F\langle IGF\rangle_{i,j}[s]=0$;
}
\item[16)]
{for each pair $(i,j)\in CPCPairSet$ and each node $u\in IG_{i,j}$, $u\ne t$,
linear equations $$IntersectLPEqSet\langle (i,j),u\rangle$$ defined below in
paragraph \ref{Par:IntersectEqs};
}
\item[17)]
{for each $i\in CommSeg\langle TapeSeg\rangle$, if $i\notin ConnPairElems$
then $\mathcal{PF}\langle K\rangle_i$ is an empty network flow;
}
\end{enumerate}
here $A_{i}$, $B_{j}$, $C_{(i,j)}$ are rational variables of sets of linear equations
$ConnLPEqSet\langle h',h''\rangle$, and $m$ is a constant defined at the beginning of
subsection \ref{SubSec:R2LP} in point 1.

Let's introduce linear program, denoted by $$\notioninform{TCPEPLP},$$
(Tape-Consistent Path Existence Problem Linear Program)
with linear equations $TCPEPLPEqSet$:
\begin{align}
\label{Eq:TCPEPLPDef}
\begin{array}{ll}
\text{\bf minimize}   & 1\\
\text{\bf subject to} & (U x=b)\ \text{and}\ (x\ge \nullmatrix)\ \text{and}\\
& \text{(x is a fractional vector).}
\end{array}
\end{align}
An explanation of linear program \eqref{Eq:TCPEPLPDef} is shown in Figure \ref{Fig:DetIfExistsPathLP};
there
\begin{enumerate}
\item[1)]
{$TASCFG$ is the shortened indication of $TArbSeqCFG$;}
\item[2)]
{$\omega\langle \lfloor p\rfloor\rangle$, wherein path $p=(u_1,\ldots,u_7)$, is a tape-consistent
sequence of the computation steps;}       
\item[3)]
{$F\langle G\rangle[u_j]=1$ for the nodes in path $p$;}
\item[4)]
{let $s_i=u_i$ and $t_i=u_{i+1}$; commodities $K_i(s_i,t_i)$ from $KSet_{\zeta'}$, such that
subpath $$(s_i,..,t_i)\in V_i,$$ are blue-colored, and commodities $K_j(s_j,t_j)$ from
$KSet_{\zeta''}$, such that  subpath $$(s_j,..,t_j)\in V_j,$$ are orange-colored
($\zeta',\zeta''\in TapeSeg$);}
\item[5)]
{$p$ is a $KSet_{\zeta'}$-path, and $p$ is also a $KSet_{\zeta''}$-path.}
\end{enumerate}
\begin{figure}
\centering
\includegraphics[height=6cm]{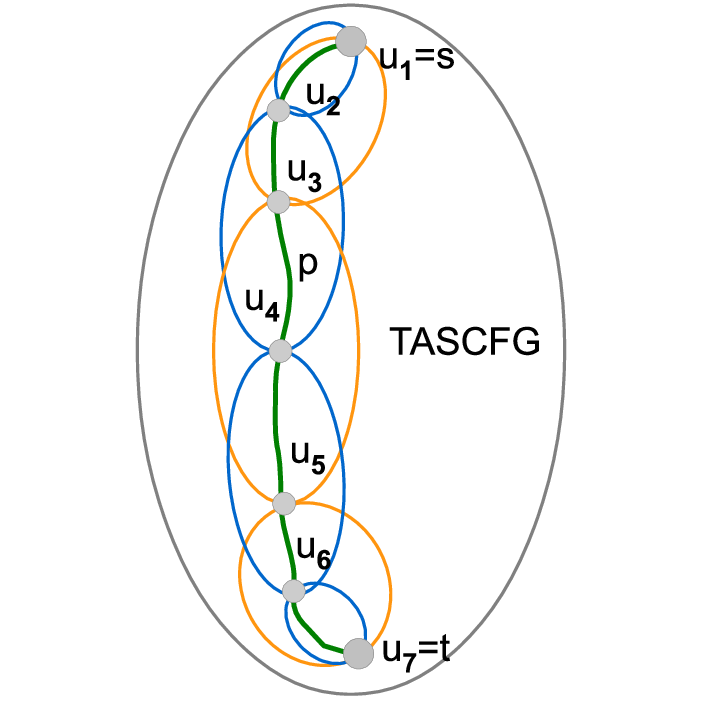}
\caption{An explanation of linear program \eqref{Eq:TCPEPLPDef}.}
\label{Fig:DetIfExistsPathLP}
\end{figure}
%
%
\mysectionc
{Deadlock configurations}
{Deadlock configurations}
\thickspace
Let's consider the solutions of linear program without linear equations 8--17.
Let node $u\in V$ and set $W\subseteq ConnPairElems$; let $F\langle K\rangle_{i}[u]>0$ for each $i\in W$.
\begin{definition}
Edge $e=(u,v)$ in graph $TArbSeqCFG$ is said to be $$KSet\langle p,W,i,out\rangle\text{-edge}$$
in the graph if
\begin{enumerate}
\item[1)]
{path $$p=(s,\ldots,u,v,\ldots,t)\in P$$ for some path set
$$P\in PathSets\langle \mathcal{PF}\langle G\rangle\rangle,$$}
\item[2)]
{$u\in \overset{\longleftrightarrow}{G_i}$ and $u\in \overset{\longleftrightarrow}{G_j}$ for some $i,j\in W$, $i\ne j$,
}
\item[3)]
{$$((F\langle \overset{\longleftrightarrow}{K}\rangle_{i}[u]>0)\wedge
(F\langle \overset{\longleftrightarrow}{K}\rangle_{j}[u]>0)),$$ and}
\item[4)]
{$$H\langle \overset{\longleftrightarrow}{K}\rangle_{i}[(u,v)]=0.$$}
\end{enumerate}
\end{definition}
\begin{definition}
Edge $e=(u,v)$ in graph $TArbSeqCFG$ is said to be $$KSet\langle p,W,in\rangle\text{-edge}$$
in the graph if
\begin{enumerate}
\item[1)]
{path $$p=(s,\ldots,u,v,\ldots,t)\in P$$ for some path set
$$P\in PathSets\langle \mathcal{PF}\langle G\rangle\rangle,$$}
\item[2)]
{$u\in \overset{\longleftrightarrow}{G_i}$ for each $i\in W$,
}
\item[3)]
{$$F\langle \overset{\longleftrightarrow}{K}\rangle_{i}[u]>0$$ for each $i\in W$, and}
\item[4)]
{$$H\langle \overset{\longleftrightarrow}{K}\rangle_{i}[(u,v)]>0$$ for each $i\in W$.}
\end{enumerate}
\end{definition}
\begin{definition}
Node $u$ in graph $TArbSeqCFG$ is said to be a deadlock node for $s$-$t$ path $p$ in graph
$TArbSeqCFG$ if each edge $e\in \delta^{+}(u)$ is $$KSet\langle p,W,i,out\rangle\text{-edge}$$
for some $i\in W$.
\end{definition}
Let's note that if there exists an optimal solution of linear program \eqref{Eq:TCPEPLPDef} then
network flow $\mathcal{PF}\langle G\rangle$ is a non-empty network flow and in fact a path flow
according to proposition \ref{Prop:AnyNetwFlowIsPathflow}.

Finding path flow $\mathcal{PF}\langle G\rangle$, which is defined by linear equations 1--7 of
linear program , can be insufficient (to determine if there exists a
tape-consistent path) in the following sense: There exist graphs $TArbSeqCFG$ such that every path set
$$P\in PathSets\langle\mathcal{PF}\langle G\rangle\rangle$$ contains tape-inconsistent
paths only.

That case is showed in Figure \ref{Fig:KSetOutEdge}; there
\begin{enumerate}
\item[1)]
{$i_1,j_1\in CommSeg_{\zeta_1}$ for some $\zeta_1 \in TapeSeg$,}
\item[2)]
{$i_2,j_2\in CommSeg_{\zeta_2}$ for some $\zeta_2 \in TapeSeg$,}
\item[3)]
{$i_1,i_2\in W$,}
\item[4)]
{$$p_1=(s,\ldots,s_{i_1},\ldots,t_{j_1},\ldots,t)$$ and
$$p_2=(s,\ldots,s_{j_1},\ldots,t_{i_1},\ldots,t)$$ are paths from a path set $P$
wherein $P\in PathSets\langle \mathcal{PF}\langle G\rangle\rangle$, 
}
\item[5)]
{$KSet\langle p_1,W,i_1,out\rangle\text{-edge}$ and $KSet\langle p_2,W,i_2,out\rangle\text{-edge}$
are red colored,
}
\item[6)]
{path flow $\mathcal{PF}\langle G\rangle$ values are blue colored,}
\item[7)]
{$p_1$ and $p_2$ are the only $s$-$t$ paths containing node $u$, and}
\item[8)]
{$u$ is a deadlock node for paths $p_1$ and $p_2$.
}
\end{enumerate}
\begin{figure}
\centering
\includegraphics[height=6cm]{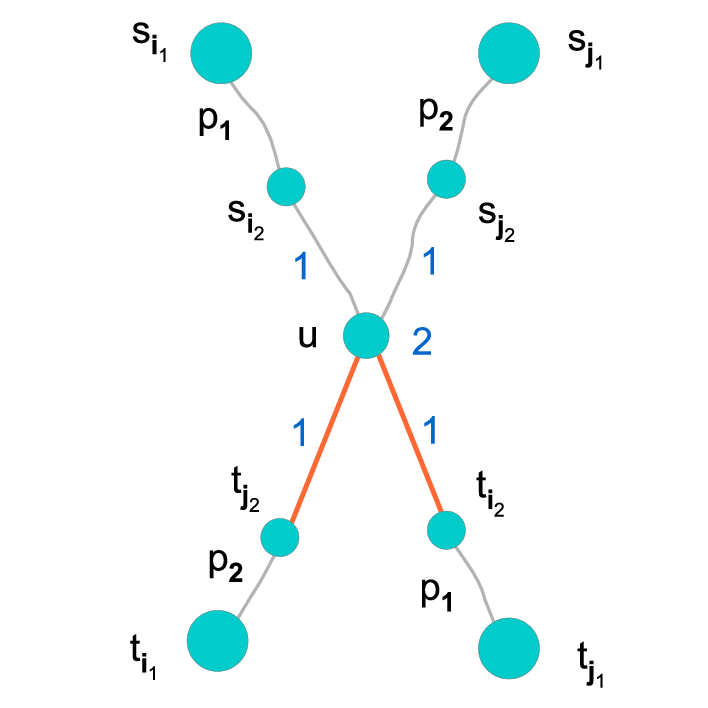}
\caption{No tape-consistent path contains node $u$.}
\label{Fig:KSetOutEdge}
\end{figure}
In the example, no tape-consistent path contains node $u$ because $u$ is a deadlock node
for path $p_1$ and $p_2$; so, if every $s$-$t$ path in graph $TArbSeqCFG$ contains node $u$,
one cannot conclude there is a tape-consistent path though a solution of linear program
\eqref{Eq:TCPEPLPDef} exists.
%
%
\mysectionc
{Intersection graphs $IG_{i,j}$}
{Intersection graphs $IG_{i,j}$}
\label{Par:IntersectGraphs}
\thickspace
Let's construct the following graphs by graphs
$\left(\overset{\longleftrightarrow}{G_i}\cap \overset{\longleftrightarrow}{G_j}\right)$
wherein $(i,j)\in CPCPairSet$:
\begin{enumerate}
\item[1)]
{all the nodes of the initial graph $\left(\overset{\longleftrightarrow}{G_i}\cap \overset{\longleftrightarrow}{G_j}\right)$,
excepting $s$ and $t$, that have no path to $s$ or $t$ are removed (iteratively, while there are such nodes in the graphs);
}
\item[2)]
{as usual, network flows for new graphs (linear equations \eqref{Eq:NetwFlowEqs1} and \eqref{Eq:NetwFlowEqs2}) are introduced.}
\end{enumerate}
\begin{notation}
Intersection graphs, constructed by graphs $\left(\overset{\longleftrightarrow}{G_i}\cap \overset{\longleftrightarrow}{G_j}\right)$
in such a way, are called by intersection graphs and denoted by $$IG_{i,j}.$$
\end{notation}
\begin{notation}
Flows in graphs $IG_{i,j}$ with properties {\normalfont 1} and {\normalfont 2} are called by intersection flows in graphs $IG_{i,j}$
and denoted by $$\mathcal{IGF}\langle K\rangle_{i,j}=(F\langle IGF\rangle_{i,j},H\langle IGF\rangle_{i,j}).$$
\end{notation}
Let's note,
\begin{enumerate}
\item[1)]
{intersection network flows $\mathcal{IGF}\langle K\rangle_{i,j}$ are empty for disconnected graphs $IG_{i,j}$;}
\item[2)]
{graphs $\left(\overset{\longleftrightarrow}{G_i}\cap \overset{\longleftrightarrow}{G_j}\right)$
and $IG_{i,j}$ are also $2$-out-regular graphs.}
\end{enumerate}
%
%
\mysectionc
{Elimination of deadlock configurations}
{Elimination of deadlock configurations}
\thickspace
%
%
To treat the case of deadlock configurations, that is to determine if there exists a
tape-consistent path in graph $TArbSeqCFG$ using linear program formulation,
one does as follows:
\begin{enumerate}
\item[1)]
{the initial control flow graph $TArbSeqCFG$ is transformed to $2$-out-regular graph;
}
\item[2)]
{set of linear equations 8--17 is added to set of linear equations $TCPEPLPEqSet$.
}
\end{enumerate}
\begin{lemma}
\label{Lem:NoKoutEdge}
Let's consider an optimal solution of linear program \notion{TCPEPLP}. Let $p$ be a $s$-$t$ path in
graph $TArbSeqCFG$, node $u\in p$, and set $W\subseteq ConnPairElems$; let 
\begin{align}
\label{Eq:NoKoutEdge}
((F\langle \overset{\longleftrightarrow}{K}\rangle_{i}[u]>0)\wedge(F\langle IGF\rangle_{i,j}[u]>0))
\end{align}
for each $i,j\in W$. In that case, there exists an edge $e\in \delta^{+}(u)$ such that it is
$KSet\langle p,W,in\rangle\text{-edge}$.
\begin{proof}
Let's suppose there is no such edge. Let edges $e_a,e_b\in \delta^{+}(u)$ be
$$KSet\langle p,W,i,out\rangle\text{-edge}$$ and $$KSet\langle p,W,j,out\rangle\text{-edge}$$
accordingly for some $i,j\in W$, $i\ne j$. In that case,
\begin{align*}
H\langle \overset{\longleftrightarrow}{K}\rangle_{i}[e_a]=0,\quad
H\langle \overset{\longleftrightarrow}{K}\rangle_{j}[e_b]=0;
\end{align*}
therefore,
\begin{align*}
F\langle IGF\rangle_{i,j}[s]=0
\end{align*}
would hold because there are only at most two out edges for node $u$. This is a contradition to
linear equations 10--12 of linear program \eqref{Eq:TCPEPLPDef}.
\end{proof}
\end{lemma}
In other words, there are no deadlock nodes for any optimal solution of linear program
\eqref{Eq:TCPEPLPDef} if equations \eqref{Eq:NoKoutEdge} hold.

%
The essensial of lemma \ref{Lem:NoKoutEdge} is explained in Figures \ref{Fig:NoKoutEdge2}--\ref{Fig:KoutEdge3}.
In Figure \ref{Fig:NoKoutEdge2},
\begin{enumerate}
\item[1)]
{nodes and edges of graphs $G_i$ and $G_j$ are orange-colored;
}
\item[2)]
{network flow values $\mathcal{PF}\langle \overset{\longleftrightarrow}{K}\rangle_i$ are green-colored;}
\item[3)]
{network flow values $\mathcal{PF}\langle \overset{\longleftrightarrow}{K}\rangle_j$ are blue-colored;}
\item[4)]
{network flow values $\mathcal{IGF}\langle K\rangle_{i,j}$ are gray-colored;}
\item[5)]
{edges $e_a,e_b\in \overset{\longleftrightarrow}{E_i}$ and $e_a,e_b\in \overset{\longleftrightarrow}{E_j}$;
}
\item[6)]
{$ind$ is an index of commodity; $ind=i$ or $ind=j$;}
\item[7)]
{$F\langle IGF\rangle_{i,j}[u]>0$;}
\item[8)]
{$H\langle IGF\rangle_{i,j}[e_a]=0$ and $H\langle IGF\rangle_{i,j}[e_b]=0$.}
\end{enumerate}
\begin{figure}
\centering
\includegraphics[height=5.5cm]{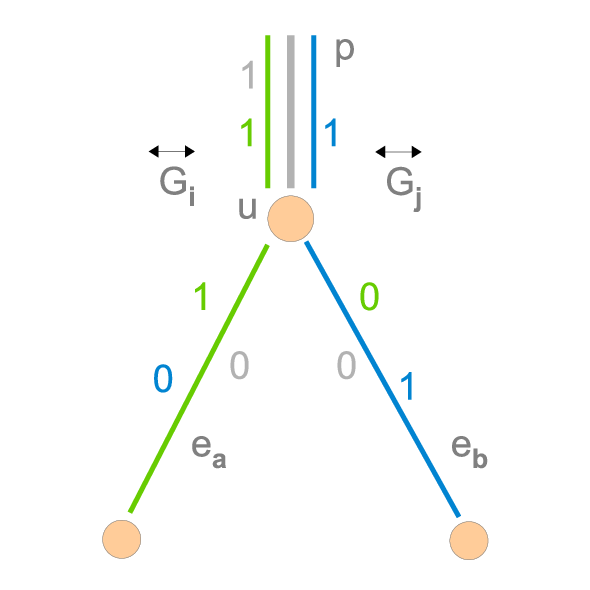}
\caption{There cannot be $KSet\langle p,W,ind,out\rangle\text{-edges}$ in case $|\delta^{+}(u)|\le 2$.}
\label{Fig:NoKoutEdge2}
\end{figure}
There cannot be both $KSet\langle p,W,i,out\rangle\text{-edge}$ and $KSet\langle p,W,j,out\rangle\text{-edge}$ because
$F\langle IGF\rangle_{i,j}[u]>0$, and one of the following should hold:
$H\langle IGF\rangle_{i,j}[e_a]>0$ or $H\langle IGF\rangle_{i,j}[e_b]>0$.

%
But there should be $KSet\langle p,W,in\rangle\text{-edge}$ as explained in Figure \ref{Fig:KinEdge2}; there
\begin{enumerate}
\item[1)]
{nodes and edges of graphs $G_i$ and $G_j$ are orange-colored;
}
\item[2)]
{network flow values $\mathcal{PF}\langle \overset{\longleftrightarrow}{K}\rangle_i$ are green-colored;}
\item[3)]
{network flow values $\mathcal{PF}\langle \overset{\longleftrightarrow}{K}\rangle_j$ are blue-colored;}
\item[4)]
{network flow values $\mathcal{IGF}\langle K\rangle_{i,j}$ are gray-colored;}
\item[5)]
{edges $e_a,e_b\in \overset{\longleftrightarrow}{E_i}$ and $e_a,e_b\in \overset{\longleftrightarrow}{E_j}$;
}
\item[6)]
{$H\langle \overset{\longleftrightarrow}{K}\rangle_{i}[e_a]>0$ and
$H\langle \overset{\longleftrightarrow}{K}\rangle_{j}[e_a]>0$;}
\item[7)]
{$H\langle \overset{\longleftrightarrow}{K}\rangle_{i}[e_b]=0$ and
$H\langle \overset{\longleftrightarrow}{K}\rangle_{j}[e_b]=0$;}
\item[8)]
{$F\langle IGF\rangle_{i,j}[u]>0$;}   
\item[9)]
{$H\langle IGF\rangle_{i,j}[e_a]>0$ and $H\langle IGF\rangle_{i,j}[e_b]=0$;}
\item[10)]
{edge $e_a=(u,v)$ is $KSet\langle p,W,in\rangle\text{-edge}$.
}
\end{enumerate}
\begin{figure}
\centering
\includegraphics[height=5.5cm]{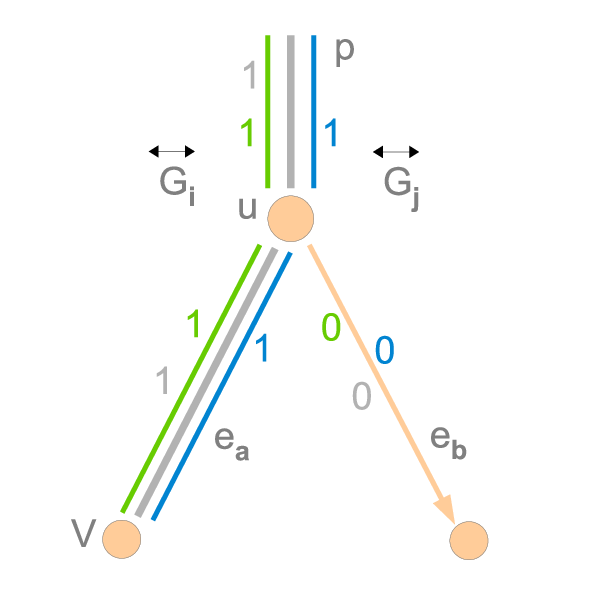}
\caption{There is $KSet\langle p,W,in\rangle\text{-edge}$, $e_a$, in case $|\delta^{+}(u)|\le 2$.}
\label{Fig:KinEdge2}
\end{figure}
%
%
Let's note that in case $|\delta^{+}(u)|\ge 3$, there can be $KSet\langle p,W,i,out\rangle\text{-edge}$
even if all the conditions of lemma \ref{Lem:NoKoutEdge} hold. It is explained in Figure \ref{Fig:KoutEdge3};
there
\begin{enumerate}
\item[1)]
{nodes and edges of graphs $G_i$, $G_j$, and $G_k$ are orange-colored;
}
\item[2)]
{network flow values $\mathcal{PF}\langle \overset{\longleftrightarrow}{K}\rangle_i$ are brown-colored;}
\item[3)]
{network flow values $\mathcal{PF}\langle \overset{\longleftrightarrow}{K}\rangle_j$ are green-colored;}
\item[4)]
{network flow values $\mathcal{PF}\langle \overset{\longleftrightarrow}{K}\rangle_k$ are blue-colored;}
\item[5)]
{network flow values $\mathcal{IGF}\langle K\rangle_{i,j}$ are gray-colored;}
\item[6)]
{edges $e_a,e_b,e_c\in \overset{\longleftrightarrow}{E_i}$,
$e_a,e_b,e_c\in \overset{\longleftrightarrow}{E_j}$, and
$e_a,e_b,e_c\in \overset{\longleftrightarrow}{E_k}$;
}
\item[7)]
{$H\langle \overset{\longleftrightarrow}{K}\rangle_{i}[e_b]=0$ and
$H\langle \overset{\longleftrightarrow}{K}\rangle_{j}[e_a]=0$;}
\item[8)]
{$F\langle IGF\rangle_{i,j}[u]>0$;}
\item[9)]
{$H\langle IGF\rangle_{i,j}[e_a]=0$, $H\langle IGF\rangle_{i,j}[e_b]=0$, and
$H\langle IGF\rangle_{i,j}[e_c]>0$;}
\item[10)]
{edge $e_b=(u,v)$ is $KSet\langle p,W,i,out\rangle\text{-edge}$.
}
\end{enumerate}
\begin{figure}
\centering
\includegraphics[height=6cm]{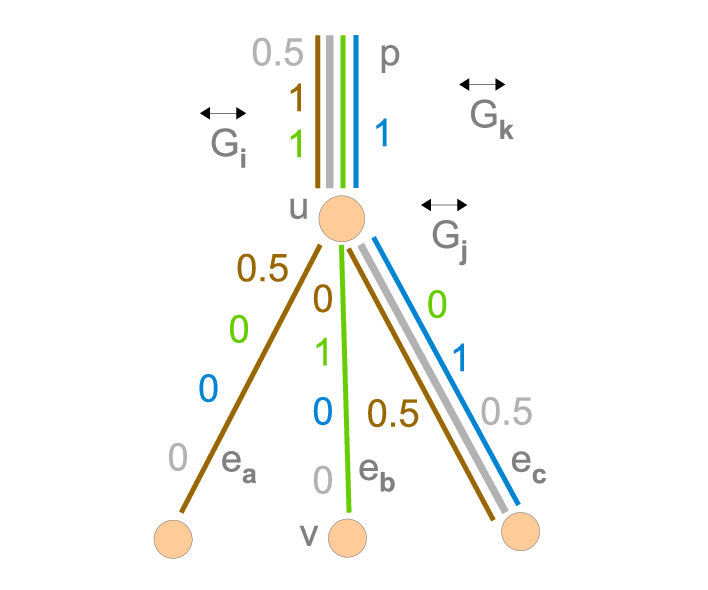}
\caption{There is $KSet\langle p,W,i,out\rangle\text{-edge}$ in case $|\delta^{+}(u)|\ge 3$.}
\label{Fig:KoutEdge3}
\end{figure}
%
%
So, it is important for the proof of proposition \ref{Prop:TCPEPLPProof} that 
$TArbSeqCFG$ is a $2$-out-regular graph ($|\delta^{+}(u)|\le 2$ for each node $u\in V$, $u\ne t$).
%
%
\mysectionc
{Elimination of intersection flow inconsistency}
{Elimination of intersection flow inconsistency}
\label{Par:IntersectEqs}
\thickspace
Let's take a look at configuration of flows in graphs $\overset{\longleftrightarrow}{G_i}$ and graphs
$IG_{i,j}$ as described in Figure \ref{Fig:NoKoutInconsist}; there
\begin{enumerate}
\item[1)]
{nodes and edges of graphs $G_i$ and $G_j$ are orange-colored;}
\item[2)]
{$\delta^{+}(u)=\{e_a,e_b\}$;}
\item[3)]
{network flow values $\mathcal{PF}\langle \overset{\longleftrightarrow}{K}\rangle_i$ are green-colored;}
\item[4)]
{network flow values $\mathcal{PF}\langle \overset{\longleftrightarrow}{K}\rangle_j$ are blue-colored;}
\item[5)]
{network flow values $\mathcal{IGF}\langle K\rangle_{i,j}$ are gray-colored;}
\item[6)]
{$F\langle IGF\rangle_{i,j}[u]>0$;}       
\item[7)]
{$H\langle \overset{\longleftrightarrow}{K}\rangle_{i}[e_a]>0$ and       
$H\langle \overset{\longleftrightarrow}{K}\rangle_{j}[e_a]>0$;}       
\item[8)]
{$H\langle \overset{\longleftrightarrow}{K}\rangle_{j}[e_b]>0$ and
$H\langle \overset{\longleftrightarrow}{K}\rangle_{j}[e_b]>0$;}       
\item[9)]
{$e_a$ is $KSet\langle p,W,in\rangle\text{-edge}$;}       
\item[10)]
{$H\langle IGF \rangle_{i,j}[e_a]=0$ and
$H\langle IGF\rangle_{i,j}[e_b]>0$.}       
\end{enumerate}
Here the issue is as follows: $e_a$ is $$KSet\langle p,W,in\rangle\text{-edge}$$ but $$H\langle IGF \rangle_{i,j}[e_a]=0.$$
\begin{figure}
\centering
\includegraphics[height=5.5cm]{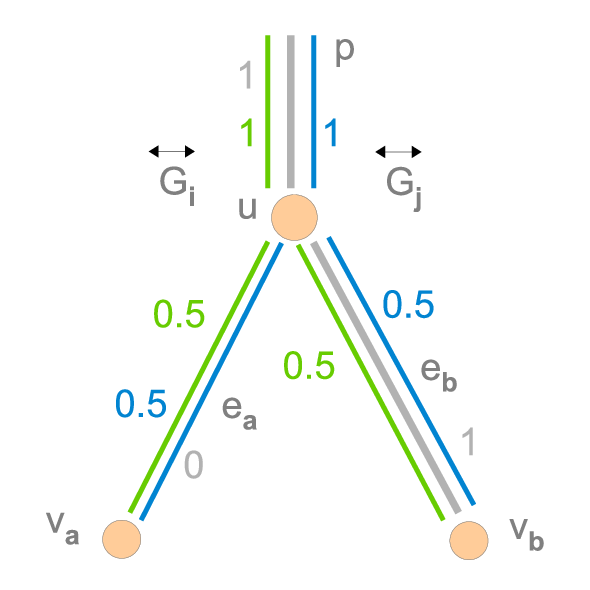}
\caption{Inconsistency in commodity flows and intersection flows.}
\label{Fig:NoKoutInconsist}
\end{figure}
\begin{notation}
Such flow configuration in graphs $\overset{\longleftrightarrow}{G_i}$, $\overset{\longleftrightarrow}{G_j}$, and graph
$IG_{i,j}$ is called by intersection flow inconsistency.
\end{notation}
One needs to eliminate intersection flow inconsistency (for proof of proposition \ref{Prop:TCPEPLPProof});
it means one needs to be sure that $$H\langle IGF\rangle_{i,j}[e]>0$$ for each
pair $(i,j)\in (W\times W)$ for any edge $e\in V$ that is $$KSet\langle p,W,in\rangle\text{-edge}$$
(for edge $e_a$ on Figure \ref{Fig:NoKoutInconsist}).
In order to meet this condition, let's introduce set of linear equations, denoted by
$$IntersectLPEqSet\langle (i,j),u\rangle,$$ for each pair $(i,j)\in CPCPairSet$ and each node
$u\in IG_{i,j}$, $u\ne t$, as follows:
\begin{enumerate}
\item[1)]
{linear equations
\begin{align*}
\alpha_1 = \frac{1}{2}\left(F\langle \overset{\longleftrightarrow}{K}\rangle_{i}[u] +
F\langle \overset{\longleftrightarrow}{K}\rangle_{j}[u]\right);
\end{align*}
}
\item[2)]
{linear equations
\begin{align*}
\beta_1 = \alpha_1 - F\langle IGF\rangle_{i,j}[u];
\end{align*}
}
\item[3)]
{linear equations
\begin{align*}
\alpha_2 = \frac{1}{2}\left(F\langle \overset{\longleftrightarrow}{K}\rangle_{i}[v_a] +
F\langle \overset{\longleftrightarrow}{K}\rangle_{j}[v_a]\right);
\end{align*}
}
\item[4)]
{linear equations
\begin{align*}
\beta_2 = \alpha_2 - F\langle IGF\rangle_{i,j}[v_a];
\end{align*}
}
\item[5)]
{linear equations
\begin{align*}
\beta_1 = \beta_2;
\end{align*}
}
\item[6)]
{linear equations
\begin{align*}
\alpha_3 = \frac{1}{2}\left(F\langle \overset{\longleftrightarrow}{K}\rangle_{i}[v_b] +
F\langle \overset{\longleftrightarrow}{K}\rangle_{j}[v_b]\right);
\end{align*}
}
\item[7)]
{linear equations
\begin{align*}
\beta_3 = \alpha_3 - F\langle IGF\rangle_{i,j}[v_b];
\end{align*}
}
\item[8)]
{linear equations
\begin{align*}
\beta_1 = \beta_3.
\end{align*}
}
\end{enumerate}
\begin{lemma}
\label{Lem:NoIntersectInconsist}
There is no intersection flow inconsistency if graphs $IG_{i,j}$, intersection flows $\mathcal{IGF}\langle K\rangle_{i,j}$, and
set of linear equations $IntersectLPEqSet\langle (i,j),u\rangle$ are additionally used in the definition of linear program
\notion{TCPEPLP}.
\begin{proof}
Let's suppose that $F\langle IGF\rangle_{i,j}[v_a] = 0$; in that case,
$$F\langle IGF\rangle_{i,j}[u] = F\langle IGF\rangle_{i,j}[v_b].$$ It means that
\begin{align*}
\beta_3 = \alpha_3 - F\langle IGF\rangle_{i,j}[u];
\end{align*}
at that,
\begin{align*}
\beta_1 = \alpha_1 - F\langle IGF\rangle_{i,j}[u],
\end{align*}
\begin{align*}
\beta_1 = \beta_3,\quad\text{and}\quad \alpha_1 = \alpha_2 + \alpha_3
\end{align*}
(because $\mathcal{PF}\langle \overset{\longleftrightarrow}{K}\rangle_i$ and
$\mathcal{PF}\langle \overset{\longleftrightarrow}{K}\rangle_j$ are network flows).
So, $\alpha_3 = \alpha_1$, and $\alpha_2=0$.

As a result, one gets
\begin{align*}
(F\langle IGF\rangle_{i,j}[v_a] = 0) \Rightarrow (\alpha_2=0),
\end{align*}
which is logically equivalent to
\begin{align*}
(\alpha_2 > 0) \Rightarrow (F\langle IGF\rangle_{i,j}[v_a] > 0).
\end{align*}
So, we get                                             
\begin{align*}
&((F\langle IGF\rangle_{i,j}[u]>0)\wedge
((F\langle \overset{\longleftrightarrow}{K}\rangle_{i}[v_a]>0) \vee
(F\langle \overset{\longleftrightarrow}{K}\rangle_{j}[v_a]>0))) \Rightarrow \\
&\quad (F\langle IGF\rangle_{i,j}[v_a] > 0);
\end{align*}
the same for node $v_b$.
\end{proof}
\end{lemma}
Moreover regarding lemma \ref{Lem:NoIntersectInconsist}, the following hold:
\begin{align*}
((H\langle IGF\rangle_{i,j}[e_a]>0) \Rightarrow
((H\langle \overset{\longleftrightarrow}{K}\rangle_{i}[e_a]>0) \wedge
(H\langle \overset{\longleftrightarrow}{K}\rangle_{j}[e_a]>0))
\end{align*}
due to linear equations 12 of linear program \eqref{Eq:TCPEPLPDef}; the same for edge $e_b$.
%
%
\begin{lemma}
\label{Lem:NoIntersectInconsistExists}
There exists the following solution of set of linear equations $$IntersectLPEqSet\langle (i,j),u\rangle:$$
\begin{enumerate}
\item[1)]
{$\alpha_1=\alpha_2$;
}
\item[2)]
{$\beta_1=0$ and $\beta_2=0$;
}
\item[3)]
{
$\alpha_3 = 0$ and $F\langle IGF\rangle_{i,j}[v_b] = 0$;
}
\item[4)]
{
$\beta_3 = 0$.
}
\end{enumerate}
\end{lemma}
It means that all the flows $\mathcal{PF}\langle \overset{\longleftrightarrow}{K}\rangle_i$ and
$\mathcal{IGF}\langle K\rangle_{i,j}$ are in edge $e_a$; the symmetric lemma can be formulated
for edge $e_b$.
%
%
\mysectionc
{Solutions of the linear program}
{Solutions of the linear program}
\thickspace
\label{Par:TCPEPLPSol}
%
%
\begin{proposition}
\label{Prop:TCPEPLPProof}
There exists a tape-consistent path in graph $TArbSeqCFG$ iff there exists an optimal solution of
linear program \notion{TCPEPLP}.
\begin{proof}
($\Rightarrow$). Let $p$ is a tape-consistent path. Let set $$B\subseteq CommSegPairs$$ be the set
of pairs $(i,j)$ such that $i\in K\langle p\rangle$, $j\in K\langle p\rangle$, and $i\ne j$;
in that case, $$B\subseteq CPCPairSet.$$ Let
\begin{enumerate}
\item[1)]
{for every $i\in K\langle p\rangle$, $$\mathcal{PF}\langle K\rangle_i=\mathcal{PF}\langle p',1\rangle$$
wherein $\left(p'=p\cap V_i\right)$;
}
\item[2)]
{for every $i\notin K\langle p\rangle$, $\mathcal{PF}\langle K\rangle_i$ is an empty network flow;
}
\item[3)]
{$F\langle IGF\rangle_{i,j}[s]=1$ for each pair $(i,j)\in B$;
}
\item[4)]
{$F\langle IGF\rangle_{i,j}[s]=0$ for each pair $(i,j)\notin B$;
}
\item[5)]
{network flow in graphs $ConnG\langle h',h''\rangle$:
\begin{enumerate}
\item[5.1)]
{$A_{i}=1$, $B_{j}=1$, and $C_{(i,j)}=1$ for each pair $(i,j)\in B$;}
\item[5.2)]
{$A_{i}=0$ and $C_{(i,j)}=0$ for each $i\notin K\langle p\rangle$;}
\item[5.3)]
{$B_{j}=0$ and $C_{(i,j)}=0$ for each $j\notin K\langle p\rangle$;}
\end{enumerate}
}
\end{enumerate}
(here $i,j\in CommSeg\langle TapeSeg\rangle$). At that, this configuration of network flows is consistent
with lemma \ref{Lem:NoIntersectInconsistExists}.

In that case, $$\mathcal{PF}\langle G\rangle=\mathcal{PF}\langle p,1\rangle$$ and
all the constraints of linear program \eqref{Eq:TCPEPLPDef} are satisfied; in particular,
all the linear equations from set $ConnLPEqSet\langle h',h''\rangle$
hold due to proposition \ref{Prop:ConnEqsProof2}.

($\Leftarrow$). Because linear equations 1--7 of linear program \eqref{Eq:TCPEPLPDef} hold,
one can take an integer $\theta\langle h\rangle$ from set $ConnElemSet\langle h\rangle$ for
$h\in HSeg$ such that
\begin{align*}
F\langle K\rangle_{i}[s_i]\ge \xi_{h}
\end{align*}
wherein $i=\theta\langle h\rangle$ ($\xi_{h}$ is defined like $\xi_{h'}$ and $\xi_{h''}$ are defined in paragraph
\ref{Par:ConnGEqs}); due to linear equations 6 of linear program \eqref{Eq:TCPEPLPDef}, such $\theta\langle h\rangle$
can be taken for each $h\in HSeg$.

Let $W$ is the set of such integers $\theta\langle h\rangle$; let's consider formulas
\begin{align}
\label{Eq:TCPEPLPProofEqFOvrGreater}
((F\langle \overset{\longleftrightarrow}{K}\rangle_{i}[u]\ge \xi_{h'})\wedge
(F\langle \overset{\longleftrightarrow}{K}\rangle_{j}[u]\ge \xi_{h''}))
\end{align}
and
\begin{align}
\label{Eq:TCPEPLPProofEqFOvrFollow}
(((F\langle \overset{\longleftrightarrow}{K}\rangle_{i}[u]\ge \xi_{h'})\wedge
(F\langle \overset{\longleftrightarrow}{K}\rangle_{j}[u]\ge \xi_{h''}))
\Rightarrow (F\langle IGF\rangle_{i,j}[s] > 0))
\end{align}
wherein node $u\in V$, $i=\theta\langle h'\rangle\in W$, and $j=\theta\langle h''\rangle\in W$ ($\xi_{h'}$
and $\xi_{h''}$ are defined in paragraph \ref{Par:ConnGEqs}).

Because $\mathcal{F}\langle ConnG\langle h',h''\rangle\rangle$ is a $1$-$1$ network flow
in graph $ConnG\langle h',h''\rangle$, formula \eqref{Eq:TCPEPLPProofEqFOvrGreater} is true
for $u=s$ and for each pair $(h',h'')\in hhPairs$. Therefore, due to linear equations 8--17 of
linear program \eqref{Eq:TCPEPLPDef} and proposition \ref{Prop:ConnEqsProof1}, formula
\eqref{Eq:TCPEPLPProofEqFOvrFollow} is true for $u=s$ and for each pair $(h',h'')\in hhPairs$.

Let path set $$P\in PathSets\langle \mathcal{PF}\langle G\rangle\rangle;$$
let's proof that there exists a $s$-$t$ path $p$ in graph $TArbSeqCFG$ such that
$p\in P$ and $$p\in AllPaths\langle \overset{\longleftrightarrow}{G_i}\rangle$$ for each $i\in W$,
using mathematical induction by the length $len$ of $s$-subpath $p'$ of path $p$:
\begin{enumerate}
\item[1)]
{Base case: $len=1$. Formulas \eqref{Eq:TCPEPLPProofEqFOvrGreater} and
\eqref{Eq:TCPEPLPProofEqFOvrFollow} are true for node $u=s$ and for each
pair $(h',h'')\in hhPairs$. 
}
\item[2)]
{Inductive step. Let node $u'\in p$ and edge $e=(u',v')\in E$; because of lemma \ref{Lem:NoKoutEdge}
and linear equations 8--16 of linear program \eqref{Eq:TCPEPLPDef}), there exists an edge
$e\in \delta^{+}(u')$ such that it is $$KSet\langle p,W,in\rangle\text{-edge},$$
and formulas \eqref{Eq:TCPEPLPProofEqFOvrGreater} and \eqref{Eq:TCPEPLPProofEqFOvrFollow}
are true for node $u=v'$ and for each pair $(h',h'')\in hhPairs$. At that, due to lemma
\ref{Lem:NoIntersectInconsist}, $$H\langle IGF\rangle_{i,j}[e]>0$$
for each $i\in W$ and each $j\in W$.
}
\end{enumerate}
Let's proof that path $p$ is a $KSet_{\zeta}$-path for each $\zeta\in TapeSeg$ using
mathematical induction by the length $len$ of $s$-subpath $p'$ of path $p$ in graph
$G\langle tcon\rangle_{\zeta}$:
\begin{enumerate}
\item[1)]
{Base case: $len=1$. There exists commodity $K_i(s_i,t_i)$ such that $i\in CommSeg_{\zeta}$, $s_i=s$,
and $i\in W$ (linear equations 2--4 of linear program \eqref{Eq:TCPEPLPDef} and
linear equations for network flow $$\mathcal{F}\langle ConnG\langle h',h''\rangle\rangle$$
in graph $ConnG\langle h',h''\rangle$ wherein $h'=1$).
}
\item[2)]
{Inductive step. Let's consider commodities $K_i(s_i,t_i)$ and $K_j(s_j,t_j)$
such that $i\in CommSeg_{\zeta}$, $s_i\in VLevel_{h'}$, $t_i\in VLevel_{h''}$,
$s_j\in VLevel_{h''}$, $h',h''\in HSeg$, and $i,j\in W$. Network flow in graph
$IG_{i,j}$ is a non-empty network flow because formulas
\eqref{Eq:TCPEPLPProofEqFOvrGreater} and \eqref{Eq:TCPEPLPProofEqFOvrFollow} are true
for node $u=s_i$ and for pair $(h',h'')$. Therefore, $t_i=s_j$ should
hold.
}
\end{enumerate}
Due to linear equations 7 of linear program \eqref{Eq:TCPEPLPDef}, there are no
`orphan' commodities $K_i(s_i,t_i)$; therefore, there are no `orphan' commodities
$K_i(s_i,t_i)$ such that $\mathcal{PF}\langle K\rangle_i$ is non-empty network flow
and $i\in W$.

So, $p$ is a $KSet_{\zeta}$-path for each $\zeta\in TapeSeg$ and, therefore, $KSet$-path in
graph $TArbSeqCFG$; it means, $p$ is a tape-consistent path in graph $TArbSeqCFG$ due to
proposition \ref{Prop:TapeConsToKSet}.
\end{proof}
\end{proposition}
The proof of proposition \ref{Prop:TCPEPLPProof} is explained in Figure \ref{Fig:TCPEPLPPRoof};
there
\begin{enumerate}
\item[1)]
{$TASCFG$ is the shortened indication of $TArbSeqCFG$;}
\item[2)]
{network flows values $$\mathcal{IGF}\langle K\rangle_{i,j}$$ are orange-colored;}
\item[3)]
{tape-consistent path $p$ is green-colored.}       
\end{enumerate}
\begin{figure}
\centering
\includegraphics[height=9cm]{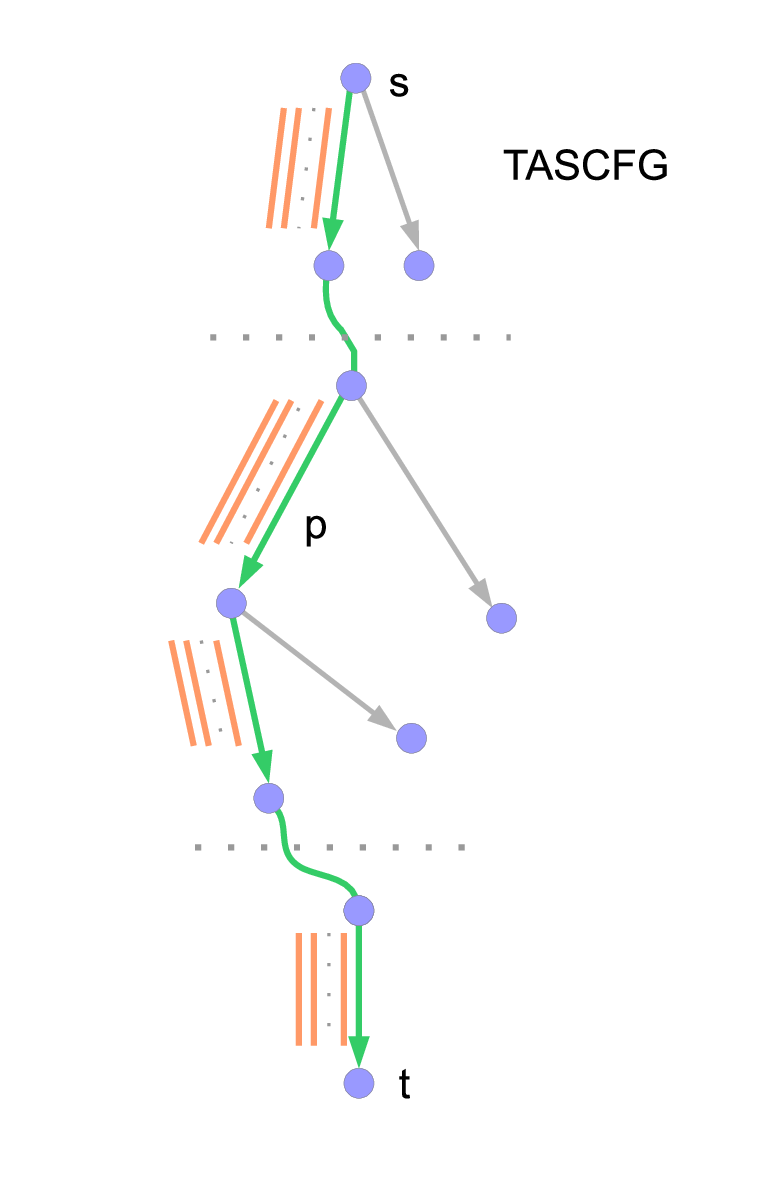}
\caption{Proof for linear program \notion{TCPEPLP}}
\label{Fig:TCPEPLPPRoof}
\end{figure}
%

%
\mysectionb
{Pseudocode of algorithm $DetermineIfExistsTConsistPath$}
{Pseudocode of algorithm $DetermineIfExistsTConsistPath$}
\thickspace
%
%
\begin{definition}
$$TapeLeft\langle \omega\rangle=\min\{\kappa^{(tape)}\ |\ t=(q,s,q',s',m,\kappa^{(tape)},
\kappa^{(step)})\in \omega\}$$ wherein $\omega$ is a sequence of the computation steps.
\end{definition}
\begin{definition}
$$TapeRight\langle \omega\rangle=\max\{\kappa^{(tape)}\ |\ t=(q,s,q',s',m,\kappa^{(tape)},
\kappa^{(step)})\in \omega\}$$ wherein $\omega$ is a sequence of the computation steps.
\end{definition}
Because $\mu$-length sequences of the computation steps are considered, there exists a $\mu$-length
tape-consistent sequence of the computation steps iff there exists $TapeSeg=[L..R]$ and
tape-consistent path $p$ such that
$$TapeLeft\langle\omega\langle \lfloor p\rfloor\rangle\rangle=L$$ and
$$TapeRight\langle\omega\langle \lfloor p\rfloor\rangle\rangle=R.$$ 

So, to find a tape-consistent path in graph $TArbSeqCFG$ it is sufficient to repeat solving
linear program \notion{TCPEPLP} for each integer segment $TapeSeg=[L..R]$ such that
$$TapeLBound\langle \mu\rangle\le L\le R\le TapeRBound\langle \mu\rangle.$$
%

\noindent\hrulefill
\begin{algorithm}
{$DetermineIfExistsTConsistPath$}
{}
\qinput  Graph $TArbSeqCFG$, $TConsistPairSet$
\qoutput If there exists a tape-consistent path in graph $TArbSeqCFG$\\
{$TArbSeqCFG:=Make2OutRegularGraph(TArbSeqCFG)$}\\
\\
\qfor each integer $L\in TapeRange\langle \mu\rangle$\\
\qdo\\
  \qfor each integer $R\in [L..TapeRBound\langle \mu\rangle]$\\
  \qdo\\
    {$TapeSeg:=[L..R]$}\\
    \\
    {compute the commodities}\\
    {remove `hiding' definitions in commodities}\\
    {compute graphs $ConnG\langle h',h''\rangle$}\\
    \\
    {compute constraints matrix $U$ of linear program \eqref{Eq:TCPEPLPDef}}\\
    {find an optimal solution of linear program \eqref{Eq:TCPEPLPDef}}\\
    \\
    \qif such optimal solution exists\\
    \qthen\\
        \qreturn $\mathbf{True}$
    \qfi\\
    \qcom{end of if}
  \qrof\\
  \qcom{end of for loop}
\qrof\\
\qcom{end of for loop}\\
\\
\qreturn $\mathbf{False}$
\end{algorithm}
\noindent\hrulefill

%
\begin{proposition}
The time complexity of deterministic algorithm $$DetermineIfExistsTConsistPath$$
is polynomial in $\mu$.
\begin{proof}
It follows from the following:
\begin{enumerate}
\item[1)]
{there exist polynomial time algorithms to solve problem \notion{LP} \cite{Kh80,K84};}
\item[2)]
{these algorithms can be applied for linear programs such that their polyhedrons are
not full-dimensional polyhedrons \cite{S98}.}
\end{enumerate}
\end{proof}
\end{proposition}
\begin{theorem}
Problem \notion{TCPE}, which is \notion{NP}-complete, is decidable in polynomial time.
\end{theorem}
%
%
\mysectionb
{Problem \notion{TCPE}$\langle 1\rangle$}
{Problem \notion{TCPE}$\langle 1\rangle$}
\thickspace
Let's consider reduction of problem \notion{TCPE} to special cases of problem \notion{CNF-SAT}.

The idea of the reduction is based on the using of Horn clauses to derive the properties
of imperative programs \cite{CL73}. Let's construct \notion{mixed-DHORN-CNF} (mixed dual HORN CNF;
similar to \notion{mixed-HORN-CNF} formulas) formula $SPF$ (Single Path Formula) as follows:
\begin{enumerate}
\item[1)]
{introduce logic variables $P_u$ for each node $u\in V$; $P_u=\mathbf{true}$ means
that $u\in p$ for a $s$-$t$ path $p$ in graph $TArbSeqCFG$;} 
\item[2)]
{$SPF=\bigwedge_{i\in [1,4]}F_i$ wherein
\begin{align*}
\begin{array}{ll}
\vspace{2pt}
(s\text{-}t\ \text{path})
&
F_1=\left(P_s\wedge P_t\right),\\
\vspace{2pt}
(\text{path})
&
F_2=\bigwedge_{v\in V}
\left(P_{u}\Rightarrow \left(\bigvee_{(u,v)\in \delta^{+}(u)}P_{v}\right)\right),\\
\vspace{2pt}
(\text{tape-consistent path})
&
F_3=\bigwedge_{(u,v)\in TConsistPairSet}
\left(P_{u}\Rightarrow P_{v}\right),\ \text{and}\\
\vspace{2pt}
(\text{single path})
&
F_4=\bigwedge_{u\in V} \left(
\bigwedge_{(u,v)\in\delta^{+}(u)}\left(P_{v}\Rightarrow\left(
\bigwedge_{(u,w)\in \delta^{+}(u),w\ne v}
\left(\neg P_{w}\right)\right)\right)\right).
\end{array}
\end{align*}
}
\end{enumerate}
\begin{definition}
Let's denote the problem of determining if formula $SPF$ is satisfiable by \notion{TCPE}$\langle 1\rangle$.
\end{definition}
So problem \notion{TCPE}$\langle 1\rangle$ is a special case of problem \notion{CNF-SAT};
in particular, problem \notion{TCPE}$\langle 1\rangle$ is a special case of problems
\notion{mixed-DHORN-SAT} and \notion{linear-CNF-SAT}. In fact, problem \notion{mixed-DHORN-SAT}
is an equivalent of problem \notion{mixed-HORN-SAT} (just replace literals with negative ones);
in general case, problems \notion{mixed-HORN-SAT} and \notion{linear-CNF-SAT} are
\notion{NP}-complete \cite{PS05,PSR05} and for now no algorithm is known to solve these
problems in polynomial time.

Problem \notion{TCPE}$\langle 1\rangle$ can be expressed by an integer linear program \notion{TCPEPLP} that is to find
path flow $\mathcal{PF}\langle p,1\rangle$ in graph $TArbSeqCFG$ (in that case, $p$ is a tape-consistent
path); so, there is reduction $\notioninform{TCPE}\langle 1\rangle\le_{m}^{P}\notioninform{ILP}$.

The reductions described in this subsection are showed in Figure \ref{Fig:TCPEPReductions}.
\begin{figure}
\centering
\includegraphics[height=4cm]{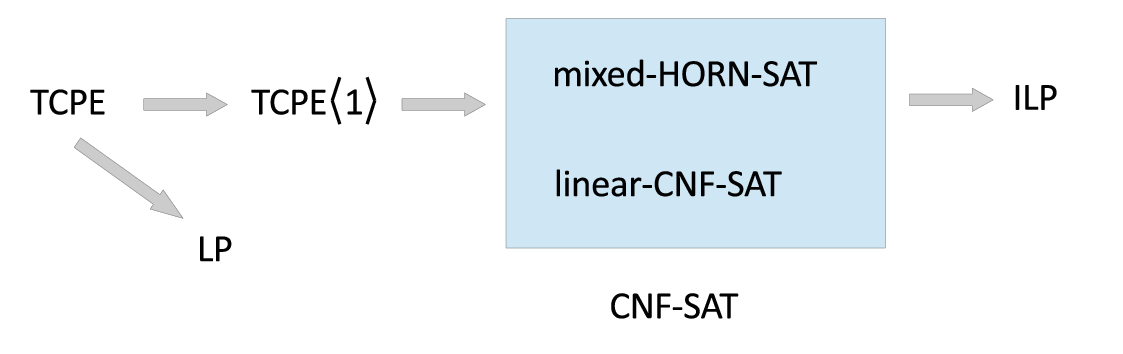}
\caption{Reductions of problem $TCPE$ to other problems.}
\label{Fig:TCPEPReductions}
\end{figure}
%

%
\mysectiona
{Main results}
{Main results}
\thickspace
In this section, machine $M\langle\exists AcceptingPath\rangle$ is used to
introduce the main results. Constant $\sigma$ (in propositions \ref{Prop:EveryLangNDisD})
is defined by equation \eqref{Eq:SigmaDef}.
%
%
\mysectionb
{Main theorem}
{Main theorem}
\thickspace
If $M$ is a deterministic multi-tape Turing machine that computes
a string function $f(x)$ and works in time $t(n)$, then one can construct
a deterministic single-tape Turing machine $M'$ that computes the
same function and works in time $\bigo\left(t(n)^2\right)$ \cite{DK00};
therefore, the following proposition holds.

\begin{theorem}
\label{Prop:EveryLangNDisD}
Every language in a finite alphabet that is decidable by a non-deterministic
single-tape Turing machine in time $t(n)$ is also decidable by a deterministic
single-tape Turing machine in time \overallcomplb, wherein $t(n)$ is an upper bound of the
time complexity of machine $M\langle NP\rangle$ and $\sigma$ is a constant depending on relation
$\Delta$ of machine $M\langle NP\rangle$.
\end{theorem}
If $t(n)$ is a polynomial, then machine $M\langle\exists AcceptingPath\rangle$
works in polynomial time in $n$ ($n=|x|$ wherein $x$ is the input of the machine);
therefore, the following main theorem holds.
\begin{theorem}
\label{Thr:PeqNP}
\begin{align*}
\notioninform{P}=\notioninform{NP}.
\end{align*}
\end{theorem}
%
%
\mysectionb
{Proof of \notion{FP}$\,=\,$\notion{FNP} based on machine $M\langle\exists AcceptingPath\rangle$}
{Proof of \notion{FP}$\,=\,$\notion{FNP} based on machine $M\langle\exists AcceptingPath\rangle$}
\label{SubSec:RetieveAcceptingPath}
\thickspace
Algorithm $RetieveAcceptingPath$ (defined in this subsection) is similar to the algorithm for problem \notion{FSAT} using an
algorithm for problem \notion{SAT}.

The possibility of finding an accepting computation path in an explicit way, using machine
$M\langle\exists AcceptingPath\rangle$, is consistent with the fact that
\notion{FP}$\,=\,$\notion{FNP} iff \notion{P}$\,=\,$\notion{NP} \cite{Papa94,DK00}.


\noindent\hrulefill
\begin{algorithm}
{$RetieveAcceptingPath$}
{}
\qinput  Graph $TArbSeqGFG$, set $TConsistPairSet$
\qoutput An accepting computation path of machine $M\langle NP\rangle$ on input $x$ or
         empty path it there are no accepting paths\\
\qcom{initialization}\\
{node $r:=s$ (the source node of the graph)}\\
{path $p:=(r)$}\\
\\
\qcom{check if there exists an accepting computation path}\\
\qcom{here $F$ is the set of the accepting states of machine $M\langle NP\rangle$}\\
{$\beta_F:=DetermineIfExistsTConsistPath(TArbSeqCFG,TConsistPairSet)$}\\
\qif $\neg (\beta_F)$\\
\qthen\\
  \qreturn (empty path $()$)
\qfi\\
\qcom{end of if}\\
\\
{store $TapeSeg$}\\
\qcom{$TapeSeg$ used in algorithm $DetermineIfExistsTConsistPath$}\\
\\
\qcom{main loop}\\
\qwhile \qtrue\\
\qdo\\
  \qif $length(p)>0$\\
  \qthen\\
    {let $p=(u_1,\ldots,u_m)$}\\
    \qif computation step $u_m.step$ contains a state $q\in F$\\
    \qthen\\
      {$p:=RemFakeNodes\langle p\rangle$}\\
      \qreturn $\omega\langle \lfloor p\rfloor\rangle$
    \qfi\\
    \qcom{end of if}
  \qfi\\
  \qcom{end of if}\\
  \\
  \qfor each edge $(r,u)\in \delta^{+}(r)$\\
  \qdo\\
    \qcom{here $t$ is the sink node of graph $TArbSeqGFG$}\\
    {let subgraph $G_u$ be $Subgraph\langle G,(u,t)\rangle$}\\
    \\
    {let path $p'$ be path $p$ concatenated with edge $(r,u)$}\\
    {let $G_{p,u}$ be subgraph $(p'\cup G_u)$}\\
    \\
    {$\beta_F:=DetermineIfExistsTConsistPath(G_{p,u},TConsistPairSet)$}\\
      {$\ $ for stored $TapeSeg$ wherein for each node $u'\in p'$ ($u'\ne s$) add linear}\\
      {$\ $ equation $F\langle G\rangle[u']=1$ to set of linear equations $TCPEPLPEqSet$}\\
      \qcom{let's by $TCPEPLPEqSet'$ denote the set of linear equations modified in this way}\\
    \\
    \qif $\beta_F$\\
    \qthen\\
      {add node $u$ to path $p$}\\
      {$r:=u$}\\
      \qbreak
    \qfi\\
    \qcom{end of if}
  \qrof\\
  \qcom{end of for loop}
\qend\\
\qcom{end of main loop}\\
\\
\qreturn (empty path $()$)
\end{algorithm}
\noindent\hrulefill

The construction of an accepting computation path by this algorithm is explained
in Figure \ref{Fig:RetrievePath} (there $TASG$ is the shortened indication of
$TArbSeqGFG$).
\begin{figure}
\centering
\includegraphics[height=6cm]{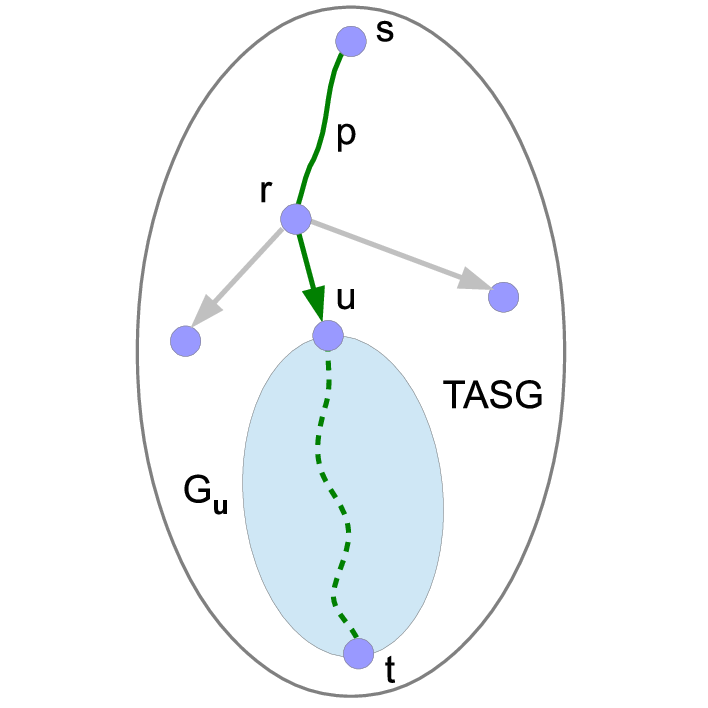}
\caption{Retrieving an accepting computation path.}
\label{Fig:RetrievePath}
\end{figure}
\begin{proposition}
Deterministic algorithm $RetieveAcceptingPath$ outputs an accepting computation path of
machine $M\langle NP\rangle$ on input $x$.
\begin{proof}
The proof is the same as the proof of proposition \ref{Prop:TCPEPLPProof}
if one considers set of linear equations $TCPEPLPEqSet'$ (defined at line 38 of algorithm
$RetieveAcceptingPath$).
\end{proof}
\end{proposition}
\begin{theorem}
For every language in a finite alphabet that is decidable by a non-deterministic
single-tape Turing machine $M\langle NP\rangle$ in time $t(n)$, an accepting computation
path of machine $M\langle NP\rangle$ on an input $x$ can be found by deterministic multi-tape
Turing machine in time \overallcomplc\ wherein $n=|x|$, $t(n)$ is an upper bound of the time complexity of
machine $M\langle NP\rangle$, and $\sigma$ is a constant depending on relation
$\Delta$ of machine $M\langle NP\rangle$.
\end{theorem}
%
%
\mysectionb
{Some consequences}
{Some consequences}
\thickspace
From the construction of machine $M\langle\exists AcceptingPath\rangle$,
one can conclude that the following proposition holds.
\begin{proposition}
Every language in a finite alphabet that is decidable by a non-deterministic
single-tape Turing machine in space $s(n)$ is also decidable by a deterministic
single-tape Turing machine in time $2^{\bigo(s(n))}$.
\end{proposition}
This result is also obtained by simulating non-deterministic computations on
a deterministic Turing machine \cite{DK00}.

In addition, proposition \ref{Prop:EveryLangNDisD} is consistent
with the fact that if \notion{P}$\,=\,$\notion{NP} then the following equality holds:
\notion{EXPTIME}$\,=\,$\notion{NEXPTIME} \cite{Papa94,DK00}.

One of the most important consequences of theorem \ref{Thr:PeqNP} is that
\notion{P}$\,=\,$\notion{PH}. 
%
%
\mysectiona
{Computer program to verify the results}
{Computer program to verify the results}
\thickspace
\notion{C\#} application to verify the results is developed using \notion{MS Visual Studio Express 2015}
and \notion{Wolfram Mathematica 9.0}; the application is available on the Internet at \cite{App}.

The solution contains the definitions of machines $M\langle NP\rangle$ and the implementations of corresponding
machines $M\langle \exists AcceptingPath\rangle$ for the following examples:
\begin{enumerate}
\item[1)]
{deterministic Turing machine that decides language 
\begin{align*}
L_1=\{ w1\ |\ w\in \{0,1\}^*\};
\end{align*}
}
\item[2)]
{non-deterministic Turing machine that decides language 
\begin{align*}
L_2=\{ x11z\ |\ x,z\in \{0,1\}^*\} \cup \{ x00z\ |\ x,z\in \{0,1\}^*\};
\end{align*}
}
\item[3)]
{non-deterministic Turing machine that decides language that has several accepting paths;
}
\item[4)]
{non-deterministic Turing machine that decides language that has a lot of accepting paths;
}
\item[5)]
{non-deterministic Turing machine that decides language from class \notion{UP}; an accepting
path of the computation tree of the machine may not exists;
}
\item[6)]
{non-deterministic Turing machine that decides language from class \notion{UP}; an accepting
path of the computation tree of the machine always exists; the machine has $2^n$ total amount of
computation paths on inputs with length n;
}
\item[7)]
{non-deterministic Turing machine that decides language from class \notion{UP};
the machine has large transition relation (approximately 900 elements);}
\item[8)]
{non-deterministic Turing machine for integer factorization (transition relation contains elements
with constant left (right) shift, with more than one size, just to decrease the length of the accepting path).}
\end{enumerate}
The application is tested on some inputs for these examples. Other examples can be easily added:
Just construct single-tape non-deterministic Turing machines $M\langle NP\rangle$ for them
(the algotithm for machines $M\langle \exists AcceptingPath\rangle$ is contained in the application).

Let's note that the solution contains the implementations of linear equations 1--7 only of
linear program \notion{TCPEPLP} because an implementation of the complete set of
the linear equations of linear program \notion{TCPEPLP} leads to too much count
of variables and equations to run ther application on modern computers. It turns out
that using linear equations 1--7 of linear program \notion{TCPEPLP} are sufficient
to run the application on small examples.


%
\mysectiona
{Conclusion}
{Conclusion}
\thinspace\thinspace
This paper presents the program of deterministic multi-tape Turing
machine $M\langle\exists AcceptingPath\rangle$ that determines in polynomial time 
if there exists an accepting computation path of polynomial time
non-deterministic single-tape Turing machine $M\langle NP\rangle$ that decides
a language $A$ over a finite alphabet (machine $M\langle \exists AcceptingPath\rangle$
is different for each machine $M\langle NP\rangle$). As a result, the equality of
classes \notion{P} and \notion{NP} is proved.

The computations presented in this paper are `not ordinary' computations in the sense that
the notion of tape-arbitrary sequences of the computation steps is used to determine if
there exists an accepting computation path of machine $M\langle NP\rangle$. The author of
this paper proposes denoting these computations by one of the following:
\begin{enumerate}
\item[1)]
{using-complement computations;}
\item[2)]
{superfluous computations.}
\end{enumerate}
But these computations is ordinary in the sense that the computations are preformed
by Turing machines; the computations presented in the present paper are not quantum
computations, for example, or other kind of computations which uses non-standard concepts.

The concept suggested in the present paper can be applied in some a way not only for
Turing machine programs but also for computer programs written on imperative
programming languages like Pascal or C\# if one adds non-deterministic commands
to such languages.
%



\end{document}